\documentclass[a4paper,12pt,leqno,tbtags,openany]{amsbook}
\pdfoutput=1
\usepackage[english]{babel}
\usepackage[square,comma,numbers,sort&compress]{natbib}
\usepackage{graphicx}
\usepackage{newcent,fouriernc}
\usepackage{hhline,xspace,threeparttable}
\usepackage{datetime}
\usepackage[usenames, dvipsnames]{color}
\usepackage[margin=10pt,font=footnotesize,labelfont=bf,figurewithin=chapter]{caption}
\usepackage{amssymb,mathtools}
\usepackage{amsmath}
\definecolor{myblue}{rgb}{0, 0.45, 0.906}
\usepackage{rotating,subfig}

\numberwithin{section}{chapter}
\numberwithin{equation}{chapter}
\numberwithin{table}{chapter}
\newcommand{\orc}{\mbox{$\ast\mathrm{d}$}}
\newcommand{\young}{\mbox{$_{\textsc{e}}$}}
\newcommand{\mean}{\mbox{$_\mathrm{g}$}}
\newcommand{\hea}[1]{\mbox{$\mbox{\rm H}(#1)$}}
\newcommand{\Hea}{\mbox{$\mbox{\rm H}$}}
\newcommand{\inv}[1]{\mbox{$#1^{\,\mathrm i}$}}
\newcommand{\bzero}[1]{\mbox{$\mbox{J}_\mathrm{o}(#1)$}}
\newcommand{\tmin}{\mbox{$\underline{t}\,$}}

\begin{document}

\frontmatter

\title{Load-depth sensing of isotropic, linear viscoelastic materials
 using rigid axisymmetric indenters}
 \markleft{P.G.TH. VAN DER VARST ET AL.}
\author{P. G. Th. van der Varst$^{\dag\ast}$}
\address{$^{\dag}$Assistant Professor (retired), Laboratory of Materials and Interface Chemistry}
\email{p.g.th.v.d.varst@xs4all.nl}
\author{A.A.F. van de Ven$^{\ddag\ast}$}
\address{$^{\ddag}$Associate Professor (retired), Department of Mathematics and Computing Science}
\email{A.A.F.v.d.Ven@tue.nl}
\author{G. de With$^{\S\ast}$}
\address{$^{\S}$ Professor (retired), Laboratory of Materials and Interface Chemistry}
\email{G.d.with@tue.nl}
\author{}
\address{$^\ast$Eindhoven University of Technology,
P.O. Box 513, 5600 MB Eindhoven, The Netherlands}



\begin{abstract}
An indentation experiment involves five variables: indenter shape, material behaviour of the substrate, contact size, applied load and indentation depth.   Only three variable are known afterwards, namely, indenter shape, plus load and depth as function of time during the experiment because the contact size is not measured and the determination of the material properties is the very aim of the test; two equations are needed to obtain a mathematically solvable system.

For elastic materials, one of these equations is the fixed -- fixed because it depends only on the indenter shape -- relation between current depth and contact size; this relation is  used to eliminate once and for all the contact size  in favor of the depth, thus yielding a \emph{single relation} between load, depth and material properties in which only the latter variable remains unknown.

For viscoelastic materials where hereditary integrals model the constitutive behaviour,  the relation between depth and contact size is much more complicated as it usually depends also on the (time-dependent) properties of the material that is investigated. Solving the inverse problem, i.e., determining the material properties from the experimental data, therefore needs both equations. Extending Sneddon's analysis of the indentation problem for elastic materials (I. N. Sneddon. \emph{Fourier transforms.} McGraw-Hill Book Company, Inc., New York, 1951, p. 450--457) to include viscoelastic materials, the two equations mentioned above are derived and analysed. To find the time dependence of the material properties the feasibility of  Golden and Graham's method of decomposing hereditary integrals (J.M. Golden and G.A.C. Graham. \emph{Boundary value problems in linear viscoelasticity}, Springer-Verlag, Berlin, 1988, p. 63--69)  is investigated and applied to two types of indentation processes. The first is a single load-unload process and the second is a sinusoidally driven stationary state. 
\end{abstract}

\pagenumbering{roman}
\maketitle
\thispagestyle{empty}

\setcounter{page}{4}
\setcounter{tocdepth}{4}
\tableofcontents
\chapter*{Extended summary}
Finding viscoelastic material properties from indentation experiments requires knowledge of the relation between five functions: the experimental data, load and  indentation depth as function of time, the shape of the indenter, the size of the contact region -- also as function of time --  and the aggregated material properties of the indenter-substrate combination, e.g., the reduced modulus if indenter and substrate are both elastic. The size of the contact region is not measured and the determination of the material properties is the ultimate aim. Therefore, two equations are always necessary to obtain a mathematically solvable system. One of these is the relation between the depth and the histories of contact size and material properties and the other links the histories of the depth, contact size and material properties to the load; the two equations are coupled in a complicated way.

For purely elastic materials, the relation between depth and contact size  is specific for the indenter and invertible and independent of the material properties, thus enabling the elimination of the contact size altogether in favour of the depth; for elastic materials only a single equation for a single unknown material parameter, the reduced modulus, remains and this reduced modulus is the proportionality factor linking the load to a function of the depth.  The elimination of the contact size is mostly not possible if the material is viscoelastic and in this case one needs to work with the two equations mentioned before.

After adaptation of Sneddon's analysis of the indentation of elastic materials \citep[p. 450--457]{Sneddon1951}, these two equations are derived for homogeneous, isotropic and viscoelastic materials for which the stress-strain law is of the linear hereditary type having $\lambda(t)$ and $\mu(t)$ as relaxation functions. Apart from the requiring that $\lambda(t)$ and $\mu(t)$ be physically allowed, no other restrictions on their behaviour are needed. The functions $\lambda(t)$ and $\mu(t)$  appear in the theory through a single function, here called $\omega(t)$.

The properties of these equations are investigated and it is shown that in the classic stress relaxation and creep experiments the depth-radius relation is of the same type as for elastic materials. However,  in most other cases this relation depends on the material properties.

The appearance of a 'nose' in the load-depth curve is found to be quite natural because the 'nose' occurs because the load always reaches its maximum (the turning point) before the depth does if the load-depth curve is smooth. If this curve is kinked at the turning point the depth may afterwards still be increasing thus rendering the slope at the start of the unloading branch negative. Although in this case a judicious choice of the history of the control variable may render this slope positive, the use of this slope -- routinely used to eliminate plastic effects when estimating the reduced modulus for elastic materials -- to estimate the initial  elastic response is for viscoelastic materials not recommended. It is better to work with jumps in the rates, as jumps in the rates of depth, load and contact size are found to be synchronized. The ratio of jumps in the load and depth is proportional to the initial elastic response $\omega(0)$ and the contact size at the jump time; this ratio can fulfil the same function as the slope in the elastic case.

If the contact radius is known to increase, the load-depth equation, though still of the hereditary type, follows from the expression for elastic materials by interchanging the elastic element (the reduced modulus) by a viscoelastic element with $\omega(t)$ as relaxation function. However, the monotonic increase requirement often limits or even forbids the use of this procedure. During dynamic experiments, where a periodic perturbation is superposed on a constant control variable (load or depth) is eventually -- that is, in the stationary state -- the contact size also periodic, so certainly not monotonically increasing.

 To tackle the problem of a non-monotonic contact size, the theory of decomposing linear hereditary integrals of
 Golden and Graham \cite[p. 63--69]{GoldenGraham1988} was used to rewrite the basis equations in a more accessible form with respect to depth, contact radius and load. The material properties, however, enter these equations in a more convoluted way as is demonstrated by showing how to find $\omega(t)$ from the experimental time series for load
and depth as generated by a single load-unload experiment.

For dynamic indentation experiments controlled by a periodic perturbation
superposed on a constant depth or load it
is shown that eventually -- in the stationary state --  the maxima of depth and contact
size occur simultaneously, whereas for load and contact size the minima coincide in time.
These results are used to derive a matrix equation for each of two control types, i.e.,
for a sinusoidal perturbation superposed on a constant depth or  a sinusoidal perturbation added to a constant load. The solution
of these equations enables the construction of the shape of the
contact size during a typical period. This shape is a function of the
particular control parameters and the set $\tilde{p}$ of material parameters
defining the assumed material function $\omega(t)$, e.g., a standard linear
solid or a Prony series. Finally, combination of the reconstructed shape
with the remaining experimental data results in a non-linear parameter
identification problem for the elements of the set  $\tilde{p}$. As examples,
the reconstructed shapes of the contact size are investigated assuming
standard-linear-solid material behaviour for $\omega(t)$.

\mainmatter
\chapter{Introduction}\section{General part}\label{sec:intro}
Indentation is an experimental tool for investigating mechanical properties at a microstructural scale. The method has a long history after its first appearance in 1859 and later standardization in 1900 by J.A. Brinell \citep{ONeill2011}. Most of the time the focus was on hardness of metals \citep{Tabor1951} by indenting the material and measuring the imprint, initially using a vernier caliper and later  a microscope. Originally the loads where relatively large but today the tendency is towards nano-sized loads and penetration depths -- so-called nanoindentation
\citep[p. 853--856]{HaqueAndSaif2008} -- thus enabling investigating materials on ever decreasing scales, down to sizes typical of thin coatings (see, e.g., \citep{VarstAndWith2001,Malzbender2002}) or suitable for studying polymer gels, biomaterials (see, e.g., \cite{LinAndHorkay2008,OyenAndCook2009}) and, possibly, even cells \citep{GuzEtAl2014}. The capacity of scanning probe microscopy equipment, particularly atomic force microscopes, to function as nanoindentation instruments has also been investigated (see e.g., \citet{VanLandinghamEtAl2001} and also \citet{Attard2007}).

A comprehensive analysis of indentation from the point of view of scaling and dimensional analysis is found in the seminal review paper of Cheng and Cheng \cite{ChengAndCheng2004} where also their earlier results on these methods are collected and applied to indented half-spaces showing elasto-plastic material behaviour including work hardening, power-law creep behaviour or linear viscoelastic behaviour. Scaling is also used by Cao and Cheng \cite{CaoEtAl2010,CaoAndChen2012} in their studies on pressing indenters of arbitrary shape into purely viscoelastic or composite (matrix: viscoelastic, filler: elastic) substrates, respectively.

The modern version of an indentation experiment is known under the names: load-depth sensing, depth-sensing indentation or instrumented indentation; one indents a material, records continuously the load and the depth of the indenter tip  and subsequently calculates material properties from the experimental data (see e.g. \cite{OliverAndPharr2004,Mencik2007} or  the monograph \cite{FischerCripps2011}). A dynamical variant of the load-depth sensing method is known as the continuous-stiffness method. Here a small sinusoidal variation is superposed on the global input variable (load or depth) and the resulting variation of the global output variable is then used to investigate the elastic or viscoelastic properties of the material \cite{PethicaOliver1989,LoubetEtAl1995,LucasEtAl1998,LiBhushan2002,OliverAndPharr2004,HuangEtAl2004,ChengEtAl2006},\cite[][p.81--83]{KnausEtAl2008}, \cite[][Chapter 7]{FischerCripps2011}. Measuring the time-dependence of the shape of the imprint and analysis of these data is sometimes even extended into the time period after load removal; so-called rebound indentation \cite{BrownEtAl2009,ArgatovAndMishuris2011,Argatov2012,ArgatovAndPopov2015}. Finally, load-depth sensing is also used as a diagnostic tool since cracking and delamination of coatings may lead to jumps, plateaus or kinks in a plot of load versus depth \cite{VarstAndWith2001,Malzbender2002}.

Mathematically, indentation is a contact problem with time-de\-pen\-dent boundary conditions; time dependent because the prescribed control variable (load or depth)  is time-dependent and, more importantly, because the contact region changes generally with time.  For viscoelastic materials the time dependence of the material behaviour comes into play also. The problem of indenting elastic materials  by axisymmetric indenters has been solved by Sneddon in closed form (see \cite[p. 450--457 ]{Sneddon1951}, \cite{Sneddon1960,Sneddon1965} and also the paper of Lebedev and Ufliand \cite{LebedevAndUfliand1958}) and the solution was found to be useful also for cases where the indenter is not a body of revolution \cite{PharrEtAl1992}. Unfortunately, the results for elastic materials cannot always be transferred to the viscoelastic case using the elastic-viscoelastic correspondence principle on account of the time-dependence of the contact region \cite{Radok1957,Hunter1960,LeeAndRadok1960}. To proceed some researchers extended the correspondence principle \cite{Graham1968,GrahamAndSabin1973,GrahamGolden1988,GrahamGolden1991} to cases with monotonic increasing or decreasing boundary regions, whereas others started from first principles or otherwise transformed elastic results assuming that the (time-dependent) contact region is known in advance \cite{Hunter1960,Graham1965,Ting1966,Yang1966,Graham1967,Ting1968,Graham1969,SabinAndGraham1980,GoldenAndGraham1987,GrahamGolden1988b,GoldenGrahamLan1994}.

Not all mathematical results concerning contact problems can be used for processing load-depth sensing data because in indention experiments the material behaviour {\em and} the size of the contact region is unknown -- determination of material behaviour is the reason why the experiment is conducted in the first place -- whereas in classic contact problems these variables are normally given data and the focus is on calculating the stress distribution under the indenter. Nonetheless, from all studies mentioned earlier it is    clear that the precise dependence of the contact region on time is of vital importance. A monotonic increasing size is the most simple situation and a monotonic decreasing\footnote{As the contact size is initially zero, a monotonic decrease is only possible after an initial jump.} contact size is more complicated but also relatively simple. The case most important in practice, i.e., the case of a contact size consecutively increasing and decreasing, is the most complicated situation even if the contact size is known in advance \cite{Ting1966,Ting1968}. In load-depth sensing experiments the size of the contact region is normally not measured and the processing of the data for viscoelastic materials -- in a general sense the topic of this paper -- is much more complicated  than the same analysis for elastic materials.

To explain the foregoing in more detail it is worthwhile to recap first briefly Sneddon's analysis (\cite[][p. 450--457]{Sneddon1951}, \cite{Sneddon1960,Sneddon1965})
of indenting an {\em ideally elastic} half-space (elastic modulus $E$; Poisson's ratio $\nu$) with a {\em rigid axisymmetric indenter} with shape profile $f(r)$ (Fig.~\ref{vfig1}, left).
The load $p$ is proportional to the elasticity factor $\omega\young =E/\{2(1-\nu^2)\}$, it depends furthermore on the depth $h$ and on the size of the contact region, i.e., the contact radius $c$ (Fig. \ref{vfig1}, right).  Mathematically the first results are
\begin{equation}
p=4\omega\young \bigl\{ch-\mathbb{L}(c)\bigr\},\quad\text{with}
\quad\mathbb{L}(c)=\int_0^cf'(r)\sqrt{c^2-r^2}\mathrm{d}r~. \label{eq:Sneddon1}
\end{equation}
The contact radius is not measured in the experiments and an additional relation is needed to eliminate the contact radius from the load formula \eqref{eq:Sneddon1}. This relation follows from the requirement that for smooth indenter profiles\footnote{The indenter tip is excluded from this smoothness requirement, i.e., it might be sharp.} the normal stress must be bounded everywhere and it follows that the  current contact radius is slaved only to the current depth and this relation furthermore only depends on the indenter shape, that is $c=\mathbb{C}(h)$ which is the inversion of $h=\mathbb{L}'(c)$. Substituting $c=\mathbb{C}(h)$  in \eqref{eq:Sneddon1} leads to an expression for the load-depth curve:
\begin{equation}
p=4\omega\young \mathbb{F}(h),\enspace\text{with $\mathbb{F}(h)=\mathbb{C}(h)h-\mathbb{L}(\mathbb{C}(h))$
and $\mathbb{F}'(h)=\mathbb{C}(h)$}~.\label{eq:Sneddon2}
\end{equation}
As $\mathbb{F}(h)$ is a known function of the depth, \eqref{eq:Sneddon2} can theoretically be used to determine the elasticity factor, $\omega\young $, from the experimental data but in practice the slope, $S=4\omega\young c$, of the load-depth curve is used for that purpose. Due to plastic deformation, which is always present to some extend, the contact radius is determined from an estimate of the contact depth, $h_\mathrm{c}$, using the data for $h$ and $p/S$ at full load, the formula $h_\mathrm{c}=h-\varepsilon p/S$ ($\varepsilon$: an indenter specific number) and the area function, $A(h_\mathrm{c})$, which equals $\pi c^2$. The resulting value for $c$ is again combined with $S$ to calculate $\omega\young$ (\cite{DoernerAndNix1986,PharrEtAl1992}, \cite[][Chapter 3]{FischerCripps2011}).

While for ideally elastic materials the relation between current contact radius and depth depends only  on the indenter shape, for ideally viscoelastic materials normally also the material properties and the deformation history come into play. Only when the contact radius is a monotonic increasing function of time (from the start up to the current time) is the situation exactly the same as for the elastic case (Chap.~\ref{sec:increasing}) and it is possible to generalize \eqref{eq:Sneddon2} using the elastic-viscoelastic analogy, a fact already noted by \citet{LeeAndRadok1960} and used during the ensuing decades as the following list of representative -- though not exhaustive -- list of references shows: \cite{WierengaAndFranken1984,LoubetEtAl1995,ChengScrivenGerberich1998,LarssonCarlsson1998,LucasEtAl1998,Shimizuetal1999,IngmanSuzdalnitskyZeifman2000,SakaiShimizuetal2001,SakaiShimizu2001,%
GiriBousfieldUnertl2001,Sakai2002,LuEtAl2003,HuangEtAl2004,KumarAndNarasimhan2004,ChengAndCheng2005a,ChengAndCheng2005b,ChengAndCheng2005c,ChengAndCheng2005d,ChengEtAl2006b,VandammeAndUlm2006,ChengAndYang2009,Cheng2011}.
So, the results obtained in this way apply only to the case of an increasing contact radius. In other cases, i.e a decreasing contact radius following a sudden or  a smooth increase, the relation between the depth and the contact radius also depends on the viscoelastic relaxation properties of the substrate and generalization of \eqref{eq:Sneddon2} is no longer allowed and a different approach is needed such as Greenwood's \cite{Greenwood2010}. This applies to conventional (Chap. \ref{sec:decreasing}, \ref{sec:AdvancingRecedingContact} \& \ref{sec:SingleLOadUnload}) and dynamic load-depth sensing (Chap. \ref{sec:perturbation}) alike.
\begin{figure}[htb]
  \center
  \includegraphics[scale=1]{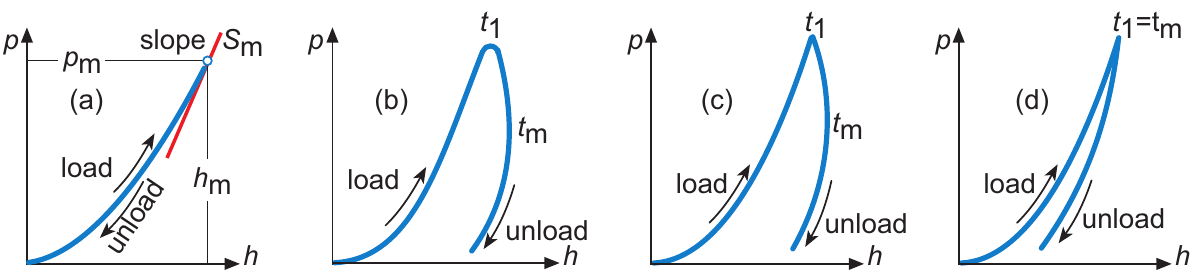}\\
  \caption{Possible plots of load $p$ versus depth $h$ for ideally elastic (a) and viscoelastic materials with smooth (b) and kinked (c \& d) contact area. Time is an increasing parameter along the curves and $t_1\leq t_\text{m}$ with $p$ maximal at $t_1$ and $h$ at $t_m$. Occurrence of a nose (b and c) if $t_1<t_m$.}\label{fig:VarstNose}
\end{figure}

Another difference between the elastic and the viscoelastic case is the occasional appearance of a 'nose' ('bulge') in the data (Fig.~\ref{fig:VarstNose}b and \ref{fig:VarstNose}c). See, for example, \citet{BriscoeEtAl1998} for indentation data about polymeric surfaces, \citet{NganAndTang2002} for polypropylene or \citet{SahinEtAl2007} for experimental data about $\beta$-Sn single crystals and \citet{ChengAndCheng2004} for numerical simulations using standard linear solid material behaviour. When a 'nose' appears the maximum of the load at time $t_1$ in Fig.~\ref{fig:VarstNose} and that of the depth at time $t_\mathrm{m}$ in Fig.~\ref{fig:VarstNose} do not appear simultaneously but after another, i.e., $t_1\leq t_\mathrm{m}$. Physically this means that between $t_1$ and $t_\mathrm{m}$ unloading has already set in, the load decreased, whereas the indenter still moves further into the substrate. The 'nose' is actually about the slope of the load-depth curve during the first stage of unloading; the slope becomes negative whereas the use of the conventional slope formula $S=4\omega\young c$ as derived from \eqref{eq:Sneddon2} needs a positive value. This raised the question whether the classic procedure to determine the initial elastic response can be used for viscoelastic materials or not.

Experimentally, it was found that the 'nose' might disappear if a suitable load schedule was used, e.g. if prior to $t_1$ a sufficiently long  load hold period is introduced \cite{BriscoeEtAl1998,SahinEtAl2007} and this was corroborated by the analytical and numerical results of \citet{NganAndTang2002,ChengEtAl2006b}. Moreover, similar studies of \citet{NganAndTang2002,ChengAndCheng2004,ChengAndCheng2005a,ChengAndCheng2005b,ChengAndCheng2005c,ChengAndCheng2005d,ChengEtAl2006b} show, for various types of indenters, that the 'nose' might also disappear when unloading is fast enough. However, generally it is not altogether clear under what circumstances the 'nose' is present (Fig.~\ref{fig:VarstNose}b \& \ref{fig:VarstNose}c) or not (Fig.~\ref{fig:VarstNose}d).

Research of \citet{FengNgan2002}, \citet{TangAndNgan2003} and \citet{NganEtAl2005} suggested that the conventional procedure for determination of elastic properties can also be used for measuring the initial elastic response of viscoelastic materials if, as suggested by \citet[][p. 609--610]{DeWith2006}, effectively  the ratio of rate jumps in load and depth data is used  instead of the slope of the unloading curve and this suggestion was later heuristically substantiated by \citet{NganAndTang2009} and used for soft-tissue characterization by \citet{TangAndNgan2012}. For this reason it seems worthwhile to study the  effect of arbitrary rate jumps in more detail.

The aim of the paper is to revisit the indentation problem particularly in view of the question how during load-depth sensing material properties of homogeneous isotropic linear viscoelastic materials influence the relation between contact radius on the one hand and load and depth on the other.

The question becomes even  more urgent for dynamic load-depth sensing, i.e., when a periodic perturbation is superposed on the global control variable (depth or load). In these cases points on the substrate surface may repeatedly move in- and outside the contact region meaning that at these points the boundary condition may change repeatedly from type -- from a prescribed stress to a prescribed displacement and back -- and the times these changes occur is not known at all. For these situation  Golden and Graham \cite[][p. 173--178]{GoldenGraham1988} and Graham and Golden \cite{GrahamGolden1988b} developed a generalized Boussinesq formula for a contact problem under varying load. The analysis performed by these authors is exclusively formulated in terms of relations  between displacements and stress and between depth and loads and the contact region is only present in a hidden form. The idea here is to obtain explicit formulae relating load, depth, contact radius and material functions; results mirroring those of Sneddon for elastic materials \eqref{eq:Sneddon1}. These results will be combined with the theory of decomposing hereditary integrals as developed by Golden and Graham
\cite[][p. 63--68]{GoldenGraham1988}. The main focus will be on the role of the contact radius.

The problem of numerically identifying the material parameters is not considered here in any detail. This is a problem in its own right because noisiness of the data or too strict assumptions on the decay rate of the material functions might lead to strong ill-conditioned identification problems; the studies performed by Sorvari and  Malinen \cite{sorvari2007numerical} and by Ciambella and coworkers \cite{CiambellaEtAl2011} can serve as an introduction to these problems.

\section{Symbols and notation}
Current time is denoted by $t$, the load by $p$, the penetration depth by $h$, the contact radius by $c$, the normal surface stress by $\sigma$, the normal surface displacement by $u$, the Heaviside unit step function by $\Hea$ and the Bessel function of the first kind and order zero by $\mathrm{J}_\mathrm{o}$.

A subscript $0$ indicates taking the limit $t\downarrow 0$, e.g., $h_0=h(0+)$. A time jump in a function, say at $t=T$, is indicated by angular brackets and the jump time as a subscript; $\langle f \rangle_T =\lim_{\delta \downarrow 0} \{f(T+\delta)-f(T-\delta)\}$. When $f$ also depends on a spatial variable, say $r$, the notation $\langle f \rangle_T(r)$ is used.  Dots denote partial differentiation to time-like arguments and primes partial differentiation to spatial variables. So, $F'=\partial F(r,t)/\partial r$ and $\dot{F}=\partial F(r,t)/\partial t$.

For the viscoelastic material behaviour the theory of the Stieltjes convolution including the corresponding notation is used (see the seminal paper by Gurtin and Sternberg \cite{GurtinAndSternberg1962} or -- for a summary -- Appendix \ref{app:convolution}). In short, the Stieltjes convolution $s(t)$ of two functions $f(t)$ and $g(t)$  is the Stieltjes integral $s(t)=\int_{-\infty}^tf(t-\tau)\,\text{d}g(\tau)$ and -- using the Stieltjes convolution operator $\orc$ -- this is notated as $s=f\orc g$. The Stieltjes inverse of $f$ is denoted by the superscript i, i.e., $f\orc\inv{f}=\Hea$.
When $f$ and $g$ depend on $t$ and also on a spatial variable $y$ the Stieltjes convolution is also a function of these two variables, i.e., $[f\orc g](y,t)$. Note that the operations {\sl convolution of functions} and {\sl substitution of a time-dependent spatial variable}, say $c(t)$, do not commute; generally $[f\orc g](y,t)|_{y=c}\neq [\{f|_{y=c}\}\orc \{g|_{y=c}\}](t)$.

\chapter[Description of the system]{Experiment, material and governing equations}\label{sec:system}
\section{Experimental setup}
\begin{figure}[hbt]
\begin{center}
\includegraphics[scale=1]{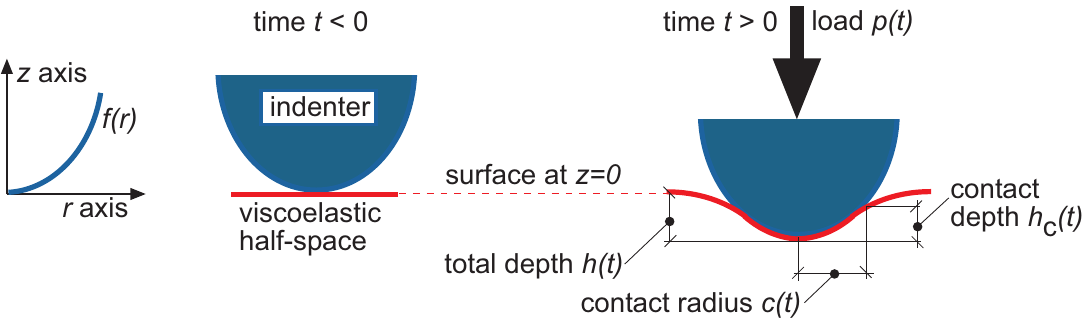}
\caption{Axi-symmetric indenter profile $f(r)$ (left) and indentation process (middle and right) indicating load $p(t)$, indentation depth $h(t)$, contact radius $c(t)$ and contact depth $h_\mathrm{c}(t)=f(c(t))$.} \label{vfig1}
\end{center}
\end{figure}

Prior to indentation, (Fig.~\ref{vfig1}, middle), the material is in its
virgin, stress free state with the indenter only touching the material; depth and load are both zero in this time period.  Starting at $t = 0$, a
{\em rigid axisymmetric indenter} with surface profile  $f(r)$, (Fig.~\ref{vfig1}-left) is pressed into a viscoelastic half-space (Fig.~\ref{vfig1}-right) and the resulting penetration depth $h$ and the necessary force $p(t)$ are recorded as function of time. These two functions -- both reckoned to be positive for historical reasons -- plus the information about the indenter shape $f(r)$ constitute the experimental data from which the material behaviour of the half space must be extracted.
Only convex and non-flat indenter shapes (Fig.~\ref{vfig1}, left) are considered; $f(r)$ is a twice continuously differentiable positive function for all $r>0$ excluding, perhaps, the point $r=0$ as for a cone. This implies also $f'(r) > 0$ and $f''(r)\geq 0$ for all $r>0$.

\section{Material behaviour}\label{sec:Material}
The half-space material is {homogeneous} and {isotropic} with a {\em linear, non-ageing viscoelastic relation} between stresses $\sigma_{ij}$ and strains $\epsilon_{ij}$ \cite{GurtinAndSternberg1962}
\begin{equation}
\sigma_{ij}=\delta_{ij}[\epsilon_{kk}\orc
\lambda]+2[\epsilon_{ij}\orc \mu]~.\label{eq:visco}
\end{equation}
The {\em stress relaxation functions} $\lambda(t)$ and $\mu(t)$ in \eqref{eq:visco} are real causal functions, i.e., they are zero for $t <0$. At $t=0$, positive jumps $\lambda_0=\lambda(0+)$ and $\mu_0=\mu(0+)$ occur with $\lambda_0$ and $\mu_0$ the Lam\'{e} constants of the initial elastic response of the material. Both function are assumed to admit a relaxation spectrum representation which implies that they are continuously differentiable to any order for $t > 0$, completely monotonic \cite{BerisAndEdwards1993,PieroAndDeseri1995} and that the material has fading memory \cite{,Anderssen2002,AnderssenAndLoy2002}; $\lambda$ and $\mu$ are positive and monotonically decreasing functions of time \cite{BreuerAndOnat1962,Anderssen2002} with non-zero positive constant asymptotic values \cite{ColemanAndNoll1961}. All these properties are sufficient for dissipative  material behaviour \cite{BreuerAndOnat1962,GurtinAndHerrera1965,PieroAndDeseri1995} and guarantee unique solutions to mixed boundary value problems \cite{BreuerAndOnat1962}.
Because $\lambda_0\neq 0$ and $\mu_0\neq 0$, the Stieltjes inverses of $\lambda$ and $\mu$ exist  \cite{GurtinAndSternberg1962} and they are creep functions, that is, they are real, causal, positive and  increasing functions for $t>0$ (see Chap. \ref{app:relax}).

Often the assumption is made that $\lambda$ and $\mu$ differ only by a multiplicative constant, i.e., Poisson's ratio is constant, whichever way it is defined \cite{Varst1992, Hilton2001,Tschoegl2002}. This assumption will not be made here because it would imply that the bulk modulus relaxes at the same rate as the shear modulus which is, according to Hilton \cite{Hilton2001}, not the case for real materials.
\subsection{Monotonicity of the relaxation and creep functions}\label{app:relax}
Let $w$ be any of the two reduced stress relaxation functions occurring in the constitutive equation \eqref{eq:visco}; i.e., $w =\mu/\mu_0$ or $w=\lambda/\lambda_0$. Then, according to the previous paragraph,    $w\in H^2$, $w_0=1$, $\dot{w}< 0$, $0<w\leq 1$,  $\ddot{w}> 0$, $w> 0$ in the limit $t \rightarrow \infty$ and the Stieltjes inverse $g$ of $w$ exists because $w_0\neq 0$ (see Appendix \ref{app:convolution}). The first mean value theorem of integration applied to $w\orc g =\Hea$ shows that a function $\zeta(t)$ exists such that $0<\zeta(t)<t$ for $t>0$ and
\begin{gather}
\{1-w(t)\}/\{t\, w(t-\zeta(t))\}=\dot{g}(\zeta(t))>0\quad \forall t>0\\
\intertext{plus}
\lim_{t\downarrow 0} \dot{g}(\zeta(t))=\dot{g}_0=-\dot{w}_0.
\end{gather}
Therefore, the reduced creep function $g=\inv{w}$ is a positive monotonic increasing function. It is also bounded if $w(\infty)> 0$ because $g(t)w(t)\leq 1$ \cite[p. 11]{GoldenGraham1988}. The same analysis also applies to the material functions $\omega$ and $\varpi$ introduced in Chap.~\ref{sec:SurfaceEquations}.
\section{Boundary conditions}
The symmetry present in the shape, material properties and loading of the system ensures that all local quantities depend only on the radial and axial spatial variables, $r$ and $z$, and the time $t$. For the same reason is the contact region, the region where the material surface conforms frictionless to the indenter, uniquely characterized by the (non-negative) contact radius $c(t)$. The {\em boundary conditions} for the normal surface displacement $u(r,t)$ and the normal surface stress $\sigma(r,t)$ are \footnotemark
\begin{equation}
u(r,t)=-h(t)+f(r)\enspace\text{for $r<c(t)$ and}\enspace \sigma(r,t)=0\enspace\text{for $r>c(t)$}.
\end{equation}
The surface shear stress must always be zero everywhere.
\footnotetext{As in Sneddon's treatment \cite{Sneddon1965}, the normal stress is required to remain bounded everywhere and the same boundary conditions will be used. This means that the radial surface displacement is free to take any value although it should in fact be constrained to remain outside the indenter too. For elastic materials it has been shown \cite{Bolshakov1997, Fu2005} that not enforcing this constraint leads to errors in the determined value of Young's modulus . However, incorporation of this constraint  is only possible using finite element methods or other approximate methods \cite{Bolshakov1997, HayAndBolshakov1999, Woigard2006}. As the focus here is on an analytical treatment this constraint is not enforced. }

\section{Normal surface stress and displacement}\label{sec:SurfaceEquations}
Sneddon's approach \cite[][p. 450--457]{Sneddon1951} to indentation of homogeneous, isotropic, linear elastic materials can easily be adapted because of the linearity, homogeneity and isotropy of the viscoelastic material combined with the existence of the  Stieltjes inverses of the viscoelastic material functions. The use of the generalized Love strain function is allowed \cite{GurtinAndSternberg1962} and a lengthy calculation shows that its Hankel transform is related to normal stress and displacement at the surface by
\begin{equation}
\sigma(r,t)=\int_0^{\infty}\!\!\!\xi \bzero{\xi r}
\psi(\xi,t)\,\text{d}\xi~,\enspace
u(r,t)=\int_0^{\infty}\!\!\!\bzero{\xi r} [\psi\orc
\varpi](\xi,t)\,\text{d}\xi~. \label{eq:basic}
\end{equation}
For arbitrary $\psi$ is mechanical equilibrium and correct asymptotic behaviour guaranteed and the shear stress is zero everywhere. The material behaviour enters  \eqref{eq:basic} through the real causal creep function $\varpi$ and
\begin{equation}
\varpi=\{\inv{(\lambda+\mu)}+\inv{\mu}\}/2,\enspace\text{with}\enspace \varpi_0=2(1-\nu_0^2)/E_0>0 ,
\end{equation}
where $E_0$ and $\nu_0$ indicate Young's modulus and Poisson's ratio for the initial elastic response, respectively. A real causal relaxation function $\omega$, the Stieltjes inverse of $\varpi$, exists because $\varpi_0>0$ (see Appendix \ref{app:convolution}). Specifically,
\begin{gather}
\omega=\inv{\varpi}=2\mu-2\{\mu\orc\mu\}\orc\{\inv{(\lambda+2\mu)}\},\enspace\text{with}\enspace
\omega_0=1/\varpi_0,\\
\intertext{and also}
 [u\orc\omega](r,t)=\int_0^{\infty}\!\!\!\bzero{\xi r} \psi(\xi,t)\,\text{d}\xi. \label{eq:basic1}
\end{gather}
The properties of $\lambda$ and $\mu$ (see Section \ref{sec:Material}) guarantee that $\omega$ is a positive decreasing function with $\omega(\infty)> 0$ and, consequently -- as shown in Chap. \ref{app:relax} --  that $\varpi$ is a positive and increasing function of time for $t>0$ and also that $\varpi(\infty)$ is positive and finite. Although the material behaviour is characterized by two material functions, the influence of these functions appears here in a lumped form\footnote{Huang and Lu \cite{HuangAndLU2007} argue that the two functions $\lambda(t)$ and $\mu(t)$ can be determined by using two different indenters, an axisymmetric and an asymmetric one.}, i.e., the function $\omega$. This function is, just like $\lambda$ and $\mu$, also completely monotonic.

At the end of this chapter, and further on,  the reduced material functions, i.e., $\phi(t)=\omega(t)/\omega_0$ and  $\varphi(t)=\varpi(t)/\varpi_0$ are also used.

\section{Automatically meeting the stress boundary condition}
Lebedev and Ufliand \cite{LebedevAndUfliand1958} and Sneddon \cite{Sneddon1960} showed that substituting for $\psi$ in \eqref{eq:basic} the equation
\begin{equation}
\psi(\xi,t)=\int_{0}^{c(t)}\!\!\theta(y,t)\cos \xi y
\,\text{d}y~ \label{eq:deftheta}
\end{equation}
results in $\sigma=0$ if $r>c$, i.e., the boundary condition for the stress is satisfied automatically irrespective of the choice of $\theta$.
The expression for $\sigma$, the first of \eqref{eq:basic}, is now found to be the sum of two terms. The first, the term $\theta(c,t)/(c^2-r^2)^{1/2}$, is a flat punch solution which is singular at the edge of the contact region. Sneddon determined  the contact radius for non-flat indenter shapes (cone, parabola, sphere, \ldots) by requiring that the {\em axial normal surface stress be bounded} everywhere \cite{Sneddon1965}. To meet this requirement, the Sneddon condition  that
\begin{equation}
\lim_{r \uparrow c(t)}\theta(r,t)=\theta(c(t),t)=0
\end{equation}
\cite{Sneddon1965} is also imposed, thus rendering the flat punch term zero.
The remaining term -- the actual \emph{stress} --  is bounded and continuous  across the contact radius \cite{Sneddon1965}:
\begin{equation}
\sigma(r,t)=\begin{dcases}- \int_r^{c(t)}\!\!\frac{\theta'(y,t)\,\text{d}y}{\sqrt{y^2-r^2}}&\quad \text{for  $r \leq c(t)$}~,\\
                            0 & \quad \text{for  $r > c(t)$}~.
                            \end{dcases}\label{eq:stress1}
\end{equation}
All attention is now focussed on the function $\theta(r,t)$.
\section{The boundary condition for the surface displacement}
For the surface displacement in \eqref{eq:basic1}, the 'Ansatz' \eqref{eq:deftheta} results in
\begin{gather}
[u\orc\omega](r,t)=\int\limits_0^r\!\!\frac{\vartheta_{c}(y,t)\,\text{d}y}{\sqrt{r^2-y^2}},\label{eq:theta1}
\intertext{\cite{Sneddon1960}, or , alternatively, in}
u(r,t)=\int\limits_0^r\!\!\frac{[\vartheta_c\orc\varpi](y,t)\,\text{d}y}{\sqrt{r^2-y^2}}. \label{eq:theta2}
\end{gather}
The function $\vartheta_c$ is defined by:
$\vartheta_{c}(r,t)=\theta(r,t)\hea{c(t)-r}$ plus -- to ensure that the stress is bounded -- that $\vartheta_c(c(t),t)=0$. Application of the boundary condition for the displacement in \eqref{eq:theta2} shows that
\begin{equation}
\int_0^r\!\!\frac{[\vartheta_c\orc\varpi](y,t)\,\text{d}y}{\sqrt{r^2-y^2}}=\{-h(t)+f(r)\}\hea{t}\enspace\text{for $0\leq r\leq c(t)$}
\end{equation}
and, after taking the limit $r\downarrow 0$ of \eqref{eq:theta1}, that
\begin{equation}
-[h\orc\omega](t)=\frac{\pi}{2}\theta(0,t).
\end{equation}
\section{The final equations}
\subsection{Equations in the complete $(r,t)$- plane}
Inversion of the equations
\eqref{eq:theta1} and \eqref{eq:theta2} \cite[][Sec. 1.1-6, Eq.~ 40]{PolyaninAndManzhirov1998} plus introduction of the function $L$ and its spatial derivative $L'$ through
\begin{equation}
L(r,t)=\int_0^r\!\! u'(y,t)\sqrt{r^2-y^2}\,\text{d}y~\enspace\text{and}\enspace
L'(r,t)=
r\int_0^r\!\!\frac{u'(y,t)\,\text{d}y}{\sqrt{r^2-y^2}}~,
\label{eq:lpl}
\end{equation}
produces two equivalent expressions which are related by a convolution
\begin{equation}
\frac{\pi}{2}\vartheta_c(r,t) =[\{L'-h\}\orc\omega](r,t)
\,\Leftrightarrow\,
L'(r,t)-h(t)=\frac{\pi}{2}[\vartheta_c\orc \varpi](r,t).\label{eq:loadequation3}
\end{equation}
Lebedev and Ufliand \cite{LebedevAndUfliand1958} and Sneddon \cite{Sneddon1965} showed by integration of \eqref{eq:stress1} that the total load\footnotemark, $p(t)$, is
\begin{equation}
p(t)=-2\pi\int_0^{\infty}\!\! r
\sigma(r,t)\,\text{d}r=-2\pi\int_0^{c(t)}\!\!\theta(y,t)\,\text{d}y~.
\label{eq:press}
\end{equation}
\footnotetext{For historical reasons the load is reckoned positive, so a minus sign is introduced because the stress is actually negative (compressive).}
Apparently, the spatial integrals of \eqref{eq:loadequation3} are also important. Define
\begin{equation}
\wp_\mathrm{c}(r,t)=2\pi\int_0^r \vartheta_c(y,t)\,\mathrm{d}y=2\pi\int_0^r \theta(y,t)\hea{c(t)-y}\,\mathrm{d}y
 \end{equation}
 which becomes with \eqref{eq:press}
 \begin{equation}
 \wp_\mathrm{c}(r,t)=-p(t)\enspace r\geq c(t).
\end{equation}
So, the spatial integrals of \eqref{eq:loadequation3} are
\begin{equation}
\wp_\mathrm{c}(r,t) =4[\{L-rh\}\orc\omega](r,t)
\,\Leftrightarrow\,
4\{L(r,t)-rh(t)\}=[\wp_\mathrm{c}\orc \varpi](r,t),\label{eq:loadequation3d}
\end{equation}
and they are also related by a convolution.
\subsection{Intermezzo: the indenter characteristic function $\mathbb{L}$}
The functions $L(r,t)$ and $L'(r,t)$ are continuously differentiable in both arguments and they will play a pivotal role in the rest of the analysis. Whenever $r\leq c(t)$, the indenter shape completely determines $L$ and $L'$ because $u'=f'$ in this case. This is the reason why the {\sl indenter characteristic function} $\mathbb{L}$, the spatial derivative\footnotemark\, $\mathbb{L}'$ and the inverse $\mathbb{C}$ of $\mathbb{L}'$ are introduced (see also Appendix \ref{app:CharacteristicFunction})
\footnotetext{The characteristic function $\mathbb{L}'(r)$, its mathematical definition originally due to \cite{Sneddon1965}, also plays a pivotal role in the theory of dimensionality reduction in contact mechanics and friction of Popov and coworkers \cite{ArgatovAndPopov2015},\cite[][Chapter  3 and 17]{PopovAndHess2015}. In this theory it is called the \emph{equivalent indenter profile}.}
\begin{equation}
\mathbb{L}(r)=\int_0^r\!\! f'(y)\sqrt{r^2-y^2}\,\text{d}y~,\enspace
\mathbb{L}'(r)=r \int_0^r\!\!\frac{f'(y)\,\text{d}y}{\sqrt{r^2-y^2}}~, \enspace \mathbb{C}(\mathbb{L}'(r))=r~.
\label{eq:sps}
\end{equation}
\begin{figure}[htb]
  \centering
  \includegraphics[scale=1]{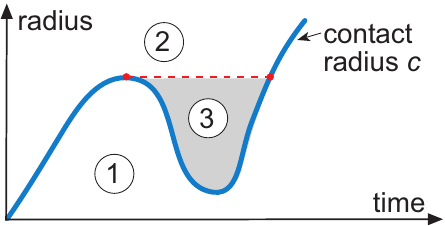}\\
  \caption{For $(r,t)\in$ region 1: $L'(r,t)=\mathbb{L}'(r)$. For $(r,t)\in$ region 2: $L'(r,t)=h(t)$. But for $(r,t) \in$ region 3 only $[\{L'-h\}\orc\omega](r,t)=0$ applies.}\label{fig:VarstUvalues}
\end{figure}
Below the contact radius curve (region 1 in Fig.~\ref{fig:VarstUvalues}), i.e., for $r<c(t)$, is  $L(r,t)=\mathbb{L}(r)$ and $L'(r,t)=\mathbb{L}'(r)$. For points above the curve (region 2 and 3 in Fig.~\ref{fig:VarstUvalues}), $[\{L'-h\}\orc\omega]=0$ because $\vartheta_c(r,t)=0$ here (see the first of \eqref{eq:loadequation3}). If -- in addition -- a line of constant radius did not cross the contact radius curve up to the present time $t$, i.e., the region 2 in Fig.~\ref{fig:VarstUvalues}, the stronger result $L'(r,\tau)=h(\tau)$ for $0<\tau<t$ follows from the second of \eqref{eq:loadequation3}. Finally, along the curve of the contact radius, continuity of $L$ and $L'$ implies $L(c(t),t)=\mathbb{L}(c(t))$ and $L'(c(t),t)=\mathbb{L}'(c(t))$.
\subsection{The equations relating depth, contact radius and load}\label{sec:ContactRadiusEquations}
These equations are found by restricting \eqref{eq:loadequation3} and \eqref{eq:loadequation3d} to the curve of the contact radius. This leads to
\begin{subequations}\label{eq:radiusequation}
\begin{equation}
[L'\orc\omega](r,t)|_{r=c(t)}=[h\orc\omega](t),\label{eqPart1:radiusequation}
\end{equation}
\begin{equation}
\mathbb{L}'(c(t))=h(t)+\frac{\pi}{2}[\vartheta_c\orc \varpi](r,t)|_{r=c(t)},\label{eqPart2:radiusequation}
\end{equation}
\end{subequations}
as two, equivalent, equations relating depth, contact radius and material properties plus two additional equations
\begin{subequations}\label{eq:loadequation3e}
\begin{equation}
p(t) =4\Bigl\{c(t)[h\orc\omega](t)-[L\orc\omega](r,t)|_{r=c(t)}\Bigr\},\label{eqPart1:loadequation3e}
\end{equation}
\begin{equation}
[\wp_\mathrm{c}\orc \varpi](r,t)|_{r=c(t)}=4\Bigl\{\mathbb{L}(c(t))-c(t)h(t)\Bigr\},\label{eqPart2:loadequation3e}
\end{equation}
\end{subequations}
involving also the load $p$ and the 'load-like' function $\wp_\mathrm{c}$; the latter two equations are also equivalent.

The two equations \eqref{eq:loadequation3} are related by a convolution, but the two derived equations in \eqref{eq:radiusequation} are not. Similarly, the two equations \eqref{eq:loadequation3d} are related by a convolution but between the two equations in  \eqref{eq:loadequation3e} no such relation exists. Convoluting with $\omega$ or $\varpi$ and substituting $r=c(t)$ does not commute. The only exception occurs when $\omega$ is a step function and this is in fact the elastic case with $\omega(t)=\omega\young \hea{t}$; convolution is in this particular case equivalent to conventional multiplication. A similar result is found for the initial -- elastic -- response of a viscoelastic material. Indeed, in the limit $t\downarrow 0$ both equations \eqref{eq:radiusequation} lead to the same result, i.e., $h_0=\mathbb{L}'(c_0)$ and from this $c_0=\mathbb{C}(h_0)$, and the two equations \eqref{eq:loadequation3e} also generate the same formula: $p_0=4\omega_0\mathbb{F}(h_0)$.

From \eqref{eqPart1:radiusequation} it is seen that the relation between depth and contact radius generally depends on the reduced relaxation function $\phi$ but not on the elasticity factor $\omega_0$ because this factor can be divided out of this equation. Division of the load equation \eqref{eqPart1:loadequation3e} by $\omega_0$ is equivalent to using $\omega_0$ as the unit of stress. The remaining equations, that is, the  equations \eqref{eqPart2:radiusequation} and  \eqref{eqPart2:loadequation3e}, are already divided by this factor because $[\vartheta_c\orc\varpi]=[(\vartheta_c/\omega_0)\orc\varphi]$ and $[\wp_\mathrm{c}\orc\varpi]=[(\wp_\mathrm{c}/\omega_0)\orc\varphi]$.
All this suggests that it is advisable to determine the material function in two steps. In a first step the elasticity factor $\omega_0$ governing the initial elastic response is determined and, subsequently, in a second step the aftereffects as represented by the reduced relaxation and creep functions $\phi$ and $\varphi$, respectively.

\chapter[Closed form solutions]{Complete and partially closed form solutions}
The formulae \eqref{eq:radiusequation} immediately produce relations between depth  and contact radius in two simple cases, namely a monotonic increasing (non-decreasing) or a monotonically decreasing\footnote{Obviously, this is only possible after an initial jump.} (non-increasing) contact radius. The classic relaxation experiments -- step-shaped depth as experimental control variable -- and creep experiment -- step-shaped load as experimental control variable -- are both found to be examples of the former case.
\section{A monotonically increasing contact radius}\label{sec:increasing}
\begin{figure}[htb]
  \centering
  \includegraphics[scale=1]{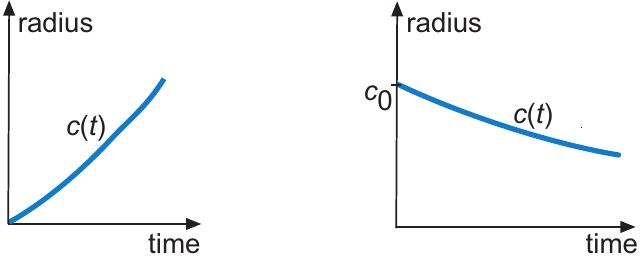}\\
  \caption{Monotonically increasing (left) or decreasing contact radius (right).}\label{fig:VarstContactRadius}
\end{figure}

For a monotonically increasing contact radius (Fig.~\ref{fig:VarstContactRadius}), the inequality $c(\tau)<c(t)$ if $0<\tau<t$ applies for arbitrary times $t$ in a connected interval $0<t<t_\mathrm{max}$. Therefore, $[\vartheta_c\orc \varpi](r,t)|_{r=c(t)}=0$ plus
$[\wp_\mathrm{c}\orc \varpi](r,t)|_{r=c(t)}=-[p\orc\varpi](t)$, also for arbitrary times in this interval. The second of  \eqref{eq:radiusequation} is now:
\begin{equation}
\mathbb{L}'(c(t))=h(t)\quad \Rightarrow \quad c(t)=\mathbb{C}(h(t))~,\label{eq:radius1}
\end{equation}
i.e., the contact radius only depends on the indenter shape and the depth\footnote{A pure example of an experimental study of this type is that of Gang Huang and Hongbing Lu \cite{HuangAndLu2006} where a linearly increasing depth is used as control variable.}. The second of the load equation \eqref{eq:loadequation3e} now takes the form
\begin{equation}
[p\orc\varpi](t)=4\Bigl\{c(t)h(t)-\mathbb{L}(c(t))\Bigr\},\label{eq:advancingcontact2a}
\end{equation}
and the contact radius can be eliminated from the right-hand side of \eqref{eq:advancingcontact2a} with the aid of the second of \eqref{eq:radius1}. Defining $\mathbb{F}(h)=h\mathbb{C}(h)-\mathbb{L}(\mathbb{C}(h))$ one eventually finds equations involving only the experimental data and the material properties, i.e.,
\begin{equation}
[p\orc\varpi](t)=4\mathbb{F}(h(t))\quad \Leftrightarrow\quad
p(t)=4[\mathbb{F}(h)\orc\omega](t)~.\label{eq:advancingcontact2b}
\end{equation}
 Specifically, for a conical and a parabolic indenter one finds the load to be proportional to $[h^2\orc\omega]$ and $[h^{3/2}\orc\omega]$, respectively, with the proportionality constants depending on the shape parameters of the indenter  (see Appendix \ref{app:CharacteristicFunction}).

The formulae \eqref{eq:advancingcontact2b} are generalizations of the elastic results using the elastic-viscoelastic analogy and this enables determination of the material function $\omega$ using conventional Laplace transformation methods with $p$ and $\mathbb{F}(h)$ as experimental input \cite{ChengAndYang2009}. An alternative way is to solve the integrals equations \eqref{eq:advancingcontact2b} directly for $\varpi$ or $\omega$ \cite{Martynova2016}. If initial jumps in the load and depth occur, i.e., $p_0>0$ and $h_0>0$, the Stieltjes inverses of $p$ and $\mathbb{F}(h)$ exist and
$\varpi=4[\mathbb{F}(h)\orc\inv{p}]$ or $\omega=[\inv{\mathbb{F}(h)}\orc p]$.
Otherwise, i.e., load and depth changing smoothly from zero, the type of the Volterra integral equations \eqref{eq:advancingcontact2b} changes from second to first kind; finding $\varpi$ or $\omega$ is more complicated \cite[Sections 8.3 \& 8.4] {PolyaninAndManzhirov1998}.

\subsection{Special case 1: the classic relaxation experiment}\label{sec:ClassicRelaxation}
A step-shaped contact radius can be seen as a (non-decreasing) limiting case of a piecewise linear contact radius.  It then follows from \eqref{eq:radius1} that a step shaped depth, i.e., $h(t)=h_0\hea{t}$, is needed to achieve this and the second of \eqref{eq:advancingcontact2b} shows that $p(t)=4\mathbb{F}(h_0)\omega(t)$. The control variable is a step in the depth and the resulting load is proportional to the relaxation function; this is the case of a {\em classic relaxation experiment}. For all members of the considered class of constitutive equations the function  $\omega(t)$ approaches a constant value if $t\rightarrow \infty$ and the same applies to the load response to the classic relaxation experiment, i.e., $p(\infty)=4\mathbb{F}(h_0)\omega(\infty)$
\subsection{Special case 2: the classic creep experiment}\label{sec:ClassicCreep}
Alternatively, in the classic creep experiment a step-shaped jump in the control variable, the load, is applied. Use of $p(t)=p_0\hea{t}$ in  \eqref{eq:advancingcontact2a} shows that
\begin{gather}
p_0\varpi(t)=4\{c(t)h(t)-\mathbb{L}(c(t))\},\\
\intertext{and after differentiation that}
p_0\dot{\varpi}(t)=4c(t)\dot{h}(t)+4\dot{c}(t)\{h(t)-\mathbb{L}'(c(t))\}.
\end{gather}
The choice $c=\mathbb{C}(h)$, i.e., $\mathbb{L}'(c)=h$, then leads to a consistent set of solutions with increasing depth $h$ (creep) and increasing contact radius because $\dot{\varpi}>0$ and the relation between $\varpi$ and depth $h$ is $\varpi(t)=4\mathbb{F}(h(t))/p_0$. As $\varpi(t)$ also approaches a constant value if $t\rightarrow \infty$ the same applies to the depth and $\varpi(\infty)=4\mathbb{F}(h(\infty))/p_0$

\section{A monotonically decreasing contact radius}\label{sec:decreasing}
In the instance of a  monotonically  decreasing contact radius (Fig.~\ref{fig:VarstContactRadius}) one has $c(\tau) > c(t)$ and, consequently, $L'(c(t),\tau)=\mathbb{L}'(c(t))$ for $0<\tau<t$, again for arbitrary time $t$. So, substitution of  $r=c(t)$ in the first of \eqref{eq:radiusequation} gives now
\begin{equation}
\mathbb{L}'(c(t))\omega(t)=h\orc\omega\enspace \Rightarrow\enspace  c(t)=\mathbb{C}\left(\frac{[h\orc\phi](t)}{\phi(t)}\right),\label{eq:radius2}
\end{equation}
i.e., the contact radius depends not only on the indenter shape and the depth but also on the reduced relaxation function $\phi$. Because also $[L\orc\omega](r,t)|_{r=c(t)}=\mathbb{L}(c(t))\omega(t)$, the load formula \eqref{eqPart1:loadequation3e} now takes the form
\begin{equation}
p(t)=4\omega(t)\{c(t)\mathbb{L}'(c(t))-\mathbb{L}(c(t))\}\enspace \Rightarrow\enspace  p(t)=4\omega(t)\mathbb{F}\left(\frac{[h\orc\phi](t)}{\phi(t)}\right)~.
\label{eq:precede3}
\end{equation}
For a cone $p\propto\omega\{[h\orc\phi]/ \phi\}^2$ and for a parabole $p\propto\omega\{[h\orc\phi]/ \phi\}^{3/2}$ (see Appendix \ref{app:CharacteristicFunction}).
The formulae \eqref{eq:precede3} are completely different from those of the previous case and the main reason is that  the contact radius no longer depends solely on the indentation depth and the indenter type but also on the material behaviour, i.e., {\em the reduced stress relaxation function} $\phi$.
The relation between the load, the contact radius and the material behaviour -- the first of \eqref{eq:precede3} -- is now mathematically the same as for purely elastic materials but the relation between the experimental data -- the second of \eqref{eq:precede3} --  and the material behaviour is completely different. Mathematically, the latter formula is a nonlinear integral equation.

It is generally difficult to determine whether the contact is receding or not only on the basis of the experimental data $h$ and $p$. Differentiation of the first of \eqref{eq:radius2} to time shows that the  contact radius is receding for $t>0$ if $[h\orc\phi]/\phi$ is a decreasing function in this time period because $\dot{c}\mathbb{L}''(c)=\text{d}\{[h\orc\phi]/\phi\}/\text{d}t$ and $\mathbb{L}''(c) > 0$. After some elementary calculations one finds
\begin{equation}
\dot{c}=\frac{1}{\mathbb{L}''(c)}\Bigl\{
\dot{h}_0-\frac{1}{\phi^2(t)}\int_0^t\,\,\bigl\{\dot{h}(\tau)\dot{\phi}(t)\phi(t-\tau)-
\ddot{h}(\tau)\phi(t)\phi(t-\tau)\bigr\}\,\text{d}\tau\Bigr\}~
\label{eq:Signa}
\end{equation}
and in the limit $t\downarrow 0$  $\dot{c}_0=\dot{h_0}/\mathbb{L}''(c_0)$ results. As expected, an initial negative value for the indenter speed is a necessary condition for a strictly receding contact radius. For other times the sign of $\dot{c}$   depends also on the rest of the term within brackets in \eqref{eq:Signa}. Considering that $\phi$ is a reduced stress relaxation function and the material has fading memory, i.e., $\phi >0$ and  $\dot{\phi}<0$, it follows that the conditions $\dot{h} < 0$ {\em and} $\ddot{h}<0$ are sufficient for receding contact.
The rate of change of the load is $\dot{p} = 4\{\dot{\phi}p/\phi+\dot{c}\omega c\mathbb{L}''(c)\}$.
As $\dot{\phi}$ and $\dot{c}$ are negative while $p/\phi$ and $\omega c \mathbb{L}''(c)$ are positive, a receding contact is accompanied by a decreasing load but it is not so that a  decreasing load always indicates a receding contact as the phenomenon of a 'nose' demonstrates.
\section{Increasing and decreasing contact radius}\label{sec:AdvancingRecedingContact}
The solutions from the two previous subsections are valid along the whole time axis. Instead of monotonically increasing or decreasing contact regions it is still possible to have $c(\tau)<c(t)$ or $c(\tau)>c(t)$ for $0<\tau<t$ except that these inequalities are not always valid but only for the current time values in certain time intervals.
\begin{figure}[htb]
  \centering
  \includegraphics[scale=1]{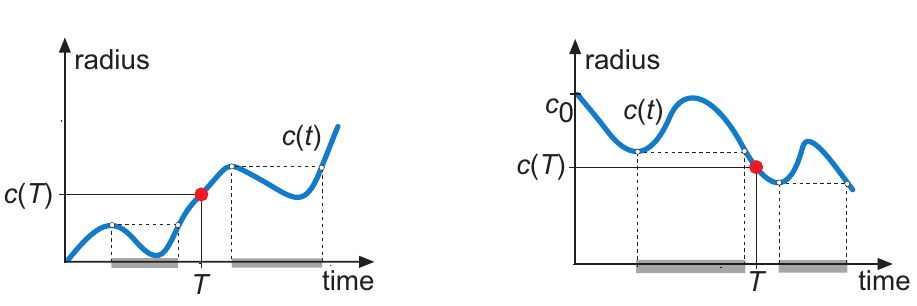}\\
  \caption{ Varying but globally increasing (left) or decreasing contact radius (right) ; $c(\tau)<c(t)$ or $c(\tau)>c(t)$ for $0<\tau<t$ only valid for times $t$, in the non-greyed regions.}\label{fig:VarstContactRadiusPartial}
  \end{figure}
For example, for those times in the non-grayed intervals, e.g. time $T$  in Figs.~\ref{fig:VarstContactRadiusPartial} (left) or \ref{fig:VarstContactRadiusPartial} (right) the results \eqref{eq:radius1} or \eqref{eq:radius2} apply but for other times, i.e., those  located in the grayed time ranges in Fig.~\ref{fig:VarstContactRadiusPartial} they do not.

Of special interest is the varying but globally increasing contact radius  because for every time value in the non-grayed intervals is the contact radius slaved only to the current depth, e.g. at time $T$ (Fig.~\ref{fig:VarstContactRadiusPartial}, left) is $c(T)=\mathbb{C}(h(T)$; during these time periods, the contact radius is linked to experimental data for $h(t)$ in just the same way as for purely elastic materials.

Essential is the answer to the question whether a line of constant radius to a particular point, say $(t,r)=(T,c(T))$, on the $c(t)$ versus $t$ curve intersected this curve at an earlier time or not. If not, either $c(\tau)<c(T)$ for $0<\tau<T$ in which case the theory of Chap.~\ref{sec:increasing} applies at time $T$ or $c(\tau)>c(T)$ for $0<\tau<T$ is valid and the theory of Chap.~\ref{sec:decreasing} can be used. The analysis becomes much more complicated if an intersection occurred at an earlier time and this might happen happens when the dynamical variant of the experiment is performed. Although in this case a monotonic contact radius might prevail, the classic analysis of dynamic load-depth sensing has serious limitations which are treated in more detail in Chap. \ref{sec:perturbation}. In dynamic load-depth sensing the  control variable varies periodically or a periodic perturbation is superposed on it. The upshot is that a point on the surface might move many times in and out of the contact region and the type of the appropriate boundary condition at these points also changes many times.  Analysis of these cases might be better performed via the decomposition method of Golden and Graham \cite[p. 63--69]{GoldenGraham1988}. More details follow later in the Chapters~\ref{sec:GenGSep},  \ref{sec:SingleLOadUnload} and \ref{sec:hereditary}.
\section{Classic dynamic load depth sensing}\label{sec:perturbation}
During dynamic load depth sensing  a small periodic perturbation is superposed on the global indenter motion and the idea is to obtain the frequency dependent transfer function linking input perturbation and the resulting effect on output variable. As the method is inherently dynamic, elastic stiffness, damping properties of the equipment and inertial effects play an import role and need to be incorporated in the analysis (see, e.g., \cite{HerbertEtAl2008}; \cite[][p. 126--131]{FischerCripps2011}). However, as the centre of attention is here the response of the substrate -- as governed by the equations \eqref{eq:loadequation3} and \eqref{eq:loadequation3d} or \eqref{eq:radiusequation} and \eqref{eq:loadequation3e} -- the influences of the equipment properties (stiffness, damping, mass)  are not taken into account.
\subsection{Basic setup}
The goal is to determine the frequency dependent storage and loss moduli of the material by measuring the amplitude and phase shift of sinusoidal response to a sinusoidal perturbation of the global control variable (depth or load). Experimentally, the choice of the global control variable -- also called carrier variable \cite{HuangEtAl2004} or main variable \cite[p. 81]{KnausEtAl2008} -- is very important for a number of reasons. Firstly, the size of the total input variable  must be such that never contact is lost during the experiment and loss of contact is signalled by a vanishing of the load. Secondly, in the classic approach one works basically with the analytical formula valid for monotonically increasing contact radius, i.e., with load-depth relation: $[p\orc\varpi](t)=4\mathbb{F}(h(t))$, but then -- necessarily -- the experimental setup must be such that the  contact radius indeed increases monotonically\footnote{See the studies by Huang et al. \cite{HuangEtAl2004} for a spherical indenter and Cheng et al. \cite{ChengEtAl2006} for a conical and spherical indenter; in both papers the ratio $\lambda(t)/\mu(t)$ was considered constant.  Knaus et al. \cite[p. 82]{KnausEtAl2008} also discussed this problem.}. The pertinent depth-contact radius connection is $h(t)=\mathbb{L}'(c(t))$ and also $\dot{h}(t)=\mathbb{L}''(c(t))\dot{c}(t)$; monotonically increasing (non-decreasing) depth is a necessary condition for monotonically increasing (non-decreasing) contact.  Thirdly, as the total response contains three effects, namely a transient part, a part due to the carrier input and a part due to the superposed perturbation the latter part must be extracted from the raw data.
\subsection{Limitations}
One way to ensure contact is to choose a step as (global) input variable with size much larger than the amplitude of the perturbation. The advantage of a step as global control variable is that in both cases -- load or depth control -- the part of the contact radius associated with the carrier step  is either constant (see Chap.~\ref{sec:ClassicRelaxation})  or increasing (see Chap.~\ref{sec:ClassicCreep}) and that asymptotically, i.e., for large time, the part of all responses associated with the carrier variable becomes constant thus facilitating the separation of the perturbational part of the response from the carrier part. Unfortunately, the behaviour of the total contact radius might be quite different from that associated with the carrier part.  If the experiment is driven in depth control with, e.g. $h(t)=h\mean\{1+\epsilon\sin(\Omega t)\}\hea{t}$ with $h\mean$ the carrier step size and $\epsilon\ll 1$ the relative amplitude of the perturbation, the indenter speed, $\dot{h}(t)=\epsilon\Omega h\mean\cos \Omega t$, changes periodically of sign and it follows that for this type of control the load-depth relation $p\orc\varpi=4\mathbb{F}(h)$ is at best -- for  very small $\epsilon$ -- only approximately true. Alternatively, for load control with $p(t)=p\mean\{1+\epsilon\sin\Omega t\}\hea{t}$ one finds asymptotically
\begin{equation}
[p\orc\varpi](t)\sim p\mean\varpi(\infty)+\epsilon p\mean\left(\widehat{\varpi}_\mathrm{st}(\Omega)\sin\Omega t + \widehat{\varpi}_\mathrm{lo}(\Omega)\cos\Omega t\right)\label{eq:AsympLoad}
\end{equation}
The frequency dependent functions $\widehat{\varpi}_\mathrm{st}(\Omega)$ and $\widehat{\varpi}_\mathrm{lo}(\Omega)$ are the storage and loss modulus associated with the creep function $\varpi$, i.e., the complex modulus $\widehat{\varpi}_\mathrm{st}(\Omega)+i\widehat{\varpi}_\mathrm{lo}(\Omega)$ is the Fourier transform of $i\Omega\varpi(t)$.
In view of the assumed load-depth relation, $p\orc\varpi=4\mathbb{F}(h)$, the factor $4\mathbb{F}(h)$ should asymptotically also behave like \eqref{eq:AsympLoad}, i.e., vary sinusoidally, and this again implies that the depth rate changes sign. Therefore, the conclusion is that also for this type of load control the load-depth relation $p\orc\varpi=4\mathbb{F}(h)$ is at best -- for  very small $\epsilon$ -- only approximately true.
 \begin{figure}[htb]
  \centering
  \includegraphics[scale=1]{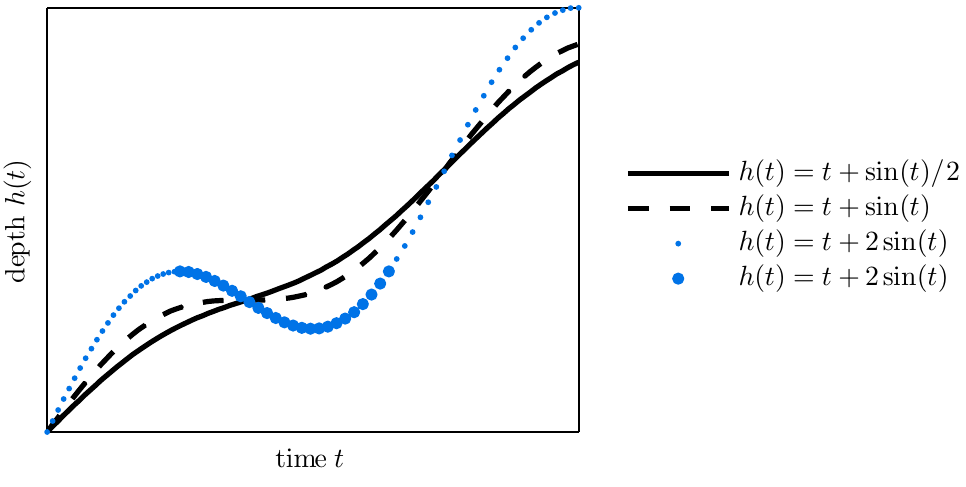}\\
  \caption{Depth control with sinusoidally perturbed ramp shaped carrier depth. Solid and dashed line: $h(t)=\mathbb{L}'(c(t))$ everywhere.
  Dotted line: depth  $h(t)=t+2 \sin t$;  $h(t)=\mathbb{L}'(c(t))$ (thin dot data points) or $h(t)\neq\mathbb{L}'(c(t))$ (thick dot data points).}\label{fig:Fluctuerend}
\end{figure}

To circumvent the problems signalled in the previous paragraph Huang et al. \cite{HuangEtAl2004} and Knaus et al. \cite[][p. 83]{KnausEtAl2008} suggested to use a ramp type function as carrier variable but the disadvantage of this approach is that  separating the perturbational part from the carrier response is much more difficult. Additionally, this limits the ramp speed, frequency and perturbation amplitude that can be used. For example, using for depth control the function $h(t)=vt +h_1\sin\Omega t$, the indenter speed $\dot{h}$ is only positive -- a necessary condition for an increasing contact -- as long as $\Omega h_1/v\leq 1$ (see Fig.~\ref{fig:Fluctuerend}). Otherwise, i.e.,  $\Omega h_1/v>1$, the relation $h(t)=\mathbb{L}'(c(t))$ is only valid in certain time periods whereas in others $h(t)$ is not equal to $\mathbb{L}'(c(t))$,  though the deviation might be small if $\Omega h_1/v$ differs only slightly from the value 1.

\section{Occurrence of a 'nose' in the load versus depth curve}\label{sec:Nose}
A monotonic increasing contact radius is always accompanied by an increasing depth because $\dot{h}=\dot{c}\mathbb{L}''(c)$ with $\mathbb{L}''(c)>0$. So, for a contact, initially increasing, the depth is also increasing, at least up to the moment that $\dot{c}$ becomes zero for the first time or jumps to a negative value when the contact radius has a kinked maximum. However, the load might have reached its maximum earlier as the (possible) appearance of a 'nose' demonstrates (Fig.~\ref{fig:VarstNose}b and Fig.~\ref{fig:VarstNose}c). Typical of a nose is that the load decreases whereas the depth and contact radius still increase.

Differentiation of the first of \eqref{eq:advancingcontact2b} to time gives
\begin{equation}
\varpi_0\dot{p}(t)+\dot{\varpi}(t)p_0+\int_0^{t}\dot{p}(\tau)\dot{\varpi}(t-\tau)\,\text{d}\tau=4c(t)\dot{h}(t)\quad 0<t<t_\mathrm{m}~.
\label{eq:pdot}
\end{equation}
Because $c(t)>0$, $\varpi_0>0$, $p_0\geq 0$, $\dot{\varpi}(t)>0$, $\dot{p}(t)>0$ for $0<t<t_1$, the integrand, $\dot{p}(\tau)\dot{\varpi}(t-\tau)$, in \eqref{eq:pdot} is non-negative for $0<\tau < t <t_1$. Therefore, the left-hand side of this equation is positive in the interval $0<t<t_1$ and it follows that the right-hand side is also positive in this time range for any $t_1< t_\mathrm{m}$.

If the load-depth curve is smooth (Fig.~\ref{fig:VarstNose}b) with $\dot{p}=0$  at $t=t_1$, the right-hand side of \eqref{eq:pdot} remains positive for some time afterwards and so does the left-hand side;  $\dot{h}(t_1)>0$ because $c(t_1)>0$. Therefore, the depth reaches a maximum at a later time and a 'nose' is always present. This is as expected for thermodynamic reasons because the substrate material is dissipative.

If the load-depth curve is kinked at the load maximum (Fig.~\ref{fig:VarstNose}c), a 'nose' might or might not be present. Just after the kink, i.e., at $t=t_1+0$, the rate $\dot{p}(t_1+0)$ is negative and  \eqref{eq:pdot} shows that $\dot{h}(t_1+0)>0$ (presence of a 'nose') whenever
\begin{equation}
\dot{\varpi}(t_1+0)p_0+\int_0^{t_1}\dot{p}(\tau)\dot{\varpi}(t_1-\tau)\,\text{d}\tau>-\varpi_0\dot{p}(t_1+0)>0~.
\label{eq:inequal}
\end{equation}
So, unloading sufficiently slow, i.e., decreasing the term $-\varpi_0\dot{p}(t_1+0)$ in \eqref{eq:inequal}, leads to a nose thus corroborating the numerical results of Cheng and Cheng \cite{ChengAndCheng2004}. Reducing the left-hand side of this inequality is possible by introducing a hold period, i.e., a period with $\dot{p}=0$ prior to the start of unloading at $t_1$. If this period is sufficiently long the inequality (\ref{eq:inequal}) is no longer met and $\dot{h}(t_1+0)<0$; the 'nose' disappears as found experimentally \cite{BriscoeEtAl1998, SahinEtAl2007}.
\section{The initial elastic response}
In Chap.~\ref{sec:ContactRadiusEquations} it was argued that the determination of the material function $\omega(t)$ better be split in two steps the first of which is the determination of the elasticity factor $\omega_0$. Theoretically this first phase is straightforward; only an initial step in the control variable and taking the limit $t\downarrow 0$ of the response is needed to generate all necessary information because initially $p_0=4\omega_0\mathbb{F}(h_0)$. However, an ideal step in the control variable is difficult to achieve experimentally and in reality a step normally consists of a ramp followed by a constant value in which case the limit $t\downarrow 0$ does not make sense. Although in principle this method is one whereby the contact radius is increasing during the ramping and the theory of Chap.~\ref{sec:increasing} applies, it is not a proper method to determine the initial elastic response.
\section{The elastic response to jumps in the load and depth rates}\label{sec:JumpConditions}
An alternative way to determine the elasticity factor $\omega_0$ is to look at the response to rate jumps as mentioned previously  in Chap.~\ref{sec:intro}.

Macroscopically, the depth $h(t)$ or the load $p(t)$ control the indentation process and -- apart from a possible jump at time zero -- these two variables are continuous functions for $t>0$ and this also applies to the resulting functions, contact radius $c(t)$ and strain $u'(r,t)$. However, as sharp transitions from loading to unloading, a sudden introduction or ending of a hold period may cause kinks in the time dependence of load and depth curve the time derivative of depth and load may well be only piecewise continuous; the rates  $\dot{h}$ and/or $\dot{p}$ exhibit jumps at certain times, say at time $T$ and rate jumps $\langle\dot{h}\rangle_T$ and/or $\langle\dot{p}\rangle_T$ are nonzero.

The sets of basic equations \eqref{eq:radiusequation} and \eqref{eq:loadequation3e} still apply for all time, but time derivatives may not exist at $T$, i.e., the time derivatives might have jumps here too. In Appendix \ref{app:JumpConditionsAanalysis} formulae are derived for the rate jumps in $h$, $c$,  and $p$ and they are recapitulated here (see \eqref{eq:jumpb} and \eqref{eq:jumpc})
 \begin{equation}
 \omega_0\frac{\langle\dot{h}\rangle_T}{\langle\dot{c}\rangle_T}
 =\frac{\pi}{2}\theta'(c(T),T),
\quad
\frac{\langle\dot{p}\rangle_T}{\langle\dot{h}\rangle_T}=4\omega_0c(T).
 \label{eq:loadanddepthjump}
 \end{equation}

 These rate jumps are determined inside the contact region up to the edge , i.e., for $r\leq c(T)$.

The presence of the quantity $\theta'$ in the results enables the determination of the signs of the rate jumps. Near the contact radius, but inside the contact zone, i.e., at
$r=(1-\delta)c$ (with $c=c(t)$ and $\delta$ a small but positive number), the stress formula \eqref{eq:stress1} gives
\begin{equation}
\sigma(c(1-\delta),t)=-\theta'(c,t)\sqrt{2\delta}\,\,\bigl\{1+\mathcal{O}(\delta)\bigr\}~.
\end{equation}
As the normal stress $\sigma$ is compressive, i.e., of negative sign,
$\theta'(c(t),t)$ is always non-negative. In addition, $c$ and $\omega_0$ are also positive, so
\begin{equation}
\text{sign}\,\langle\dot{p}_T\rangle=\text{sign}\,\langle\dot{h}_T\rangle=
\text{sign}\,\langle\dot{c}_T\rangle~.
\label{eq:velocityjumps}
\end{equation}

All these results lead to the conclusion that jumps in the loading rate $\dot{p}$, the indentation rate $\dot{h}$ and the rate of change $\dot{c}$ of the contact radius  always \emph{occur simultaneously and are of equal sign.}

The second of \eqref{eq:loadanddepthjump} shows that the response to jumps in the rates is elastic and this corroborates earlier results of Ngan et al. \cite{NganEtAl2005}, de With \cite[][p. 609--610]{DeWith2006}, Ngan and Tang \cite{NganAndTang2009} and Tang and Ngan \cite{TangAndNgan2012}. It parallels the equation $S=4\omega\young c$ for the contact stiffness for elastic materials with $\langle\dot{p}\rangle_T/\langle\dot{h}\rangle_T$ functioning as the contact stiffness.

The response to jumps in the rates is always elastic; it is immaterial at which point of the load-depth curve  the rate jumps occur, it is immaterial whether a nose is present afterwards or not and it is even immaterial whether the load-depth curve kinks upwards or not. However, the contact radius $c(T)$ appearing in the equation might be dependent on the reduced relaxation function $\phi$ (see Chap.~\ref{sec:decreasing} \& \ref{sec:AdvancingRecedingContact}). So, the practical use of this equation is limited to the situation where the contact radius is independent of the material properties, i.e., to those situations where the contact radius at the jump time exceeds all previous values because only in these cases  $c(T)=\mathbb{C}(h(T))$  (Fig.~\ref{fig:VarstContactRadius}a, Fig.~\ref{fig:VarstContactRadius}c and Chap.~\ref{sec:increasing}).
The rate jump ratio can be determined directly from  measured rates \cite{NganAndTang2009} or indirectly, e.g. by applying a correction to the unload stiffness after a specific load schedule \cite{FengNgan2002,TangAndNgan2003,NganEtAl2005} or by a sufficiently fast unload in which case the rate jump ratio equals approximately the unload stiffness \cite{ChengAndCheng2005a,ChengAndCheng2005b,ChengEtAl2006b}.

\subsection{Intermezzo: the influence of plastic deformation}\label{sec:PlasticInfluence}
\subsubsection{The theory of the effective indenter}
\begin{figure}[htb]
\centering
\includegraphics[scale=1]{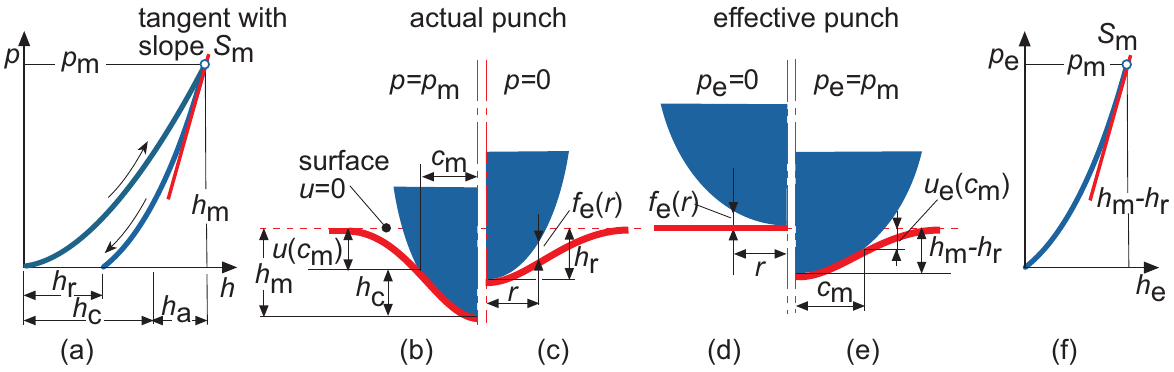}\\
\caption{Elasto-plastic material. Leftmost: load-depth curve, first loading and unload curve (a). Middle, left: actual configuration at maximum load (b) and at zero load after unloading (c). Middle, right: 'effective' configuration at zero load (d) and at maximum load (e). Rightmost: load-depth curve  'effective indenter' (f). Variables pertaining to the 'effective' configuration are supplied with the subscript 'e'.} \label{fig:VarstElastoPlastic}
\end{figure}

In practice, an additional complication arises because plastic deformation, however slight, is always present and this phenomenon obscures the relation between contact radius and depth. During the first loading the geometry of substrate changes  from a flat surface -- an assumption on which the theory is based -- to one having a permanent impression.
The complication caused by plasticity also occurs when elastic materials are indented. In this case, this problem is solved by using only data of the unload curve -- possibly after load cycling a number of times, i.e., preconditioning the substrate -- because along this curve only elastic properties are probed (Fig.~\ref{fig:VarstElastoPlastic}a). The slope $S_\mathrm{m}$ of the tangent at the start of the unload curve still equals $4\omega\young c_\mathrm{m}$ \cite{ChengAndCheng1997} but the value of $c_\mathrm{m}$ must be estimated in a different way, that is to say different as compared to the purely elastic case. To do so, several methods have been derived in the past (see, e.g., \cite[Chapter  3]{FischerCripps2011}) but modern methods are based on the idea of the 'effective indenter'
\cite{OliverAndPharr1992,BolshakovEtAl1994,WoigardAndDargenton1997,PharrAndBolshakov2002,OliverAndPharr2004}. The geometry change is shifted from the substrate to the indenter because the 'effective indenter' is considered to be pressed into a flat, purely elastic substrate; the basic idea is graphically elucidated in Fig.~\ref{fig:VarstElastoPlastic}. The shape profile of the 'effective indenter' is  assumed to be such that the \emph{load-depth curve of the 'effective indenter' matches the experimental data of the actual unload curve} (Fig.~\ref{fig:VarstElastoPlastic}f).
At full load\footnote{Variables pertaining to the effective configuration are supplied with the subscript 'e'}, i.e., at $p_\mathrm{e}=p_\mathrm{m}$, $S_\mathrm{e}=S_\mathrm{m}$ and therefore $c_\mathrm{e}=c_\mathrm{m}$. Two additional assumptions are vital; firstly, the displacements at the edge of the contact regions are the same when the contact radii are the same, i.e $u_\mathrm{e}(c_\mathrm{m})=u(c_\mathrm{m})$, and, secondly, the 'effective indenter' shape is a power law as this implies that $u_\mathrm{e}(c)\propto -p_\mathrm{e}/S_\mathrm{e}$. The upshot of all this is that at full load:
\begin{equation}
h_\mathrm{c}=h_\mathrm{m}+u(c_\mathrm{m})=h_\mathrm{m}-\varepsilon p_\mathrm{m}/S_\mathrm{m} \label{eq:PlasticCorrection0}
\end{equation}
in which $\varepsilon$ is an indenter specific constant and the contact radius is now determined using the area function mentioned in Chap.~\ref{sec:intro}.
\subsubsection{Correcting for plasticity effects}
To evaluate, experimentally, the initial elastic response of polypropylene and amorphous selenium and, numerically, the accuracy of the estimated contact radius of a power law creeping material,  Tang and Ngan \cite{TangAndNgan2003}, Ngan et al. \cite{NganEtAl2005}, Ngan and Tang  \cite{NganAndTang2009} used the same procedure by substituting their creep corrected contact stiffness -- effectively equal to the rate jump ratio because of the applied load schedule -- for the slope in \eqref{eq:PlasticCorrection0}, i.e., they used
\begin{equation}
h_\mathrm{c}(T)=h(T)-\varepsilon p(T)\langle\dot{h}\rangle_T/\langle\dot{p}\rangle_T\label{eq:PlasticCorrection1}
\end{equation}
to approximate the contact depth.

The validity of this formula is analysed in Appendix \ref{app:Plastic} by application of the 'effective indenter' idea to viscoelastic materials, using the same assumptions as for the elastic case and and considering a situation where the contact radius increases up to the jump  time $T$ and exceeds there all previous values. It is shown that, instead of \eqref{eq:PlasticCorrection1}, the final result, equation \eqref{eq:effectiveVisco5b}, is found to be
\begin{equation}
h_\mathrm{c}(T)=h(T)-\varepsilon p(T)\frac{\langle\dot{h}\rangle_T}{\langle\dot{p}\rangle_T}
\mathcal{R}(T).\label{eq:effectiveVisco5}
\end{equation}
The factor $\mathcal{R}$ (see \eqref{eq:effectiveVisco5a}) is
\begin{equation}
 \mathcal{R}(T)=1+\int_0^T\frac{p(\tau)}{p(T)}\dot{\varphi}(T-\tau)\mathrm{d}\tau
\label{eq:effectiveVisco5c}
\end{equation}
and for  elastic materials  $\mathcal{R}(T)$ is always 1. For viscoelastic materials $\mathcal{R}$ depends on the load history prior to $T$ and the reduced creep function $\varphi$. For the considered load schedule, i.e., $p$ starting from zero and monotonically increasing up to $T$ one finds -- on account of the second mean value theorem\footnote{In literature also known as the third mean value theorem \cite[][p.236--237]{Spivak1967} or the Weierstrass form of Bonnet's theorem \cite[][p. 138]{Widder1947}}  for integrals \cite[p. 1095]{GradshteynAndRyzhik1980} -- that $\mathcal{R}(T)=\varphi(\xi(T))$ with $0<\xi(T)<T$, so $1<\mathcal{R}(T)<\varphi(T)$.

\chapter[Decomposing hereditary integrals]{ Decomposing the hereditary integrals in the indentation equations}\label{sec:GenGSep}
\section{General part}\label{sec:GenGSep1}
In the normal experimental situation an indentation experiment starts at zero contact radius and ends at zero contact radius. In the mean time the contact radius increases and decreases only once or a few times -- conventional load-depth sensing -- or, alternatively, a large number of times as is the case in the dynamic variant of the experiment; a point  at the substrate surface might move many times in and out the contact region. If this point is currently located in the contact region one knows the normal displacement here because in the contact region the surface conforms to the indenter; locally a displacement boundary condition applies at this point in time. If, perhaps some time later or earlier, the same surface location is part of the free surface, the stress must be zero here and a stress boundary condition is now valid.

\begin{figure}[htb]
  \centering
  \includegraphics[scale=1]{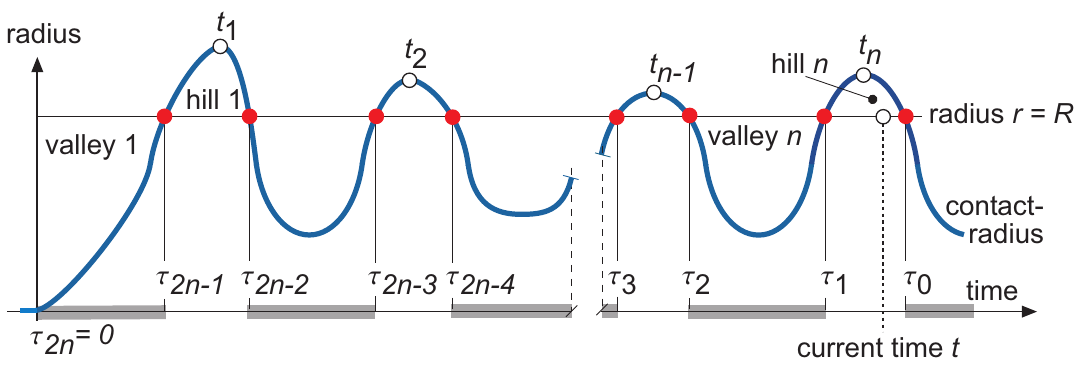}\\
  \caption{History of varying contact radius showing that points at a fixed distance $r=R$ move in and out the contact region many times. During the grayed time intervals  $\varrho_c(R,t)=0$ and $q_c(R,t)=\ell(t)$ and during the non-grayed ones $\mathcal{L}'(R,t)=\mathbb{L}'(R)-h(t)$ and $\mathcal{L}(R,t)=\mathbb{L}(R)-Rh(t)$. The current time is here located below 'hill' $n$.}\label{fig:VarstFluctuatingContacta}
\end{figure}

Consider the basic equations \eqref{eq:loadequation3} and \eqref{eq:loadequation3d} at a fixed value of $r$, i.e., along a line of constant radius (Fig.~\ref{fig:VarstFluctuatingContacta}), say $r=R$. To simplify the notation, these formulae are divided by $\omega_0$ and the following variables are defined
\begin{equation}
\begin{matrix}
\rho_c(r,t)=\dfrac{\pi\vartheta_c(r,t)}{2\omega_0}, & \ell(t)=\dfrac{p(t)}{\omega_0}, \\[8pt]
q_\mathrm{c}(r,t)=\dfrac{\wp_\mathrm{c}(r,t)}{\omega_0}, &\mathcal{L}(r,t)=L(r,t)-rh(t).
\end{matrix}\label{eq:Scaled}
\end{equation}
The expressions \eqref{eq:loadequation3} and \eqref{eq:loadequation3d} now take the form
\begin{gather}
\varrho_c(R,t)=[\mathcal{L}'\orc\phi](R,t)
\genfrac{}{}{0pt}{}{\text{\scriptsize{convolution}}}{\leftrightarrows}
\mathcal{L}'(R,t)=[\varrho_c\orc\varphi](R,t),\label{eq:GenGRadius}\\
q_c(R,t)=4[\mathcal{L}\orc\phi](R,t)
\genfrac{}{}{0pt}{}{\text{\scriptsize{convolution}}}{\leftrightarrows}
4\mathcal{L}(R,t)=[q_c\orc\varphi](R,t).\label{eq:GenGLoad}
\end{gather}
Viewed along a line of constant $r$ these equations can be envisaged as two different examples of a single mathematical problem, namely the determination of two time-dependent functions\footnote{That in indentation the idea is to determine the material function $\phi$ or $\varphi$ is for the moment left out of consideration.} $a(t)$ and $b(t)$, related by $a(t)=[b\orc\phi](t)$ \emph{and} $b(t)=[a\orc\varphi](t)$, whereby the time axis consists of a set of disjoint, but abutting, intervals -- the grayed and non grayed time intervals in Fig.~\ref{fig:VarstFluctuatingContacta} -- where either $a(t)$ or $b(t)$ is given.  For the equations \eqref{eq:GenGRadius}  $a(t)=\varrho_c(R,t)$ and $b(t)=\mathcal{L}'(R,t)$ and for the equations \eqref{eq:GenGLoad}: $a(t)=q_c(R,t)$ and $b(t)=4\mathcal{L}(R,t)$.

The mathematical key feature is that given and unknown functions occur intermixed on the time axis. Hence, the integrand  in the convolution integral $[a\orc\phi](t)$ is not known on the complete integration range because the integration variable  passes through various intervals on which the function $a$ is not known. However, precisely on these intervals  is the integrand of the dual equation $b\orc\varphi$ defined because $b$ is known here. Alternatively, on all intervals where the integrand of $b\orc\varphi$ is not known, is that of $a\orc\phi$ given.  To tackle this problem, Golden and Graham \cite[p. 63--69]{GoldenGraham1988} decompose the convolution integrals in such a way that every interval where $a$ (or $b$) is not known turns out to be a sum of integrals over intervals with known integrands. How the decomposition technique actually works is explained in Appendix \ref{sec:GenGAppendix}.

The governing equations \eqref{eq:GenGRadius} and \eqref{eq:GenGLoad} are decomposed using this technique. The decomposed equations have the property that the intermixing of the depth, contact radius and load, present  in the original equations eventually disappears (see  Chap.~\ref{sec:SimplifiedLoad}).   Later, the results are applied to the case of classic indentation (Chap.~\ref{sec:SingleLOadUnload}), i.e., simple loading and unloading,  and to dynamic indentation experiments  (Chap.~\ref{sec:hereditary}) where a sinusoidal perturbation is superposed on a constant control variable (depth or load).

\section{Viewing the time axis as a sequence of 'hill' and 'valley' intervals}
A traveller, metaphorically  moving along the line $r=R$ (Fig.~\ref{fig:VarstFluctuatingContacta}),  passes in the course of time through ''hills', i.e., time intervals for which $c(t)>R$, and over ''valleys'', intervals for which $c(t)< R$.
The 'hills', are numbered progressively in the positive time direction with $t_1$, $t_2$, \ldots, the times of the local maxima in these regions, and the 'valley' directly preceding a 'hills' is assigned the same number (Fig.~\ref{fig:VarstFluctuatingContacta}). The current value $t$ of time is always located in the union of some 'valley' and the next 'hill' interval, say in the union of 'valley' $n$ and 'hill' $n$, and the time $\tau_0$ is -- by definition -- the time where 'hill' $n$ changes into 'valley' $n+1$. The values of boundary times
where the passage from 'valley' to 'hill' and vice versa  occurs, $\tau_1$, $\tau_2, \ldots$ are again numbered progressively but now in the \emph{negative} time direction until $\tau_{2n-1}$, which is basically the first actual intersection point. Finally, the sequence is completed by adding $\tau_{2n}$ and choosing this time equal to zero. Obviously, the time values $\tau_j$
depend on the radius $R$  because -- by construction $c(\tau_j)=R$ -- but the subscript value may also change depending on the value of the current time one considers because this also determines the value of $n$ and the starting point for the $\tau$-numbering.   Moving with the current time from 'valley' $n$ to 'hill' $n$ does \emph{not change} the numbering but moving from a 'hill' $n$ to a 'valley' $n+1$ does. Consequently
\begin{subequations}\label{eq:TauRelations}
\begin{equation}
\tau_j(R,n)=\tau_{j+2}(R,n+1)\enspace\text{for $j=0, 1,\ldots,2n-1$,}\label{eqPart1:TauRelations}
\end{equation}
\begin{equation}
\mathcal{L}(R,\tau_j)=\mathbb{L}(R)-Rh(\tau_j)\enspace\text{for $j=0, 1,\ldots,2n-1$,}\label{eqPart2:TauRelations}
\end{equation}
\end{subequations}
whereas for the subscript value $j=2n$ only $\tau_{2n}(R,n)=0$ is defined.
\section{The depth equation}\label{sec:ContactRadius}
\subsection{Current time in a 'valley' interval: $\tau_2(R,n)<t<\tau_1(R,n)$}
For $t$ in  'valley' interval $n$, \eqref{eq:BNIs1} applies for $n=1$ and \eqref{eq:BNIsn} for $n\geq 2$. Take $a(t)=\varrho_c(R,t)$, $b(t)=\mathcal{L}'(R,t)$, $\tau_2(R,1)=0$ in these equations and note that $\varrho_c(R,s)=0$ in the 'valley' intervals and $\mathcal{L}'(R,s)=\mathbb{L}'(R)-h(s)$ below the 'hills. The result is
\begin{equation}
\mathcal{L}'(R,t)=\begin{dcases} 0&\quad n=1,\\
\sum\limits_{k=1}^{n-1}\int\limits_{\tau_{2k+1}(R,n)}^{\tau_{2k}(R,n)}
N_{2k}(t,s,R,n)\Bigl\{\mathbb{L}'(R)-h(s)\Bigr\}\,\mathrm{d}s&\quad n \geq 2.
\end{dcases}\label{eq:GenGRadius1}
\end{equation}
The material properties are contained in the functions $N_j$ (see \eqref{eq:NEven} and \eqref{eq:NUneven}) starting with
\begin{equation}
N_1(t,s,R,n)=\dot{\varphi}(t-s),
\end{equation}
and proceeding with
\begin{multline}
N_{2k}(t,s,R,n)=
N_{2k-1}(t,s,R,n)\\
\quad+\int\limits_s^{\tau_{2k}(R,n)}N_{2k-1}(t,u,R,n)\dot{\phi}(u-s)\,\mathrm{d}u,\enspace k\geq1,
\end{multline}
plus
\begin{multline}
N_{2k+1}(t,s,R,n)=
N_{2k}(t,s,R,n)\\
\quad+\int\limits_s^{\tau_{2k+1}(R,n)}N_{2k}(t,u,R,n)\dot{\varphi}(u-s)\,\mathrm{d}u,\enspace k\geq 1.
\end{multline}
These functions act as kernel for the integral operators $\mathcal{V}_k$ and $\mathcal{W}_k$ defined by
\begin{gather}
\mathcal{V}_{k}(t,R,n)=\int\limits_{\tau_{2k+1}(R,n)}^{\tau_{2k}(R.n)}N_{2k}(t,s,R,n)\,\mathrm{d}s,\label{eq:VkFunction}\\
\mathcal{W}_{k}(t,R,n)=\int\limits_{\tau_{2k+1}(R,n)}^{\tau_{2k}(R.n)}N_{2k}(t,s,R,n)h(s)\,\mathrm{d}s.\label{eq:WkFunction}
\end{gather}
With these definitions \eqref{eq:GenGRadius1} becomes
\begin{equation}
\mathcal{L}'(R,t)=\begin{dcases} 0&\quad n=1,\\
\mathbb{L}'(R)\sum\limits_{k=1}^{n-1}\mathcal{V}_k(t,R,n)-\sum_{k=1}^{n-1}\mathcal{W}_k(t,R,n)&\quad n \geq 2.
\end{dcases}\label{eq:GenGRadius2}
\end{equation}
The formulae \eqref{eq:GenGRadius2} represent a 'decomposed' version of \eqref{eq:loadequation3} but to find the equivalent of \eqref{eq:radiusequation} -- the equations relating depth, contact radius and material properties along the curve of the contact radius -- the limits $t\uparrow \tau_1(R,n)$ and $t\downarrow \tau_2(R,n)$ must be taken because the fixed radius $R$ equals the contact radius at these times.

The limit $t\uparrow \tau_1(R,n)$, the end of a 'valley', yields for $n=1$
\begin{equation}
\mathbb{L}'(R)=h(\tau_1)), \enspace \tau_1=\tau_1(R,1),\label{eqPart1:GenGRadius4}
\end{equation}
and for $n\geq 2$
\begin{multline}
\mathbb{L}'(R)\left(1-\sum\limits_{k=1}^{n-1}\mathcal{V}_k(\tau_1,R,n)\right)=h(\tau_1)\\
    -\sum\limits_{k=1}^{n-1}\mathcal{W}_k(\tau_1,R,n),\enspace \tau_1=\tau_1(R,n).\label{eq:GenGRadius4}
\end{multline}
Since $R=c(\tau_1(R,1))$, the limit for $n=1$ can also be written as
\begin{equation}
h(\tau_1)=\mathbb{L}'(c(\tau_1)),\enspace \tau_1=\tau_1(R,1).\label{eq:GenGRadius3}
\end{equation}
This expression is identical to \eqref{eq:radius1}, the result found earlier for an increasing contact radius .

The limit $t\downarrow \tau_2(R,n)$, the start of a 'valley', only makes sense for $n\geq 2$ because $\tau_2(R,1)$ is always zero, whatever the value of $R$ might be. This limit gives
\begin{multline}
\mathbb{L}'(R)\left(1-\sum\limits_{k=1}^{n-1}\mathcal{V}_k(\tau_2,R,n)\right)=h(\tau_2)\\
-\sum\limits_{k=1}^{n-1}\mathcal{W}_k(\tau_2,R,n),
\enspace\tau_2=\tau_2(R,n),\enspace n\geq 2.\label{eq:GenGRadius4a}
\end{multline}
\subsection{Current time in a 'hill' interval: $\tau_1(R,n)<t<\tau_0(R,n)$}
For $t$ below 'hill' $n$, application of \eqref{eq:ANIsn} with $a(t)=\varrho_c(R,t)$ and $b(t)=\mathcal{L}'(R,t)$ to \eqref{eq:GenGRadius}, with $\varrho_c(R,s)=0$ above a 'valley' and $\mathcal{L}'(R,s)=\mathbb{L}'(R)-h(s)$ below a 'hill', shows that for $\tau_1(R,n)<t<\tau_0(R,n)$ and $n=1$
\begin{equation}
\varrho_c(R,t)=\mathcal{L}'(R,t)+\int\limits_{\tau_1(R,1)}^t T_0(t,s)\mathcal{L}'(R,s)\,\mathrm{d}s, \label{eqPart1:GenGRadius5}
\end{equation}
and, if $n\geq 2$,
\begin{multline}
\varrho_c(R,t)=\mathcal{L}'(R,t)+\int\limits_{\tau_1(R,n)}^t T_0(t,s)\mathcal{L}'(R,s)\,\mathrm{d}s \\
                                 +\sum\limits_{k=1}^{n-1}\int\limits_{\tau_{2k+1}(R,n)}^{\tau_{2k}(R,n)}T_{2k}(t,s,R,n)\mathcal{L}'(R,s)\,\mathrm{d}s.
\label{eq:GenGRadius5}
\end{multline}
The material properties now enter the analysis through the $T_j$ functions (see \eqref{eq:TEven} and \eqref{eq:TUneven}), starting with
\begin{equation}
T_0(t,s,R,n)=\dot{\phi}(t-s),
\end{equation}
and proceeding with
\begin{multline}
T_{2k+1}(t,s,R,n)=T_{2k}(t,s,n)\\
+\int\limits_s^{\tau_{2k+1}(R,n)}T_{2k}(t,u,R,n)\dot{\varphi}(u-s)\,\mathrm{d}u\enspace k=0,1,2,\ldots,
\end{multline}
plus
\begin{multline}
T_{2k}(t,s,R,n)=T_{2k-1}(t,s,R,n)\\
+\int\limits_s^{\tau_{2k}(R,n)}T_{2k-1}(t,u,R,n)\dot{\phi}(u-s)\,\mathrm{d}u\enspace k=1,2,\ldots
\end{multline}
Define  the integral operators $\mathcal{U}_k$ and $\mathcal{H}_k$  by
\begin{gather}
\mathcal{U}_{k}(t,R,n)=\int\limits_{\tau_{2k+1}(R,n)}^{\tau_{2k}(R.n)}T_{2k}(t,s,R,n)\,\mathrm{d}s,\label{eq:UkFunction}\\
\mathcal{H}_{k}(t,R,n)=\int\limits_{\tau_{2k+1}(R,n)}^{\tau_{2k}(R.n)}T_{2k}(t,s,R,n)h(s)\,\mathrm{d}s,\label{eq:HkFunction}
\end{gather}
and take  the limits $t\downarrow \tau_1(R,n)$ and $ t\uparrow \tau_0(R,n)$ of  \eqref{eq:GenGRadius5} to obtain at the start of the 'hill'
\begin{equation}
\mathbb{L}'(R)=h(\tau_1),\enspace \tau_1=\tau_1(R,1),\enspace n=1,
\label{eqPart1:GenGRadius6}
\end{equation}
\begin{multline}
\mathbb{L}'(R)\left(1+\sum\limits_{k=1}^{n-1}\mathcal{U}_k(\tau_1,R,n)\right)=h(\tau_1)\\
        +\sum\limits_{k=1}^{n-1}\mathcal{H}_k(\tau_1,R,n),
        \enspace \tau_1=\tau_1(R,n),\enspace n\geq 2,~\label{eq:GenGRadius6}
        \end{multline}
and at the end
\begin{multline}
\mathbb{L}'(R)\left(1+\mathcal{U}_0(\tau_0,R,1)\right)=h(\tau_0)\\
+\mathcal{H}_0(\tau_0,R,1),\enspace \tau_0=\tau_0(R,1),\enspace n=1,
\label{eqPart1:GenGRadius7}
\end{multline}
\begin{multline}
\mathbb{L}'(R)\left(1+\sum\limits_{k=0}^{n-1}\mathcal{U}_k(\tau_0,R,n)\right)=h(\tau_0)\\
+\sum\limits_{k=0}^{n-1}\mathcal{H}_k(\tau_0,R,n),\enspace \tau_0=\tau_0(R,n),\enspace n\geq 2.
\label{eq:GenGRadius7}
\end{multline}
The equations \eqref{eqPart1:GenGRadius6} to \eqref{eq:GenGRadius7} represent the 'decomposed' equivalent of the equations \eqref{eqPart1:radiusequation}  and \eqref{eqPart2:radiusequation} relating depth, contact radius and material properties along the curve of the contact radius.
\section{The load equation}\label{sec:loadformulae}
For the application of the decomposition technique to the load equation \eqref{eq:GenGLoad}, the choice for the functions $a$ and $b$ from Appendix \ref{sec:GenGAppendix} is: $a(t)=q_c(R,t)$ and $b(t)=4\mathcal{L}(R,t)$.

The procedure now is basically the same as in the previous section, so explanatory remarks will be kept to a minimum.
\subsection{Current time in a 'valley' interval: $\tau_2(R,n)<t<\tau_1(R,n)$}
Substitution of $a(s)=q_c(R,s)$ and $b(s)=4\mathcal{L}(R,s)$ in \eqref{eq:BNIs1} and \eqref{eq:BNIsn} and noting that in all 'valley' intervals $q_c(R,s)=-\ell(s)$ yields expressions  for $4\mathcal{L}(R,t)$. For $n=1$ it is found that
\begin{equation}
4\mathcal{L}(R,t)=-\ell(t)-\int\limits_0^t N_1(t,s,R,1)\ell(s)\mathrm{d}s\enspace n=1,\label{eq:GenGLoad0}
\end{equation}
and for larger values of $n$ that
\begin{multline}
4\mathcal{L}(R,t)=-\ell(t)-\int\limits_{\tau_2(R,n)}^t N_1(t,s,R,n)\ell(s)\,\mathrm{d}s+\\
                4\sum\limits_{k=1}^{n-1}
                    \int\limits_{\tau_{2k+1}(R,n)}^{\tau_{2k}(R,n)}N_{2k}(t,s,R,n)\mathcal{L}(R,s)\mathrm{d}s\\
                -\sum\limits_{k=1}^{n-1}\int\limits_{\tau_{2k+2}(R,n)}^{\tau_{2k+1}(R,n)}N_{2k+1}(t,s,n)\ell(s)\mathrm{d}s,\enspace n\geq 2.\label{eq:GenGLoad1}
\end{multline}
Define
\begin{equation}
\mathcal{Q}_k(t,R,n)=\int\limits_{\tau_{2k+2}(R,n)}^{\tau_{2k+1}(R,n)}\ell(s)N_{2k+1}(t,s,R,n)\,\mathrm{d}s,\label{eq:QkFunction}
\end{equation}
and note that for the integrals  over the 'hill' intervals $\mathcal{L}(R,s)=\mathbb{L}(R)-Rh(s)$. The  limit $t\uparrow \tau_1(R,n)$ of \eqref{eq:GenGLoad0} and \eqref{eq:GenGLoad1} then leads for $n=1$ to
\begin{equation}
\ell(\tau_1)+\mathcal{Q}_0(\tau_1,R,1)=4\left\{Rh(\tau_1)-\mathbb{L}(R)\right\},\enspace\tau_1=\tau_1(R,1),\label{eq:GenGLoad2}
\end{equation}
and for larger values of $n$ to
\begin{multline}
\ell(\tau_1)+\sum\limits_{k=0}^{n-1}\mathcal{Q}_k(\tau_1,R,n)=4R\left\{h(\tau_1)-\sum\limits_{k=1}^{n-1}\mathcal{W}_k(\tau_1,R,n)\right\}\\
-4\mathbb{L}(R)\left(1-\sum\limits_{k=1}^{n-1}\mathcal{V}_k(\tau_1,R,n)\right),\enspace\tau_1=\tau_1(R,n),\enspace n\geq 2.
\label{eq:GenGLoad3}
\end{multline}
As expected, the result for $n=1$, i.e., equation \eqref{eq:GenGLoad2}  supplied with $R=c(\tau_1)$, equals the load equation \eqref{eq:advancingcontact2a} for increasing contact.

As before, the limit $t\downarrow \tau_2(R,n)$ only makes sense for $n\geq 2$ and this results in
\begin{multline}
\ell(\tau_2)+\sum\limits_{k=1}^{n-1}\mathcal{Q}_k(\tau_2,R,n)=4R\left(h(\tau_2)-\sum\limits_{k=1}^{n-1}\mathcal{W}_k(\tau_2,R,n)\right)\\
-4\mathbb{L}(R)\left(1-\sum\limits_{k=1}^{n-1}\mathcal{V}_k(\tau_2,R,n)\right),\enspace\tau_2=\tau_2(R,n).
\label{eq:GenGLoad4}
\end{multline}
\subsection{Current time in a 'hill' interval: $\tau_1(R,n)<t<\tau_0(R,n)$}
Substitution of $a(t)=q_c(R,t)$ and $b(t)=4\mathcal{L}(R,t)$ in \eqref{eq:ANIs1} and \eqref{eq:ANIsn} results for $n=1$ in
\begin{multline}
q_c(R,t)=4\mathcal{L}(R,t)+\int\limits_{\tau_1(R,1)}^t4T_0(t,s,R,1)\mathcal{L}(R,s)\,\mathrm{d}s\\
-\int\limits_0^{\tau_1(R,1)}T_1(t,s,R,1)\ell(s)\,\mathrm{d}s\label{eq:GeGLoad3}
\end{multline}
and for $n\geq 2$ in
\begin{multline}
q_c(R,t)=4\mathcal{L}(R,t)+\int\limits_{\tau_1(R,n)}^t4T_0(t,s,R,n)\mathcal{L}(R,s)\,\mathrm{d}s\\
-\sum\limits_{k=1}^{n}\int\limits_{\tau_{2k}(R,n)}^{\tau_{2k-1}(R,n)}T_{2k-1}(t,s,R,n)\ell(s)\,\mathrm{d}s\\
+\sum\limits_{k=1}^{n-1}\int\limits_{\tau_{2k+1}(R,n)}^{\tau_{2k}(R,n)}4T_{2k}(t,s,R,n)\mathcal{L}(R,s)\,\mathrm{d}s
\label{eq:GenGLOad4}
\end{multline}
Define
\begin{equation}
\mathcal{P}_k(t,R,n)=\int\limits_{\tau_{2k}(R,n)}^{\tau_{2k-1}(R,n)}\ell(s)T_{2k-1}(t,s,R,n)\,\mathrm{d}s,\label{eq:PkFunction}
\end{equation}
and take again the limits $t\uparrow \tau_0$ and $t\downarrow \tau_1$.  At $\tau_0$ the result is
\begin{multline}
\ell(\tau_0)-\mathcal{P}_1(\tau_0,R,1)=
4R\Bigl\{h(\tau_0)+\mathcal{H}_0(\tau_0,R,1)\Bigr\}\\
-4\mathbb{L}(R)\Bigl\{1+\mathcal{U}_0(\tau_0,R,1)\Bigr\},\enspace\tau_0=\tau_0(R,1),\enspace n=1,\label{eq:GeGLoad5a}
\end{multline}
\begin{multline}
\ell(\tau_0)-\sum_{k=1}^n\mathcal{P}_k(\tau_0,R,n)=4R\left(h(\tau_0)+\sum_{k=0}^{n-1}\mathcal{H}_k(\tau_0,R,n)\right)\\
-4\mathbb{L}(R)\left(1+\sum_{k=0}^{n-1}\mathcal{U}_k(\tau_0,R,n)\right),\enspace\tau_0=\tau_0(R,n),\enspace n\geq 2,\label{eq:GenGLoad6a}
\end{multline}
and at $\tau_1$
\begin{equation}
\ell(\tau_1)-\mathcal{P}_1(\tau_1,R,1)=4\bigl\{Rh(\tau_1)-\mathbb{L}(R)\bigr\},\enspace\tau_1=\tau_1(R,1),\enspace n=1, \label{eq:GeGLoad5}
\end{equation}
\begin{multline}
\ell(\tau_1)-\sum_{k=1}^n\mathcal{P}_k(\tau_1,R,n)=4R\left(h(\tau_1)+\sum_{k=1}^{n-1}\mathcal{H}_k(\tau_1,R,n)\right)\\
-4\mathbb{L}(R)\left(1+\sum_{k=1}^{n-1}\mathcal{U}_k(\tau_1,R,n)\right),\enspace\tau_1=\tau_1(R,n),\enspace n\geq 2.\label{eq:GenGLoad6}
\end{multline}

\section{Eliminating the depth from the load equations}\label{sec:SimplifiedLoad}
In the right-hand sides of the load equations \eqref{eq:GenGLoad3} and \eqref{eq:GenGLoad4} for the 'valleys' and \eqref{eq:GenGLoad6a} plus \eqref{eq:GenGLoad6} for the 'hills' the terms dependent on the depth can be eliminated using the corresponding equations linking depth and contact radius. For example, for time $\tau_1(R,n)$ in 'valley' $n$ multiplying \eqref{eq:GenGRadius4} by $R$ shows that
\begin{equation}
R\left(h(\tau_1)-\sum\limits_{k=1}^{n-1}\mathcal{W}_k(\tau_1,R,n)\right)=
R\mathbb{L}'(R)\left(1-\sum\limits_{k=1}^{n-1}\mathcal{V}_k(\tau_1,R,n)\right).
\end{equation}
Substitution of this result in \eqref{eq:GenGLoad3} with $\mathbb{G}(R)=R\mathbb{L}'(R)-\mathbb{L}(R)$ (see \eqref{eq:HFunction}) gives the simplified version
\begin{multline}
\ell(\tau_1)+\sum\limits_{k=0}^{n-1}\mathcal{Q}_k(\tau_1,R,n)=\\
4\mathbb{G}(R)\left(1-\sum\limits_{k=1}^{n-1}\mathcal{V}_k(\tau_1,R,n)\right),\enspace\tau_1=\tau_1(R,n),\enspace n\geq 2.
\label{eq:GenGLoad3A}
\end{multline}
Similarly, the other equation \eqref{eq:GenGLoad4} valid in a 'valley' and the two 'hill'-equations \eqref{eq:GenGLoad6a} plus \eqref{eq:GenGLoad6} simplify to
\begin{multline}
\ell(\tau_2)+\sum\limits_{k=1}^{n-1}\mathcal{Q}_k(\tau_2,R,n)=\\
4\mathbb{G}(R)\left(1-\sum\limits_{k=1}^{n-1}\mathcal{V}_k(\tau_2,R,n)\right),\enspace\tau_2=\tau_2(R,n),\enspace n\geq 2,
\label{eq:GenGLoad4aa}
\end{multline}
\begin{multline}
\ell(\tau_0)-\sum\limits_{k=1}^n\mathcal{P}_k(\tau_0,R,n)=\\
4\mathbb{G}(R)\left(1+\sum_{k=0}^{n-1}\mathcal{U}_k(\tau_0,R,n)\right),\enspace\tau_0=\tau_0(R,n),\enspace n\geq 2,\label{eq:GenGLoad6aa}
\end{multline}
\begin{multline}
\ell(\tau_1)-\sum\limits_{k=1}^n\mathcal{P}_k(\tau_1,R,n)=\\
4\mathbb{G}(R)\left(1+\sum_{k=1}^{n-1}\mathcal{U}_k(\tau_1,R,n)\right),\enspace\tau_1=\tau_1(R,n),\enspace n\geq 2.\label{eq:GenGLoad6b}
\end{multline}

\section{Notes on continuity}
The functions $\phi$ and $\varphi$ are each other Stieltjes inverses and it follows from $\phi\orc\varphi=\Hea$ that
the time derivative of $\phi\orc\varphi$ is zero for $t>0$ as \eqref{eq:DotFStarG} shows. From the definitions of $T_1$ and $N_1$ and the transformation rules for the $T_j$ and the $N_j$ it can then be shown that
\begin{gather}
 T_j(\tau_1(R,n),s,R,n)+N_j(\tau_1(R,n),R,s,n)=0\qquad j=1,2,3,\ldots,\label{eq:Continuity1}\\
T_i(\tau_0(R,n),R,s,n)+N_{i+2}(\tau_0(R,n),s,R,n+1)=0\quad i=0,1,3,\ldots\label{eq:Continuity2}
\end{gather}
For $i=0$, \eqref{eq:Continuity2} with $\tau_0(R,n)=\tau_2(R,n+1)$ reveals that
\begin{equation}
N_2(\tau_2(R,n+1),R,s,n+1)=-\dot{\phi}(\tau_0(R,n)-s).\label{eq:Continuity3}
\end{equation}
The limits to $t\rightarrow\tau_1$, always a point at an increasing part of the contact radius, were taken from inside a 'valley' interval -- limit from below -- or from inside a 'hill' interval -- limit from above. Property \eqref{eq:Continuity1} ensures that this yields the same equations, as it should be.
The limits to a point at a decreasing part of the contact radius, were also taken from inside a 'valley' interval -- limit from above -- or from inside a 'hill' interval -- limit from below. However, due to the numbering convention, this is  'valley' interval number $n+1$ 'hill' interval number $n$. Since $\tau_0(R,n)=\tau_2(R,n+1)$, property \eqref{eq:Continuity2} now ensures that these limits also lead to the same equations.
\section{Dependance of the kernels $T_j$ and $N_j$  on their arguments}\label{sec:ArgumentsOfTandN}
The kernel functions $T_j$ and $N_j$ both transform in the same way and the dependence of these functions on the current time $t$ and the time-like variable $s$ is basically the same  for the two function classes -- apart from obvious differences with respect to the starting functions $T_0$ and $N_1$. The generic form of the transformation is for $j=0,1,2,\ldots$ and $\tau_1<t<\tau_0$
\begin{gather} T_{j+1}(t,s,R,n)=T_j(t,s,R,n)+\int\limits_s^{\tau_{j+1}}\upsilon(u-s)T_j(t,u,R,n)\mathrm{d}u, \enspace \label{eq:DefOfT}\\
T_0(t,s)=\dot{\phi}(t-s),\enspace\text{and}\enspace  \upsilon(x)=\begin{cases}\dot{\phi}(x)&\quad\text{$j$=odd,}\\
					\dot{\varphi}(x)&\quad\text{$j$=even.}
					\end{cases}
\end{gather}
The notation $T_{j+1}(t,s,R,n)$ is actually a shorthand notation\footnote{Not included in the list of arguments, but nonetheless  present, is the dependence of $T_j$ and $N_j$ on a set $\tilde{p}$ of material constants present in the functions $\dot{\phi}$ and $\dot{\varphi}$, e.g. $\tilde{p}=\{q,\kappa\}$ for material behaviour according to the standard linear solid because $\dot{\phi}(t)=-q \kappa \exp(-\kappa t)$ as is shown in \ref{sec:SLE}.} to express that $T_j$ depends also on $\tau_1,\tau_2,\dots,\tau_{j+1}$ because the value of $R$ and the choice of $n$ -- the number of the interval the current time $t$ is actually located in -- determines the values of all $\tau$'s.

The purpose of this part of this section is to prove -- by induction -- that\footnote{The particular choice $t = \tau_0$ does not restrict the conclusions.}
\begin{equation}
T_{j}(\tau_0,s,R,n)=T_{j}(\tau_0-s,\tau_1-s,\ldots,\tau_{j}-s;P(0,j))\label{eq:ArgumentsOfTt0}
\end{equation}
with $P(0,j)$ a parameter set of differences $\Delta_{ij}=\tau_i-\tau_j$ defined by
\begin{align}
P(j,j)&=\emptyset,\\
P(l,j)&=
\Bigl\{\,\Delta_{ik}\,|\,\, i\in\{l,\ldots,j-1\},\,\, k\in\{l+1,\ldots j\},\,\,i\neq k\Bigr\},\enspace j\geq l+1.
\label{eq:SetOfTauDiffs}
\end{align}
The statement in \eqref{eq:ArgumentsOfTt0} is true for $j=0$. Suppose that it is true for $j>1$. This means that $T_j(\tau_0,s,R,n)$ is some function, say $g$, of the listed arguments and parameters:
\begin{equation}
T_{j}(\tau_0,s,R,n)=g(\tau_0-s,\tau_1-s,\ldots,\tau_{j}-s;P(0,j))
\end{equation}
Next, note that the solution of an inhomogeneous partial differential equation in $j+2$ independent variables $x ,y_0,y_1,y_2,\ldots,y_{j}$ of the form
\begin{equation}
\frac{\partial F}{\partial x}-\sum\limits_{k=0}^{j}\frac{\partial F}{\partial y_k}=\upsilon(x)g(y_0,y_1,\ldots,y_{j};P(0,j))
\end{equation}
is some function defined on the $j+2$ dimensional space spanned by the variables $x$, $y_0$, $y_1$, $y_2,\ldots,y_{j}$. Additionally, a description of $F$ also contains the set parameters $P(0,j)$. Consider the restriction of  $F$ to the straight line defined by the parameter representation
\begin{equation}
x(u)=u-s,\enspace y_0(u)=\tau_0-u,\enspace\dots\enspace, y_j(u)=\tau_j-u,
\end{equation}
and determine the total derivative  $\mathrm{d}F/\mathrm{d}u$, along this line. It is found that
\begin{equation}
\begin{split}
\frac{\mathrm{d} F}{\mathrm{d}u}&=
\frac{\partial F}{\partial x}\frac{\mathrm{d}x(u)}{\mathrm{d}u}+\sum\limits_{k=0}^{j}\frac{\partial F}{\partial y_k}\frac{\mathrm{d}y_k(u)}{\mathrm{d}u}\\
&=\upsilon(u-s)g(\tau_0-u,\ldots,\tau_j-u;P(0,j))
\end{split}
\end{equation}
because $\mathrm{d}x(u)/\mathrm{d}u=1$ and $\mathrm{d}y_k(u)/\mathrm{d}u=-1$ for all $k$-values. The integrand in \eqref{eq:DefOfT} is apparently
the derivative of some function $F$ and integration over $u$ then leads to
\begin{multline}
\int\limits_s^{\tau_{j+1}}\upsilon(x(u))g(y_0(u),\ldots,y_{j}(u);P(0,j))\mathrm{d}u=\\
F(\tau_{j+1}-s,\tau_0-\tau_{j+1},\ldots,\tau_j-\tau_{j+1};P(0,j)\\
-F(0,\tau_0-s,\ldots,\tau_j-s;P(0,j)
\end{multline}
Comparison of this result with \eqref{eq:DefOfT} shows that to the existing arguments and parameters of $T_{j}$ the argument  $\tau_{j+1}-s$ and  the parameters $\tau_0-\tau_{j+1},\ldots,\tau_j-\tau_{j+1}$, respectively, have to be added to obtain those of $T_{j+1}$. This proves that
\begin{equation}
T_{j+1}(\tau_0,s,R,n)=T_{j+1}(\tau_0-s,\tau_1-s,\ldots,\tau_{j+1}-s;P(0,j+1)).
\end{equation}
If the current time is taken to be $\tau_1$ instead of $\tau_0$ one finds
\begin{equation}
T_{j+1}(\tau_1,s,R,n)=T_{j+1}(\tau_1-s,\ldots,\tau_{j+1}-s;P(1,j+1)).\label{eq:ArgumentsOfTt1}
\end{equation}
For the functions $N_j$ with $j\geq 1$ one finds in the same way
\begin{gather}
N_j(\tau_1,s,R,n)=N_j(\tau_1-s,\ldots,\tau_j-s;P(1,j)),\label{eq:ArgumentsOfNt1}\\
N_j(\tau_2,s,R,n)=N_j(\tau_2-s,\ldots,\tau_j-s;P(2,j)).\label{eq:ArgumentsOfNt2}
\end{gather}

The present analysis can be pushed further if more specific assumptions on the nature of the functions $\phi$ and $\varphi$ are made. For example,  by assuming standard-linear-element material behaviour closed form expressions for the $T_j$ and $N_j$ functions were derived in Appendix \ref{sec:SLEkernels}. Often the material behaviour is modelled as a Prony series. For this material type the functions $T_j$ and $N_j$ also have 'Prony-like' properties as is shown in Appendix \ref{sec:Pronykernels}.

\chapter[Single load and unload: decomposition method]{A single load and unload: decomposed hereditary integrals }\label{sec:SingleLOadUnload}
\begin{figure}[htb]
\centering
\includegraphics[scale=1]{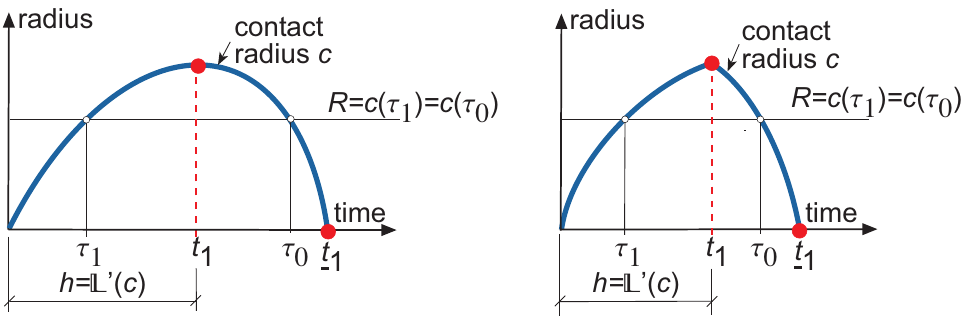}\\
\caption{Contact radius for single load and unload process. Smooth (left) or kinked (right) at the maximum at $t_1$. Contact is lost at time $\tmin_1$.}\label{fig:VarstUpAndDown}%
\end{figure}

In classic indentation the indenter is pressed into the material and subsequently retracted until contact is lost (Fig.~\ref{fig:VarstUpAndDown}). Whatever the precise shape of the curve of the contact radius actually is, it is clear that the contact radius $c(t)$  increases until it reaches a maximum at time $t_1$ and decreases afterwards until at some time $\underline{t}_1$ contact is lost, i.e., $c(\underline{t}_1)=0$.
This is the situation $n=1$ from the previous paragraphs.

The  focus is here on the determination of the reduced functions $\phi$ or $\varphi$, i.e., it is assumed that the elasticity factor is known and is removed from the equations by working with $\varrho_c$, $\ell$ and $q_c$ (see the definitions \eqref{eq:Scaled} in Chap.~\ref{sec:GenGSep1})

\section{The advancing phase}
The period $0<\tau_1<t_1$ is always a 'valley' and \eqref{eq:GenGRadius3} applies for the relation between the depth and the contact radius, i.e., the contact radius is slaved to the depth  because from $R=c(\tau_1)$ one finds
\begin{equation}
h(\tau_1)=\mathbb{L}'(c(\tau_1))\Rightarrow c(\tau_1)=\mathbb{C}(h(\tau_1))\enspace 0<\tau_1<t_1.\label{eq:GenGRadius3a}
\end{equation}
This equation applies until $\dot{h}$ becomes zero for the first time or exhibits a jump (see (Fig.~\ref{fig:VarstUpAndDown}, left and right, respectively). The load is now given by \eqref{eq:GenGLoad2}
\begin{equation}
\ell(\tau_1)+\mathcal{Q}_0(\tau_1,R,1)=4\left\{Rh(\tau_1)-\mathbb{L}(R)\right\},\enspace\tau_1=\tau_1(R,1).\label{eq:GenGLoad2a}
\end{equation}
In view of \eqref{eq:GenGRadius3a}, the right-hand side of \eqref{eq:GenGLoad2a} becomes $4\{h(\tau_1)\mathbb{C}(h(\tau_1))-\mathbb{L}(\mathbb{C}(h(\tau_1)))\}=
4\mathbb{F}(h(\tau_1))$, and with the definition of $\mathcal{Q}_0$, according to \eqref{eq:QkFunction}, the left-hand side is found to equal $[\ell\orc\varphi](\tau_1)$. This applies for all times in the interval$(0,t_1)$, so \eqref{eq:advancingcontact2b} is recovered in the form
\begin{gather}
[\ell\orc\varphi](t)=4\mathbb{F}(h(t))\enspace 0<t<t_1,\label{eq:advancingcontact2c}
\intertext{or, equivalently,}
\ell(t)=4[\mathbb{F}(h)\orc\phi](t)\enspace 0<t<t_1.\label{eq:advancingcontact2d}
\end{gather}
From these equations data for the reduced creep and relaxation function on the time interval $(0,t_1)$ are obtained.
\section{The receding phase}
The period $\tau_1<t<\tau_0$ is a 'hill' interval  and to determine the contact radius the 'hill' equation  \eqref{eq:GenGRadius7} at $\tau_0=\tau_0(R,1)$ must be used:
\begin{equation}
\mathbb{L}'(R)\left(1+\mathcal{U}_0(\tau_0,R,1)\right)=h(\tau_0)+\mathcal{H}_0(\tau_0,R,1).\label{eq:GenGRadius7a}
\end{equation}
Use of $\mathbb{L}'(R)=h(\tau_1)$ and the definitions \eqref{eq:UkFunction} and \eqref{eq:HkFunction} of $\mathcal{U}_0$ and  $\mathcal{H}_0$, respectively, in  \eqref{eq:GenGRadius7a} shows that its left-hand side equals
\begin{equation}
\mathbb{L}'(R)\{1+\mathcal{U}_0(\tau_0,R,1)\}=h(\tau_1)\phi(\tau_0-\tau_1),\label{eq:GenGRadius7b}
\end{equation}
and that after one partial integration its right-hand side becomes
\begin{equation}
h(\tau_0)+\mathcal{H}_0(\tau_0,R,1)=h(\tau_1)\phi(\tau_0-\tau_1)+\int\limits_{\tau_1}^{\tau_0}\phi(\tau_0-s)\dot{h}(s)\,\mathrm{d}s.
\label{eq:GenGRadius7c}
\end{equation}
So, \eqref{eq:GenGRadius7a} simplifies to an equation connecting times  of equal contact radius before and after the first maximum;
\begin{equation}
\int\limits_{\tau_1}^{\tau_0}\phi(\tau_0-s)\dot{h}(s)\,\mathrm{d}s=0, \enspace 0<\tau_1<t_1,\enspace \tau_0>t_1.\label{eq:GenGRadius7d}
\end{equation}
From this equation $\tau_1$ must be solved as function of $\tau_0$, i.e., $\tau_1=\Upsilon(\tau_0)$, and this is possible as long as $0<\tau_0-\tau_1<t_1$ because $\phi$ is at the end of the advancing phase only known on the the interval $(0,t_1)$.
The contact radius has, by definition, a local maximum at $t_1$ and therefore $\Upsilon(t_1)=t_1$. As $\phi$ never changes sign, only a change in sign of $\dot{h}$ in  the interval $(\tau_1,\tau_0)$ can cause \eqref{eq:GenGRadius7d} to be zero, the result being that the depth rate $\dot{h}$ must change sign at $t_1$. This is an important point because it shows that the first maximum of the contact radius and the depth coincide.

The time interval on which the contact radius can be calculated is thus extended from the interval $(0,t_1)$ to $(0,\hat{t})$, where $\hat{t}$ is defined by the equation $\hat{t}-\Upsilon(\hat{t})=t_1$, and the contact radius satisfies
\begin{equation}
c(t)=\begin{cases}\mathbb{C}(h(t)) & 0<t<t_1,\\
                  \mathbb{C}(h(\Upsilon(t))) & t_1<t<\hat{t}.
                  \end{cases}\label{eq:ConctRecedingPhase}
\end{equation}

For the load in the receding phase, \eqref{eq:GeGLoad5a} applies
\begin{multline}
\ell(\tau_0)-\mathcal{P}_1(\tau_0,R,1)=
4R\Bigl\{h(\tau_0)+\mathcal{H}_0(\tau_0,R,1)\Bigr\}\\
-4\mathbb{L}(R)\Bigl\{1+\mathcal{U}_0(\tau_0,R,1)\Bigr\},\enspace\tau_0=\tau_0(R,1),\enspace n=1.\label{eq:GeGLoad5b}
\end{multline}
With \eqref{eq:GenGRadius7b}, \eqref{eq:GenGRadius7c}, \eqref{eq:GenGRadius7d}, $R=\mathbb{C}(h(\tau_1))$ and $\mathbb{L}'(R)=h(\tau_1)$ it can be shown that the right-hand side of \eqref{eq:GeGLoad5b} equals $4\phi(\tau_0-\tau_1)\mathbb{F}(h(\tau_1))$ and thus
\begin{equation}
\ell(\tau_0)=\mathcal{P}_1(\tau_0,R,1)+4\phi(\tau_0-\tau_1)\mathbb{F}(h(\tau_1)),\enspace\tau_0=\tau_0(R,1),\enspace n=1.\label{eq:GeGLoad5b1}
\end{equation}
With the definition of $\mathcal{P}_1(\tau_0,R,1)$ given in \eqref{eq:PkFunction} and according to \eqref{eq:GenGa2} with the choices $a(t)=q_c(t)$ and $b(t)=4\mathcal{L}(R,t)$ at $t=\tau_0$ it turns out that
\begin{multline}
\mathcal{P}_1(\tau_0,R,1)=\int\limits_0^{\tau_1(R,1)}\ell(s)T_1(\tau_0,s,R,1)\,\mathrm{d}s\\
=-4\int\limits_0^{\tau_1(R,1)}\mathcal{L}(R,s)T_0(\tau_0,s,R,1)\,\mathrm{d}s.
\label{eq:GeGLoad5b2}
\end{multline}
Since $T_0=\dot{\phi}$, one partial integration in the last integral leads to
\begin{multline}
\mathcal{P}_1(\tau_0,R,1)=4\phi(\tau_0-\tau_1)\mathcal{L}(R,\tau_1)-4\mathcal{L}(R,0)\\
+4\int\limits_0^{\tau_1(R,1)}\dot{\mathcal{L}}(R,s)\phi(\tau_0-s)\,\mathrm{d}s.\label{eq:GeGLoad5b3}
\end{multline}
Now, $\mathcal{L}(R,\tau_1)=-\mathbb{F}(h(\tau_1))$ and $\mathcal{L}(R,0)=0$. Moreover, for $0<s<\tau_1$ is
\begin{equation}\label{eq:GeGLoad5b4}
\begin{split}
\dot{\mathcal{L}}(R,s)&=\frac{\partial}{\partial s}\left(\int\limits_0^{c(s)}\mathcal{L}'(r,s)\,\mathrm{d}r+\int\limits_{c(s)}^R\mathcal{L}'(r,s)\,\mathrm{d}r\right)\\
&=\frac{\partial}{\partial s}\mathcal{L}(c(s),s)
=-\frac{\partial}{\partial s}\mathbb{F}(h(s))=\mathbb{C}(h(s))\dot{h}(s),
\end{split}
\end{equation}
because $\mathcal{L}(0,s)=0$, $\mathcal{L}'(r,s)=0$ if $c(s)<r<R$ (region 2 in Fig.~\ref{fig:VarstUvalues}) and $\mathcal{L}(c(s),s)=-\mathbb{F}(h(s))$ for $0<s<\tau_1$ because $h=\mathbb{L}'(\mathbb{C}(h))$ in this time range. Substitution of \eqref{eq:GeGLoad5b4} in \eqref{eq:GeGLoad5b3} and the result in \eqref{eq:GeGLoad5b1} yields \begin{equation}
\ell(t)=-4\int\limits_0^{\tau_1}\phi(\tau_0-s)\mathbb{C}(h(s)\dot{h}(s)\,\mathrm{d}s.
\end{equation}
The relation $\tau_1=\Upsilon(\tau_0)$ together with the replacement of $\tau_0$ by the current time $t$ changes the latter equation in
\begin{equation}
\ell(t)=-4\int\limits_0^{\Upsilon(t)}\phi(t-s)\mathbb{C}(h(s)\dot{h}(s)\,\mathrm{d}s,\enspace t_1<t.\label{eq:GeGLoad5c}
\end{equation}
At this stage, the function $\Upsilon(t)$ is known for $t_1<t<\hat{t}$; the data $\ell(t)$ and $h(t)$ enable computation of the relaxation and creep function on the interval $(t_1,\hat{t})$ thus extending the total time interval to $(0,\hat{t})$. If $\hat{t}$ is smaller than the end time of the experiment,  the domain of $\Upsilon(t)$ can be extended using again
\eqref{eq:GenGRadius7d} and, subsequently, reusing \eqref{eq:GeGLoad5c} to calculate the relaxation and creep function for times beyond $\hat{t}$.
\section{Example: standard-linear-solid material behaviour}
Equation \eqref{eq:GeGLoad5b} shows that finding $\phi$ from the data is relatively straightforward once  $\Upsilon(t)$ has been found. Determination of this function from \eqref{eq:GenGRadius7d} constitutes the bottleneck in the procedure because it depends on the material properties.

To demonstrate this consider a process where a cone shaped indenter is pressed with constant speed $v_1$ into a half space until a maximum depth is reached at $t_1$.
\begin{figure}[htb]
  \centering
  \includegraphics[scale=1]{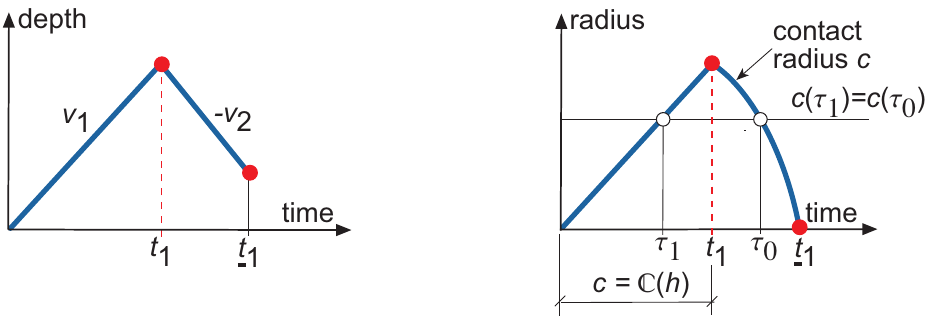}\\
  \caption{Left: linearly increasing depth with speed $v_1$ and linearly decreasing depth with speed $-v_2$. Right: possible shape of contact radius for a cone shaped indenter. At time $\underline{t}_1$, the indenter is not fully retracted (right) but looses nonetheless contact (left).}\label{fig:VarstUpAndDown2b}
\end{figure}
Afterwards the indenter is retracted with speed $-v_2$ until at some time $\underline{t}_1$ contact is lost. In the advancing phase the relation between depth and contact radius only depends on the indenter shape, a cone in this case, and  Table \ref{tab:CharFun} shows that in this phase:
 \begin{equation}
 c(t)=\mathbb{C}(h(t))=\frac{2h(t)}{k_1\pi}, \quad 0\leq t\leq t_1,\quad\text{$k_1$: constant.}
\end{equation}
 \begin{figure}
  \centering
  \includegraphics[scale=1]{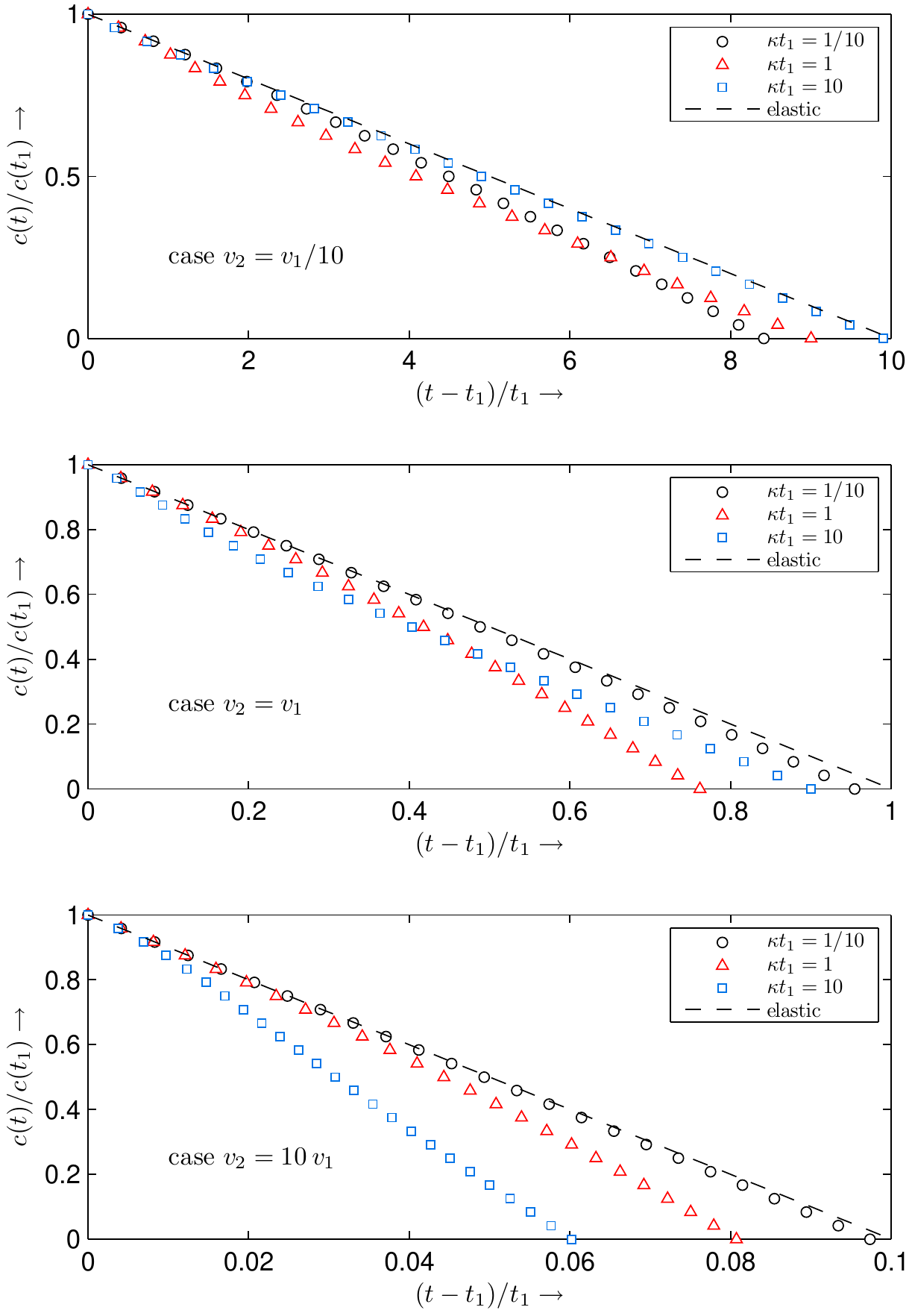}\\
  \caption{Contact radius in the receding phase $t>t_1$. Conical indenter, relaxation function according to 'standard linear element' with $q=1/2$.The lines 'elastic' are added for comparison with the purely elastic case.}\label{fig:SLEplusKegel}
\end{figure}
 Later -- the receding phase -- the contact radius also depends on the material properties. Assuming a reduced relaxation function as described by that of a standard linear solid (see Appendix~\ref{sec:SLE}), i.e.,
\begin{equation}
 \phi(t)=(1-q)+q\exp(-\kappa t),\enspace 0<q<1,\enspace\kappa>0,\enspace t>0,
\end{equation}
 the equation \eqref{eq:GenGRadius7d} becomes
 \begin{equation}
 v_1\int\limits_{\tau_1}^{t_1}(1-q)+q\exp(-\kappa(\tau_0-\tau))\mathrm{d}\tau
 =v_2\int\limits_{t_1}^{\tau_0}(1-q)+q\exp(-\kappa(\tau_0-\tau))\mathrm{d}\tau\label{eq:UpsilonSLE}
 \end{equation}
The introduction of the scaled variables
$ s=\kappa \tau$, $\beta=\kappa \tau_0$, $\alpha=\kappa \tau_1$, $s_m=\kappa t_1$, $a=q/(1-q)$, and $b=v_2/v_1$ plus division by $v_1(1-q)$, simplifies \eqref{eq:UpsilonSLE} to
 \begin{equation}
 \int\limits_{\alpha}^{s_m}1+a\exp(-(\beta-s))\mathrm{d}s
 =b\int\limits_{s_m}^{\beta}1+a\exp(-(\beta-s))\mathrm{d}s.\label{eq:UpsilonSLEa}
 \end{equation}
From \eqref{eq:UpsilonSLEa} the variable $\alpha$ -- actually the function $\kappa\Upsilon$ -- can be solved as function of $\beta$. The solution involves the main branch $\mathrm{W}_0$ of the Lambert W function\footnote{See for instance \cite{CorlessEtAl1996} for the history of this function, its occurrence in science and technology and the question how to calculate it.}, i.e., the solution of the equation $x=\mathrm{W}_0(x)\exp\mathrm{W}_0(x)$ for $x>-\exp(-1).$
 Specifically
 \begin{equation}
 \alpha(\beta,s_m,a,b)=\beta +g(\beta,s_m,a,b)-W_0\left(a\exp\{g(\beta,s_m,a,b\}\right),\label{eq:UpsilonSLEb}
 \end{equation}
 where $g$ is defined by
 \begin{equation}
 g(\beta,s_m,a,b)=(1+b)\{a\exp(s_m-\beta)+s_m-\beta\}-ab.\label{eq:UpsilonSLEc}
 \end{equation}
 Fig. \ref{fig:SLEplusKegel} shows the results of such a computation\footnote{Alternatively, one can calculate the inverse, i.e., $\beta$ as function of $\alpha$ and the parameters $s_m$, $a$ and $b$. The Lambert W function also appears in the solution in this case.} using that $c(\beta)=c(\alpha)$.
 Three speed ratios $v_2/v_1$ and three ratios $\kappa t_1$ of the characteristic rise time of the depth and the characteristic relaxation time of the standard linear element were considered. The parameter $a=q/(1-q)$ was chosen to be equal to 1 which corresponds to $\phi(\infty)=1/2$. All graphs show that during the retraction phase of the indenter which in itself leads to a receding contact, the receding speed of the contact is amplified by the viscoelastic properties of the material. Specifically, stress relaxation of the material leads to a faster receding contact during the retraction phase compared to the same experiment performed on a purely elastic material. The magnitude of this effect depends on the magnitude of the ratio $\kappa t_1$, i.e., the ratio of the relevant time scales in the experiment. So,  the contact radius becomes zero before the indenter is completely retracted as is visible in the graphs in Fig. \ref{fig:SLEplusKegel} because the 'elastic' comparison graph also indicates the position of the cone-shaped indenter because  $c(t)/c(t_1)$ equals $h(t)/h(t_1)$.

\chapter[Dynamic load depth sensing: decomposition method]{Dynamic load-depth sensing: decomposed hereditary integrals}\label{sec:hereditary}
The previous chapter showed that the classic analysis of dynamic load-depth sensing experiment is actually limited to those cases were a monotonic contact radius is always or almost always present. The practical important case of a fluctuating perturbation superposed on an applied step shaped carrier is not in this category and an intermediate step is needed here. In this step, a link is established between the input variable (depth or load) on the one hand and the contact radius on the other. This link and how it depends on the material properties is the topic of the Secs \ref{sec:Preliminaries} -- \ref{sec:DynamicGeneralVisco2}.
The actual determination of the viscoelastic material parameters, the final step (Chap.~\ref{sec:FinalStep}), uses this link. This part is only treated at a general level.
\section{The stationary state}\label{sec:Preliminaries}
For the present analysis\footnote{As in Chap.~\ref{sec:SingleLOadUnload} it is also assumed that the elasticity factor is already known and removed from the equations by working with $\varrho_c$, $\ell$ and $q_c$ (see the definitions \eqref{eq:Scaled} in Chap.~\ref{sec:GenGSep1}).}  only a certain type of control variables is considered, namely a (relatively small) sinusoidal perturbation superposed on a step shaped carrier variable and it is also restricted to the \emph{stationary state,}
 \begin{figure}[htb]
  \centering
  \includegraphics[scale=1]{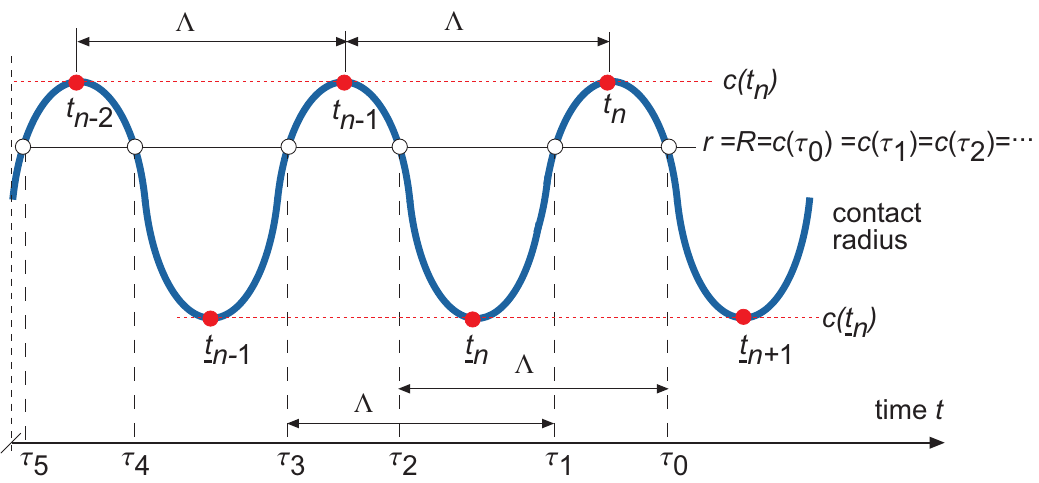}\\
  \caption{Periodic contact radius during the stationary phase. The times $t_n$ and $\tmin_n$ indicate the times of the maxima and the minima of the contact radius, respectively.}\label{fig:PeriodicContact}
\end{figure}
i.e., the region of time for which depth, contact radius and and load vary periodically (Fig.~\ref{fig:PeriodicContact}) around a steady state. The \emph{steady state} is -- by definition -- the asymptotic response if only the step shaped carrier variable is used as input.
For the steady state the results of Chap.~\ref{sec:increasing} apply with
\begin{equation}
h(t)=h\mean\hea{t},\enspace c(t)=\mathbb{C}(h\mean),\enspace\ell(t)\sim 4\mathbb{F}(h\mean)\phi(\infty)\label{eq:DepthControl1}
\end{equation}
for depth control (Chap.~\ref{sec:ClassicRelaxation}), whereas for load control (Chap.~ \ref{sec:ClassicCreep})
\begin{equation}
\ell(t)=\ell\mean\hea{t},\enspace\mathbb{F}(h(t))\sim \ell\mean\varphi(\infty)/4,\enspace
 c(t)\sim \mathbb{C}(h(\infty)).\label{eq:LoadControl1}
\end{equation}
Well into the stationary phase the contact radius changes periodically and the number of 'hills' and 'valleys' that were crossed in the past is very large. The equations from Chap.~\ref{sec:GenGSep} relating depth, contact radius and load during an arbitrary 'valley' plus subsequent 'hill' period also involve all previous 'valleys' and 'hills' because the sums  in these formulae do so. However, increasing the values of  the counting subscripts of terms  in these sums means going further back in the deformation or loading history and decreasing influence of these terms on the current response is to be expected, i.e., the contribution of the integral operators $\mathcal{U}_j$, $\mathcal{V}_j$, $\mathcal{H}_j$ $\mathcal{W}_j$, $\mathcal{P}_j$ and $\mathcal{Q}_j$ to the sums decreases if the value of $j$ increases. So, it may well be assumed that the control is switched on at $t=-\infty$ and that the number of terms in the sums is infinite.

The period $\Lambda=2\pi/\Omega$ of the control variable  equals that of the other dependent variables. So, the intersection times with even index are related to $\tau_0$ and those with odd index to $\tau_1$ (see Fig.~\ref{fig:PeriodicContact}). Specifically, for $ j=0,1,2,\ldots$,
\begin{equation}
 \tau_{2j}(R,n)=\tau_0(R,n)-j\Lambda, \enspace \tau_{2j+1}(R,n)=\tau_1(R,n)-j\Lambda.\label{eq:IntersectionTimes}
  \end{equation}
In the stationary phase all 'hills' and all 'valleys' are equivalent,
\begin{equation}
\tau_0(R,n-j)=\tau_0(R,n,\quad \tau_1(R,n-j)=\tau_1(R,n),
\end{equation}
implying that now $\tau_0$ and $\tau_1$ are independent of the 'hill' number $n$, i.e., $\tau_0=\tau_0(R)$ and $\tau_1=\tau_1(R)$. For the analysis it suffices to take as control variable a single sinusoidal function with period $\Lambda$. However, from here on the time will be scaled on $\Omega$, so the period becomes $2\pi$.

The analysis in \ref{sec:Extrema} shows that during the stationary phase the maxima of the contact radius and the depth coincide, whereas the minima of the contact radius coincide with those of  the load. The relation between the maxima of the contact and the depth is independent of the material behaviour because according to \eqref{eq:Htn} and \eqref{eq:Vtn} in Appendix \ref{sec:MaxOfDepthAndRadius} is
\begin{equation}
h(t_n)=\mathbb{L}'(c(t_n)), \quad \dot{c}(t_n)=0,\quad\dot{h}(t_n)=0.\label{eq:Htn1}
\end{equation}
The relation between the minima of the contact and the load is not so simple because it depends also on the material properties; at the times $\tmin_n$, when the contact is minimal, load and contact radius are related by (see \eqref{eq:LoadAtMin} and \eqref{eq:LoadRateAtMin} in Appendix \ref{sec:MinOfLoadAndRadius})
\begin{equation}
\ell(\tmin_n)=4\mathbb{G}(c(\tmin_n))\phi(\infty)\quad \dot{c}(\tmin_n)=0,\quad \dot{\ell}(\tmin_n)=0.\label{eq:LoadAtMin1}
\end{equation}
 These results suggest to choose $h(t)=h_\mathrm{max}\{1+h_1(\cos t-1)\}$ as input for depth controlled experiments and $\ell(t)=\ell_\mathrm{min}\{1-\ell_1(\cos t-1)\}$  as input function for load controls. The advantage of these choices is that the location in time of the extrema of the contact radius are known from the recorded data for depth and load. The maxima $h_\mathrm{max}$ and $c_\mathrm{max}=\mathbb{C}(h_\mathrm{max})$ of depth and contact radius both occur at times $2n\pi$ if the depth is controlled and the minimum of the measured load coincides with the minimum of the contact radius. Similarly, if the load is chosen as control variable, the minima $\ell_\mathrm{min}$ of the load and contact radius also occur at $2n\pi$ and the times of the maximum contact radius follow now from the recorded depth data.
\begin{figure}[htb]
  \centering
  \includegraphics[scale=1]{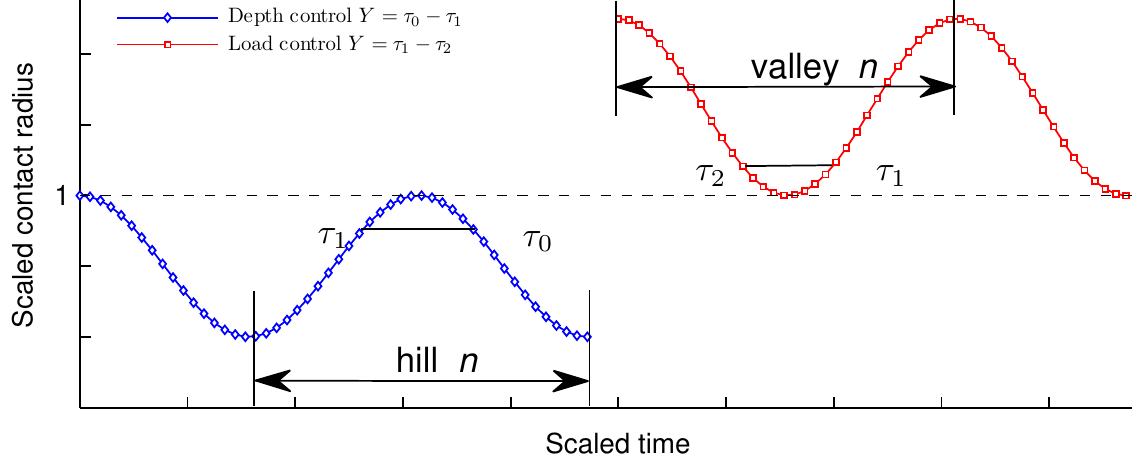}
  \caption{\emph{Depth control (left).} Radius scaled by $c_\mathrm{max}$. The aim is to determine $\tau_0$ as function of $Y=\tau_0-\tau_1$ for the descending  part of a representative 'hill'. \emph{Load control (right).} Radius is scaled by $c_\mathrm{min}$. Aim is to determine $\tau_1$ as function of $Y=\tau_1-\tau_2$ for the ascending part of a representative 'valley'.}\label{fig:SetupForMatrixEquations}
\end{figure}

The idea is to reconstruct the shape of a representative period of the scaled contact radius (Fig.~\ref{fig:SetupForMatrixEquations}) using the equations
 derived  in Chap.~\ref{sec:GenGSep}, i.e., the equations that relate depth and contact radius or load and contact radius during an arbitrary 'valley' and subsequent 'hill' period.  For  depth controlled indentation  the 'hill' equations \eqref{eq:GenGRadius6} and \eqref{eq:GenGRadius7} are used whereas for load control the 'valley' equations  \eqref{eq:GenGLoad3A} and \eqref{eq:GenGLoad4aa} are used as  starting point. The sums appearing in these formulae involved all previous 'valleys' and 'hills', the number of which is here taken to be infinite as the system is considered to be switched on at $t=-\infty$.

\section[Depth control: general case]{Depth controlled contact radius: general viscoelastic material behaviour}\label{sec:DynamicsGeneralVisco}
The depth function, $h(t)=h_\mathrm{max}\{1+h_1(\cos t-1)\}$ is maximal at $t=t_n$, where is further irrelevant except that $\tau_0$ and $\tau_1$ are defined as the first times after and before $t_n$, respectively, where $c(t)=R$.
In the stationary state the equations \eqref{eq:GenGRadius7} and   \eqref{eq:GenGRadius6})  relating depth $h$ and radius $R$   are
\begin{gather}
        h(\tau_0)+\sum\limits_{j=0}^\infty\mathcal{H}_j(\tau_0)=\mathbb{L}'(R)\left(1+\sum\limits_{j=0}^\infty\mathcal{U}_j(\tau_0)\right),
        \enspace\tau_0=\tau_0(R),
        \label{eq:GenGRadius7A}\\
        h(\tau_1)+\sum\limits_{j=1}^\infty\mathcal{H}_j(\tau_1)=\mathbb{L}'(R)\left(1+\sum\limits_{j=1}^\infty\mathcal{U}_j(\tau_1)\right),\enspace
        \tau_1=\tau_1(R),\label{eq:GenGRadius6A}
 \end{gather}
 and they are valid along a descending and ascending part of an arbitrary 'hill', respectively.

 Substitution of $h(t)$ in the definition of $\mathcal{H}_j(t)$, \eqref{eq:HkFunction}, yields
 \begin{equation}
 \mathcal{H}_j(t)=h_\mathrm{max}\{(1-h_1)\mathcal{U}_j(t)+h_1\mathcal{F}_j(t)\},
 \label{eq:Hjfie}
 \end{equation}
 with\footnote{The function $\mathcal{F}_j(t)$ also depends on the particular value of $R$ and on the material constants $\tilde{p}$, just like the other integrals -- $\mathcal{U}_j$, $\mathcal{V}_j$, $\mathcal{H}_j$ $\mathcal{W}_j$, $\mathcal{P}_j$ and $\mathcal{Q}_j$ -- do. To obtain concise formulae this dependence is only explicitly written down if confusion may arise.\label{note:1}}
 \begin{equation}
 \mathcal{F}_j(t)=\int\limits_{\tau_{2j+1}}^{\tau_{2j}}T_{2j}(t,s,R)\cos s \,\mathrm{d}s.
 \label{eq:Fjfie}
 \end{equation}
 Use of these relations in \eqref{eq:GenGRadius7A} and \eqref{eq:GenGRadius6A} leads, after division by $h_\mathrm{max}=\mathbb{L}'(c_\mathrm{max})$, to the matrix equation
 \begin{equation}
h_1\begin{bmatrix}\cos\tau_0+\sum\limits_{0}^\infty \mathcal{F}_j(\tau_0)\\
\cos\tau_1+\sum\limits_{1}^\infty \mathcal{F}_j(\tau_1)
\end{bmatrix}=\mathbb{A}(R,c_\mathrm{max},h_1)
\begin{bmatrix}1+\sum\limits_{0}^\infty \mathcal{U}_j(\tau_0)\\
1+\sum\limits_{1}^\infty \mathcal{U}_j(\tau_1)
\end{bmatrix} ,
\label{eq:DepthMatEqBasis}
\end{equation}
with
\begin{equation}
\mathbb{A}(R,c_\mathrm{max},h_1)=
\left(\frac{\mathbb{L}'(R)}{\mathbb{L}'(c_\mathrm{max})}-1+h_1\right) .
\label{eq:defAA}
\end{equation}

For all conical, parabolic or other 'power-law shaped' indenters the ratio $\mathbb{L}'(R)/\mathbb{L}'(c_\mathrm{max})$ depends exclusively on the scaled radius $R/c_\mathrm{max}$, e.g.,
\begin{equation}
\mathbb{L}'(R)/\mathbb{L}'(c_\mathrm{max})=\begin{cases}R/c_\mathrm{max}&\enspace\text{(cone),}\\
\left(R/c_\mathrm{max}\right)^2&\enspace\text{(parabola).}\end{cases}
\end{equation}
\begin{figure}[htb]
  \centering
  \includegraphics[scale=1]{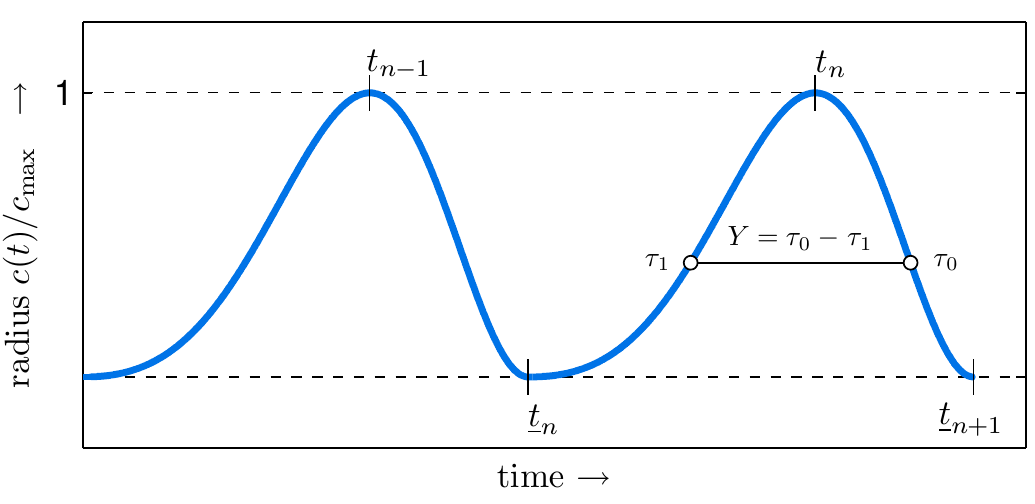}
  \caption{Depth controlled stationary state. Maxima of depth and contact radius coincide at $t_n$. Calculation of $\tau_0-t_n$ as function of $Y=\tau_0-\tau_1$ along the descending branch $t_n<\tau_0<\tmin_{n+1}$ of a typical period.}\label{fig:GenGplotV1a}
\end{figure}
\indent In  Fig.~\ref{fig:GenGplotV1a}, the time difference $Y=\tau_0-\tau_1$ ranges from 0 to $2\pi$ when the contact radius decreases from $c_\mathrm{max}$ to $c_\mathrm{min}$ and it does not matter which period of the stationary phase is considered. So, for an arbitrary 'hill' with maximum at $t_n$ one can use $\tau_0=t_n+X$ along the descending 'slope' and the idea is to determine first $X$ as function of the difference $Y=\tau_0-\tau_1$ and this solution must have the properties $X=0$ if $Y=0$ and $X=\tmin_{n+1}-t_n$ if $Y=2\pi$ . To do so, the equations \eqref{eq:DepthMatEqBasis} need to be rewritten in terms of the new variables $Y$ and $X$. Obviously, also the parameters used to describe the material behaviour, i.e., the set $\tilde{p}$, appear in the function $X$ and thus both $\tau_0$ and $\tau_1$ appear to be a function of $Y$ and $\tilde{p}$: $\tau_0=t_n+X(Y,\tilde{p})$ and $\tau_1=t_n+X(Y,\tilde{p})-Y$ but this dependence will not always be explicitly indicated further on.

As $t_n$ is  a multiple of $2\pi$, $\tau_0=2n\pi +X$ and $\tau_1=2n\pi+(X-Y)$, it follows that $\cos \tau_0=\cos X$ and $\cos\tau_1=\cos(X-Y)$.
The sum terms $\mathcal{U}_j$ and $\mathcal{F}_j$ in \eqref{eq:DepthMatEqBasis} involve integrals of $T_{2j}(\tau_i,s)$ and $T_{2j}(\tau_i,s)\cos s$ for $i=0,1$. It is shown in Chap.~\ref{sec:ArgumentsOfTandN} (see \eqref{eq:ArgumentsOfTt0} and \eqref{eq:ArgumentsOfTt1}) that the $T_{2j}$-kernels are \emph{always} of the form\footnotemark
\begin{gather}
T_{2j}(\tau_0,s)=T_{2j}(\tau_0-s,\tau_1-s,\ldots,\tau_{2j}-s;P(0,2j)),\enspace j=0,1,2,\ldots,\\
T_{2j}(\tau_1,s)=T_{2j}(\tau_1-s,\tau_2-s,\ldots,\tau_{2j}-s;P(1,2j)),\enspace j=0,1,2,\ldots.
\end{gather}
The elements of the sets $P(0,2j)$ and $P(1,2j)$ are composed of differences between any $\tau_k$ and another $\tau_i$ (see \eqref{eq:SetOfTauDiffs}). \footnotetext{In \ref{sec:SLEkernels} more specific formula are derived for the functions $T_{j}$ when the material is assumed to behave like a standard linear element -- actually the most simple case. For more general material behaviour often a Prony series (Appendix \ref{sec:Prony}) is used and formulae facilitating the computation of the kernels $T_j$ and $N_j$ for this case were derived in Appendix \ref{sec:Pronykernels}.\label{fnote:TandNProperties}}
For periodic contact radii -- the period is $2\pi$ in the present case -- the elements of the sets $P(0,2j)$ and $P(1,2j)$ are equal to either a multiple of $2\pi$ or to the sum of $Y$ and such a multiple of $2\pi$; the sets both depend on $Y$ but not on $X$. As the integration variable $s$ of the integrals in the sums range from $\tau_{2j+1}$ to $\tau_{2j}$ with $\tau_{2j}-\tau_{2j+1}=Y$, the substitution $s=\tau_{2j+1}+u$ then shows that the dependence of the kernel functions $T_{2j}$ on the variable $s$ involves only the independent variables $u$ and $Y$; the kernels $T_{2j}$ and -- by extension -- the sum terms $\mathcal{U}_j$ do not depend on $X$. However, the sum terms $\mathcal{F}_j$ do because the same substitution, i.e., $s=\tau_{2j+1}+u$, transforms the function $\cos s$ in the integrands into $\cos(X-Y+u)$ because $\tau_{2j+1}=\tau_1-2j\pi=t_n+X-Y-2j\pi$ and $t_n$ is a multiple of $2\pi$. All these considerations lead to the conclusion that  \emph{always} -- whatever the assumed functional form of $\phi$  --  the equations \eqref{eq:DepthMatEqBasis} can be written as
\begin{equation}
h_1\begin{bmatrix}m_{11}(Y)&m_{12}(Y)\\
m_{21}(Y)&m_{22}(Y)
\end{bmatrix}\begin{bmatrix}\sin(X)\\
\cos(X)
\end{bmatrix}=\mathbb{A}(R,c_\mathrm{max},h_1)
\begin{bmatrix}b_1(Y)\\
b_2(Y)
\end{bmatrix}.\label{eq:DepthMatEqBasis1}
\end{equation}
Multiplication of \eqref{eq:DepthMatEqBasis1} from the left with the row $[b_2\,\,\,-b_1]$ eliminates the factor $\mathbb{A}$ from the resulting equation from which now $X$ as function of $Y$ can be found. The problem with this direct approach is that the mathematics is somewhat tricky as the solution is multivalued and physical arguments must be used to resolve these matters during the solution process. Therefore it makes sense to push the analysis one step further by taking
\begin{equation}
X=\frac{Y}{2}+\alpha
\end{equation}
and solving for $\alpha$ instead of $X$. The rationale for this shift is that the contact radius cannot have the same type of symmetry as the depth. Whereas the depth has mirror symmetry around all maxima at $t_n$, the contact radius does not have this symmetry and the function $\alpha$ takes this difference into account\footnote{A contact radius with mirror symmetry around the maximum \emph{always} has the property $X=Y/2$.} . The new set of equations arrived at is
\begin{equation}
h_1\begin{bmatrix}m_{11}&m_{12}\\
m_{21}&m_{22}
\end{bmatrix}\begin{bmatrix}\cos\left(\frac{Y}{2}\right)&\sin\left(\frac{Y}{2}\right)\\
-\sin\left(\frac{Y}{2}\right)&\cos\left(\frac{Y}{2}\right)
\end{bmatrix}\begin{bmatrix}\sin(\alpha)\\
\cos(\alpha)
\end{bmatrix}=\mathbb{A}
\begin{bmatrix}b_1\\
b_2
\end{bmatrix}.\label{eq:DepthMatEqBasis2}
\end{equation}
Now, multiplication of \eqref{eq:DepthMatEqBasis2} from the left by the row $[b_2\,\,\,-b_1]$ shows  that the final equation is
\begin{equation}\beta_1(Y)\sin(\alpha)+\beta_2(Y)\cos(\alpha)=0,\label{eq:DepthMatEqBasis3}
\end{equation}
with
\begin{multline}
\beta_1(Y)=\cos\left(\frac{Y}{2}\right)\Bigl\{b_2(Y)m_{11}(Y)-b_1(Y)m_{21}(Y)\Bigr\},\\
+\sin\left(\frac{Y}{2}\right)\Bigl\{-b_2(Y)m_{12}(Y)+b_1(Y)m_{22}(Y)\Bigr\},
\end{multline}and
\begin{multline}
\beta_2(Y)=\cos\left(\frac{Y}{2}\right)\Bigl\{b_2(Y)m_{12}(Y)-b_1m_{22}(Y)\Bigr\}\\
+\sin\left(\frac{Y}{2}\right)\Bigl\{b_2(Y)m_{11}(Y)-b_1(Y)m_{21}(Y)\Bigr\}.
\end{multline}
The solution of \eqref{eq:DepthMatEqBasis3} shows that
\begin{equation}
X(Y)=\frac{Y}{2}+\alpha(Y)=\frac{Y}{2}
+\arctan\left(-\frac{\beta_2(Y)}{\beta_1(Y)}\right).\label{eq:AlphaSol}
\end{equation}
From the sum of the two components of \eqref{eq:DepthMatEqBasis1}, $\mathbb{A}$ can be found, and then with \eqref{eq:AlphaSol}, $\mathbb{L}'(c_\mathrm{max})=h_\mathrm{max}$ and the definition \eqref{eq:defAA} of $\mathbb{A}$ leads to
\begin{multline}
\mathbb{L}'(R)=h_\mathrm{max}\Biggl(1-h_1\\
\times\Biggl(1-
\frac{\sum\limits_{i=1}^2\Bigl\{\sin(X(Y))\,m_{i1}(Y)+\cos(X(Y))\,m_{i2}(Y)\Bigr\}}
{b_1(Y)+b_2(Y)}\Biggr)\Biggr).\label{eq:RadiusSol}
\end{multline}
With $R=\mathbb{C}(\mathbb{L}'(R))$, $\mathbb{L}'(R)$ can be inverted, generating $R$ as function of $Y$.

The results \eqref{eq:AlphaSol} and \eqref{eq:RadiusSol} constitute a parameterized description -- with $Y$ as the parameter ranging from 0 to $2\pi$ -- of the contact radius graph in the $t$-$r$ plane during a single period
, i.e.,
\begin{equation}
(t,c(t))=\begin{dcases}(\tau_0,c(\tau_0))=(t_n+\frac{Y}{2}+\alpha(Y),R(Y))\enspace \text{(descending part),}\\
(\tau_1,c(\tau_1))=(t_n-\frac{Y}{2}+\alpha(Y),R(Y))\enspace \text{(ascending part).}
\end{dcases}\label{eq:ParameterizedCurve1}
\end{equation}
For interpretation of this result it should be kept in mind that the set $\tilde{p}$ of material parameters  and the set $\mathbf{h}=\{h_\mathrm{max},h_1\}$
of depth parameters  also appear in this description because $X=X(Y;\tilde{p})$ and $R(Y;\tilde{p},\mathbf{h})$.

The contact between the indenter and the substrate is lost whenever the contact radius becomes zero before the indenter is retracted. For the present setup,  the minimum of the contact radius occurs at $Y=2\pi$ and contact is maintained if  $\mathbb{L}'(R=2\pi)>0$. After defining the function\footnote{For clarity, the dependence on the material parameters $\tilde{p}$ is again included.}
 \begin{equation}
\mathcal{D}(\tilde{p})= \frac{\sum\limits_{i=1}^2\Bigl\{\sin(X(2\pi,\tilde{p}))\,m_{i1}(2\pi,\tilde{p})+\cos(X(2\pi,\tilde{p}))\,m_{i2}(2\pi,\tilde{p})\Bigr\}}
{b_1(2\pi,\tilde{p})+b_2(2\pi,\tilde{p})},\label{eq:BbbSolution}
 \end{equation}
 it follows from  \eqref{eq:RadiusSol} that
\begin{eqnarray*}
  \mathcal{D}(\tilde{p})&>&1-1/h_1\text{: $\Rightarrow\mathbb{L}'(0)>0$ $\Rightarrow$ $c>0$: contact,} \\
  \mathcal{D}(\tilde{p})&=& 1-1/h_1\text{: $\Rightarrow\mathbb{L}'(0)=0$ $\Rightarrow$ $c=0$: loss of contact at minimum,} \\
 \mathcal{D}(\tilde{p})&<&1-1/h_1\text{: $\Rightarrow\mathbb{L}'(0)<0$ $\Rightarrow$ $c<0$: impossible, contact was lost.}
\end{eqnarray*}
The last inequality suggests that too large a value of $h_1$  might  even prohibit the build up  of a stationary state  as  during the experiment contact is probably repeatedly lost.

Finally, as the result \eqref{eq:AlphaSol} for $X(Y,\tilde{p})$ is independent of the depth parameter $h_\mathrm{max}$ and $h_1$, it follows that $\tau_0-t_n$ depends only on the material parameters $\tilde{p}$. At  $\tmin_{n+1}=t_n+X(2\pi,\tilde{p})$ the contact radius reaches a minimum and here also the load reaches a minimum (see \ref{sec:MinOfLoadAndRadius}). On the other hand, the depth reaches a minimum at $t_n+\pi$. Consequently, the time difference $\pi-X(2\pi,\tilde{p})$ is a measure for the phase shift between depth and load and this phase shift depends only on the parameter set $\tilde{p}$. However, note that the time-like members of the set $\tilde{p}$ also involve the  unit of time $2\pi/\Omega$ of the actually prescribed frequency; from a physics point of view, the set $\tilde{p}$ also depends on $\Omega$ and the phase shift does so as well.
\subsection[sls]{Depth control: the 'standard linear solid'}\label{Sec:DepthControlledSLE}
For material behaviour modelled as the so-called  'standard linear solid' it was shown in Appendix \ref{sec:SLEkernels}   that the functions $T_{2j}(t,s)$ are proportional to $T_0=\dot{\phi}$ because the rates  $\dot{\phi}$ and $\dot{\varphi}$ entering the transformation formulae are simple exponentials. Therefore, substitution of \eqref{eq:T2k} and \eqref{eq:GammaM} in the definitions of $\mathcal{U}_j$  and $\mathcal{F}_j$ yields
\begin{gather}
\mathcal{U}_j(t)=-q\kappa\exp\left(-\kappa t+q\kappa\gamma_{2j}\right) \int\limits_{\tau_{2j+1}}^{\tau_{2j}}\exp(\kappa s)\,\mathrm{d}s,\label{eq:UjFunctieA}\\
\mathcal{F}_j(t)=-q\kappa\exp\left(-\kappa t+q\kappa\gamma_{2j}\right) \int\limits_{\tau_{2j+1}}^{\tau_{2j}}\exp(\kappa s)\cos(s)\,\mathrm{d}s\label{eq:FjFunctieA}
\end{gather}
with  $\gamma_{2j}=-j(\tau_0-\tau_1)+2j\pi $ because of the periodicity of the contact radius. These integrals were evaluated in Appendix \ref{app:MatrixCoefs} for $t=\tau_0=t_n+X$ and $t=\tau_1=\tau_0-Y$ using also $\tau_{2j+1}=\tau_1-2j\pi$ and $\tau_{2j}=\tau_0-2j\pi$;  formulae for the matrix coefficients $m_{ij}$ and $b_i$ appearing in \eqref{eq:DepthMatEqBasis2} can also be found there. These expressions will be  used in the next paragraph to investigate the influence of the material parameters $q$ and $\kappa$  on the shape of the contact radius, the time to go from a maximum to the next minimum (receding time) and the value of $h_1$ for which contact is lost.
\begin{figure}[htb]
  \centering
  \includegraphics[scale=1]{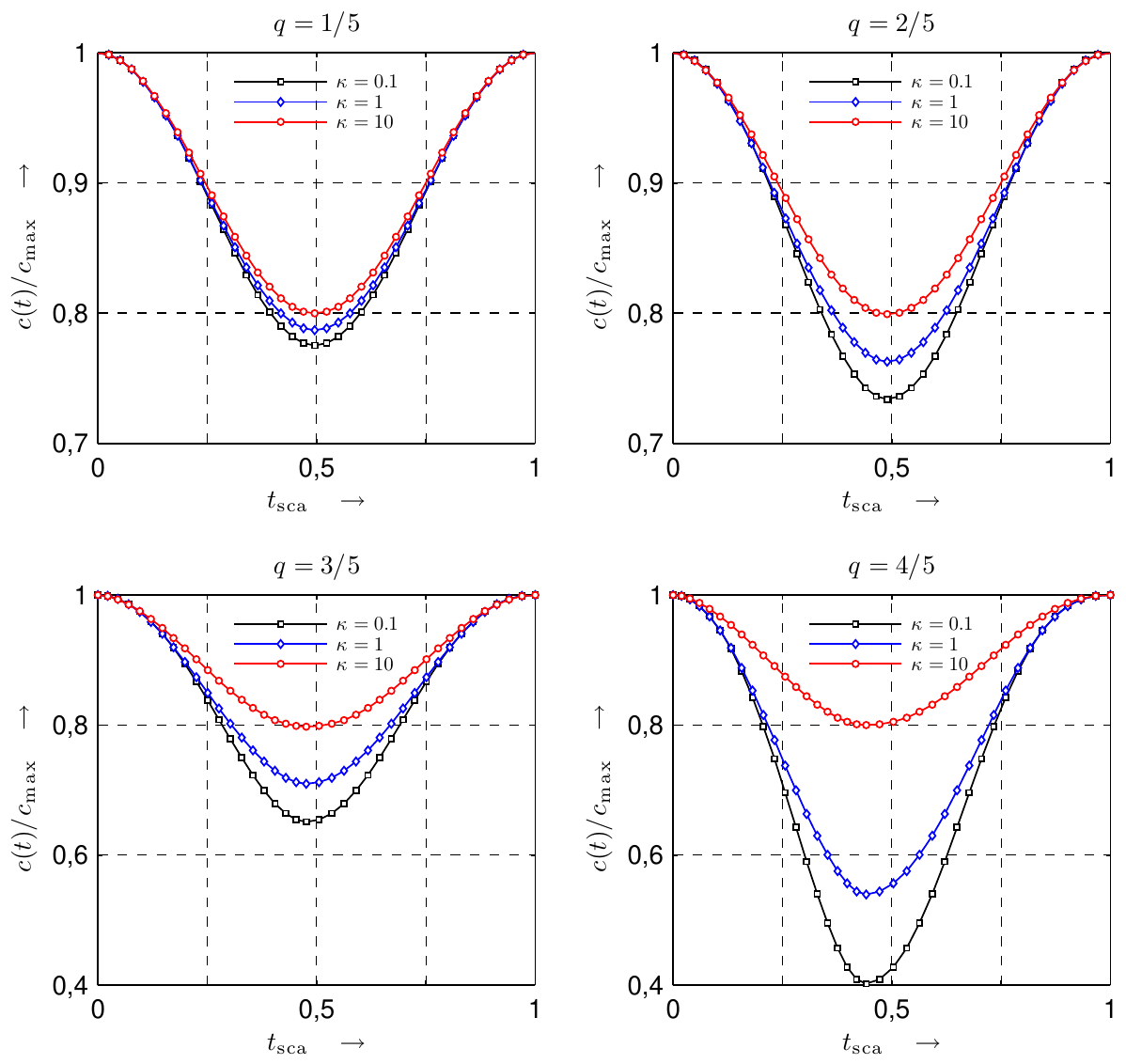}\\
  \caption{Depth controlled stationary state, conical indenter. Relative contact radius $c(t)/c_\mathrm{max}$  as function of the scaled time $t_\mathrm{sca}=(t-t_n)/2\pi$ for a complete period $t_n<t<t_{n+1}$. Parameters: $h_1=1/10$, $q=1/5$, 2/5, 3/5, 4/5 and $\kappa=1/10$, 1 and 10. For an elastic material, the minimal value is always $1-2 h_1$ at $t_\mathrm{sca}=1/2.$  Taking the root of the ordinate yields the graphs for a  \emph{parabolic} indenter.}\label{fig:FixedRadiusVersusTime}
\end{figure}
\subsubsection{Influence of the parameters, $q$, $\kappa$ and $h_1$ on the contact radius}
Fig. \ref{fig:FixedRadiusVersusTime} shows,  for twelve  different parameter combinations $(q,\kappa)$,  graphs of the contact radius of a conical indenter as function of time during an arbitrary period in the stationary state. The relative size $h_1$ of the sinusoidal perturbations is always 0.1 in these graphs.
The parameter $q$ equals $\phi(0)-\phi(\infty)$, i.e., the difference between initial and long term elastic response. So for small $q$ elastic effects are expected to dominate whereas for $q$-values approaching 1, viscous effects become more important. Four values for $q$ were investigated, namely $q=1/5$, $2/5$, $3/5$ and $4/5$.  If $\kappa<<1$ the time scale of the control variable is small with respect to that of the relaxation and for $\kappa>>1$ the reverse is true. The three investigated values for $\kappa$ were $1/10$, $1$ and $10$.
\begin{figure}[htb]
  \centering
  \includegraphics[scale=1]{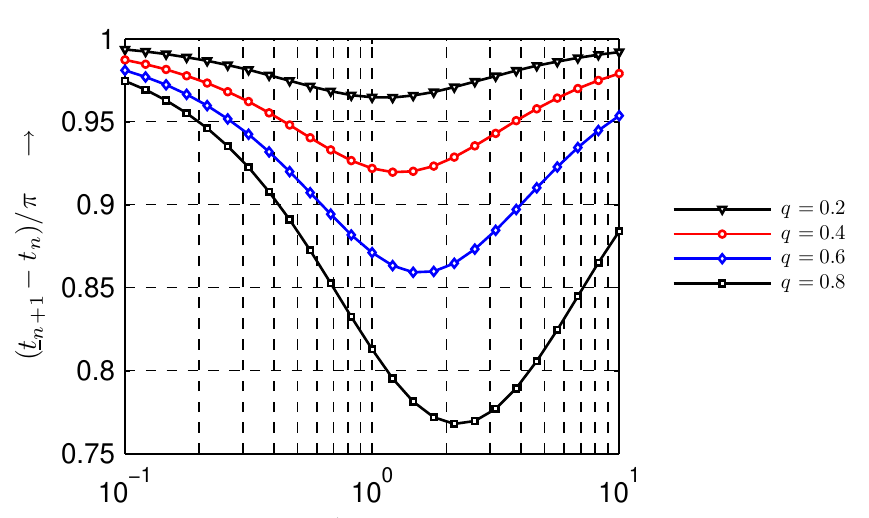}\\
  \caption{Depth controlled stationary state, conical indenter. Influence of the material properties of the standard linear solid on the receding time of the contact radius for sinusoidally perturbed depth (case $h_1=1/10$). }\label{fig:FixedTminVersusKappa}
\end{figure}
\begin{figure}[htb]
  \centering
  \includegraphics[scale=1]{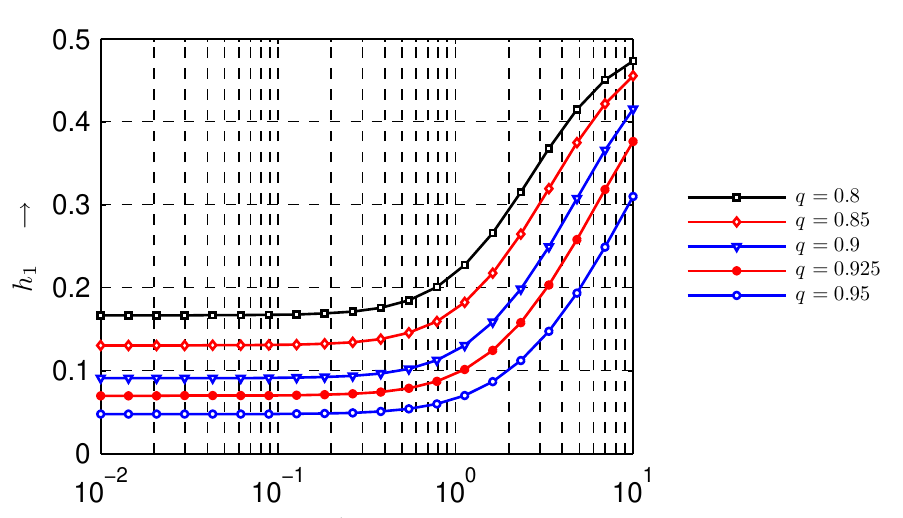}\\
  \caption{Depth controlled stationary state, conical indenter. Value of the (relative) perturbation magnitude $h_1$ for which the minimal contact radius is zero with  indenter still not retracted. }\label{fig:LossOfContact}
\end{figure}
The graphs of Fig.~\ref{fig:FixedRadiusVersusTime} remain valid for all  axisymmetric and convex indenters if the ordinate is interpreted as $\mathbb{L}'(c(t))/h_\mathrm{m}$; for instance when the ordinate is replaced by its root then the graph for a parabolic indenter is found.

In the stationary state, the magnitude and time of occurrence of the maxima are completely determined by the depth. At first sight it would seem that  viscoelasticity has more effect on the magnitude of the minimal contact radius than on the times $\tmin_n$  the minima are attained. A plot of the relative shift of the time of the contact minimum in Fig.~\ref{fig:FixedTminVersusKappa} shows that the time needed to go from a maximum to a minimum decreases if $q$ approaches 1. The influence of $\kappa$ is more involved because -- for fixed $q$ -- an initial decrease of the period of time  to reach a minimum is followed by an increase if $\kappa$ increases. The largest effect occurs if $\kappa$ is of order 1, that is, if the physical relaxation time and the period of the actually applied depth function are of the same order of magnitude. Generally, however, it can be concluded that the contact radius reaches its minimum before the depth does ($\tmin_{n+1}-t_n<\pi$).

The behaviour of the function $\mathcal{D}$ from \eqref{eq:BbbSolution} is also illuminating as far as the influence of viscoelasticity is concerned.
Loss of contact certainly occurs whenever $\kappa$, $q$ and $h_1$ combine in such a way that at the time of minimal contact, i.e., at $Y=2 \pi$,
$\mathcal{D}(q,\kappa)=1-1/h_1$. The transition value of $h_1$, i.e., $h_{\mathrm{1,trans}}=1-\mathcal{D}(q,\kappa)$ is shown in Fig.~\ref{fig:LossOfContact}. For any fixed value $q$, this graph reveals that permanent contact prevails if $h_1$ and $\kappa$ yield a $(h_1,\kappa)$ point below the curve in this plot.

\section[Depth control: general case]{Load controlled contact radius: general viscoelastic material behaviour}\label{sec:DynamicGeneralVisco2}
As explained in Chap. ~\ref{sec:Preliminaries}, the minimum of the contact radius is  taken as a reference value because in a load driven stationary state the minima of load and contact radius coincide, i.e., $\ell(\tmin_n)=\ell_\mathrm{min}=4\mathbb{G}(c_\mathrm{min})\phi(\infty)$. Moreover, it was suggested to use the load function $\ell(t)=\ell_\mathrm{min}\{1-\ell_1(\cos(t)-1)\}$.
The equations relating  load and contact radius are \eqref{eq:GenGLoad3A} and \eqref{eq:GenGLoad4aa} and they are valid in an arbitrary 'valley' (see Fig.~\ref{fig:PeriodicContact}),
\begin{gather}
\ell(\tau_1)+\sum\limits_{k=0}^\infty\mathcal{Q}_k(\tau_1)=4\mathbb{G}(R)\left(1-\sum\limits_{k=1}^\infty\mathcal{V}_k(\tau_1\right),\enspace\tau_1=\tau_1(R,n),
\label{eq:GenGLoad3A1}\\
\ell(\tau_2)+\sum\limits_{k=1}^\infty\mathcal{Q}_k(\tau_2)=4\mathbb{G}(R)\left(1-\sum\limits_{k=1}^\infty\mathcal{V}_k(\tau_2)\right),\enspace\tau_2=\tau_2(R,n).
\label{eq:GenGLoad4aa2}
\end{gather}
After the introduction of two new functions, $\mathcal{N}_k(t)$ and $\mathcal{S}_k(t)$, by\footnote{For the same reasons as  mentioned earlier in footnote \ref{note:1}, the dependence of $\mathcal{N}_k$, $\mathcal{S}_k$ and $\mathcal{V}_k(t)$ on $R$, $n$  and $\tilde{p}$ is implied.}
 \begin{equation}
 \mathcal{N}_k(t) =\int\limits_{\tau_{2j+2}}^{\tau_{2j+1}}N_{2k+1}(t,s)\,\mathrm{d}s,\quad
 \mathcal{S}_k(t)=\int\limits_{\tau_{2j+2}}^{\tau_{2j+1}}N_{2k+1}(t,s)\cos s \,\mathrm{d}s,\label{eq:NandSFunctions}
 \end{equation}
and with $\ell(t)$ as above, $\mathcal{Q}_k(t)$ becomes
\begin{equation}
\mathcal{Q}_k(t)=\ell_\mathrm{min}\{(1+\ell_1)\mathcal{N}_k(t)-
\ell_1\mathcal{S}_k(t)\}.
 \end{equation}
With this, and after division by $\ell_\mathrm{min}=4\mathbb{G}(c_\mathrm{min})\phi(\infty)$, the following matrix equation is obtained
\begin{multline}
(1+\ell_1)\begin{bmatrix}1+\sum\limits_{k=0}^\infty \mathcal{N}_k(\tau_1)\\
1+\sum\limits_{k=1}^\infty \mathcal{N}_k(\tau_2)\end{bmatrix}-\ell_1\begin{bmatrix}\cos\tau_1 +\sum\limits_{k=0}^\infty \mathcal{S}_k(\tau_1)\\
\cos\tau_2+\sum\limits_{k=1}^\infty \mathcal{S}_k(\tau_2)
\end{bmatrix}\\
=\frac{\mathbb{G}(R)}{\mathbb{G}(c_\mathrm{min})\phi(\infty)}
\begin{bmatrix}1-\sum\limits_{k=1}^\infty \mathcal{V}_k(\tau_1)\\
1-\sum\limits_{k=1}^\infty\mathcal{V}_k(\tau_2)\end{bmatrix}.\label{eq:LoadMatEqBasis}
\end{multline}
In the limit $\ell_1\to 0$,  $R\downarrow c_\mathrm{min}$, and then this limit applied to  \eqref{eq:LoadMatEqBasis} shows that two factors in this equation are related by
\begin{equation}
\frac{1}{\phi(\infty)}\begin{bmatrix}1-\sum\limits_{k=1}^\infty \mathcal{V}_k(\tau_1)\\
1-\sum\limits_{k=1}^\infty \mathcal{V}_k(\tau_2)\end{bmatrix}=\begin{bmatrix}1+\sum\limits_{k=0}^\infty \mathcal{N}_k(\tau_1)\\
1+\sum\limits_{k=1}^\infty \mathcal{N}_k(\tau_2)\end{bmatrix}.\label{eq:ColumnRelations}
\end{equation}
Combination of \eqref{eq:LoadMatEqBasis} and \eqref{eq:ColumnRelations} leads to
\begin{equation}
-\ell_1\begin{bmatrix}\cos\tau_1 +\sum\limits_{k=0}^\infty \mathcal{S}_k(\tau_1)\\
\cos\tau_2+\sum\limits_{k=1}^\infty \mathcal{S}_k(\tau_2)
\end{bmatrix}=\mathbb{B}(R,c_\mathrm{min},\ell_1)\begin{bmatrix}1+\sum\limits_{k=0}^\infty \mathcal{N}_k(\tau_1)\\
1+\sum\limits_{k=1}^\infty \mathcal{N}_k(\tau_2)\end{bmatrix},\label{eq:LoadMatEqBasis1}
\end{equation}
with  $\mathbb{B}$ defined by:
\begin{equation}
\mathbb{B}(R,c_\mathrm{min},\ell_1)=\frac{\mathbb{G}(R)}{\mathbb{G}(c_\mathrm{min})}-1-\ell_1.
\end{equation}
For a cone, a parabola or a 'power-law shaped' indenter, the ratio of $\mathbb{G}(R)$ and $\mathbb{G}(c_\mathrm{min})$ depends only on the ratio
$R/c_\mathrm{min}$; specifically
\begin{equation}
\frac{\mathbb{G}(R)}{\mathbb{G}(c_\mathrm{min})}=\begin{cases}\left(R/c_\mathrm{min}\right)^2&\enspace\text{(cone),}\\
\left(R/c_\mathrm{min}\right)^3&\enspace\text{(parabola).}\end{cases}
\end{equation}
\begin{figure}[htb]
  \centering
  \includegraphics[scale=1]{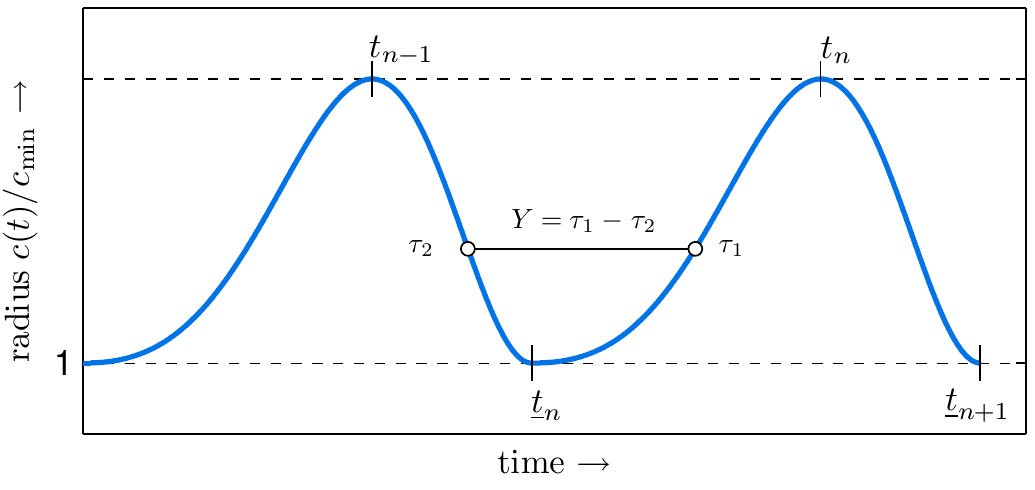}\\
  \caption{Load controlled stationary state. The minima of load and contact radius coincide at $\tmin_n$. Calculation  of $\tau_1-\tmin_n$ as function of $Y=\tau_1-\tau_2$ along the ascending branch $\tmin_n<\tau_1<t_{n+1}$ of a typical period.}\label{fig:GenGplotV1}
\end{figure}
The time difference  $Y=\tau_1-\tau_2$ (Fig.~\ref{fig:GenGplotV1} )  ranges from 0 to $2\pi$, and is zero at an arbitrary time $\tmin_n$  at which the contact radius is minimal (Fig.~\ref{fig:GenGplotV1}). Therefore, with $\tau_1=\tmin_n+X$ and $\tau_2=\tmin_n+X-Y$, the idea is again to derive equations linking $X$ and $Y$. For the same reasons as mentioned previously in Chap.~\ref{sec:DynamicsGeneralVisco}, the actual calculation is again through the intermediate variable $\alpha$ -- defined by $\alpha=X-Y/2$ -- which measures the deviation from the mirror symmetry the contact radius exhibits around the minima at $\tmin_n$; for this, the equations \eqref{eq:LoadMatEqBasis1} need to be rewritten in terms of the new variables.

The times $\tmin_n$ are multiples of $2\pi$ and  $\cos(\tau_1)=\cos(X)$ plus  $\cos\tau_2=\cos(X-Y)=\sin X\sin Y+\cos X \cos Y$.

The terms $\mathcal{N}_k$ and $\mathcal{S}_k$ in  the sums in \eqref{eq:LoadMatEqBasis1} involve integrals of $N_{2k+1}(\tau_i,s)$ and $N_{2k+1}(\tau_i,s)\cos s$ for $i=1,2$. In Chap.~\ref{sec:ArgumentsOfTandN} (see \eqref{eq:ArgumentsOfNt1}--\eqref{eq:ArgumentsOfNt2}) it is shown that the $N_{l}$ kernels are -- just like the $T_j$'s -- \emph{always} of the form\footnote{See also footnote \ref{fnote:TandNProperties}.}
\begin{gather}
N_{l}(\tau_1,s)=N_{2k+1}(\tau_1-s,\tau_2-s,\ldots,\tau_{l}-s;P(1,l)),\\
N_{l}(\tau_2,s)=N_{2k+1}(\tau_2-s,\tau_3-s,\ldots,\tau_{l}-s;P(2,l)).
\end{gather}
Due to the $2\pi$ periodicity of the contact radius, the elements of the sets $P(1,l)$ and $P(2,l)$ are equal to either a multiple of $2\pi$ or to the sum of such a multiple of $2\pi$ and $Y$;  both sets depend on $Y$ but not on $X$. The integration variable $s$ of the $\mathcal{N}_{2k}$ and $\mathcal{S}_{2k}$ integrals ranges from $\tau_{2k+2}$ to $\tau_{2k+1}$ with $\tau_{2k+1}-\tau_{2k+2}=Y$. The substitution $s=\tau_{2j+2}+u$ then shows that the dependence of $N_{2k+1}$ on the variable $s$ involves only the independent variables $u$ and $Y$; the $N_{2k+1}$-functions and -- by extension -- the sum terms $\mathcal{N}_k$ do not depend on  $X$. However, the terms in  $\sum\mathcal{S}_k$ do, because the substitution $s=\tau_{2j+1}+u$ transforms the function $\cos s$ in the integrand into $\cos(X-Y+u)$. Just as in Chap.~\ref{sec:DynamicsGeneralVisco} it is concluded that the equations \eqref{eq:LoadMatEqBasis1} can be written as
\begin{equation}
-\ell_1\begin{bmatrix}\hat{m}_{11}(Y)&\hat{m}_{12}(Y)\\
\hat{m}_{21}(Y)&\hat{m}_{22}(Y)
\end{bmatrix}\begin{bmatrix}\sin(X)\\
\cos(X)
\end{bmatrix}=\mathbb{B}(R,c_\mathrm{min},\ell_1)
\begin{bmatrix}\hat{b}_1(Y)\\
\hat{b}_2(Y)
\end{bmatrix}.\label{eq:LoadMatEqBasis2}
\end{equation}
and they can be solved in the same way as described there, i.e., by setting $X=Y/2+\alpha(Y)$. Now,
\begin{multline}
\hat{\beta}_1(Y)=\cos\left(\frac{Y}{2}\right)\Bigl\{\hat{b}_2(Y)\hat{m}_{11}(Y)-\hat{b}_1(Y)\hat{m}_{21}(Y)\Bigr\},\\
+\sin\left(\frac{Y}{2}\right)\Bigl\{-\hat{b}_2(Y)\hat{m}_{12}(Y)+\hat{b}_1(Y)\hat{m}_{22}(Y)\Bigr\}
\end{multline}plus
\begin{multline}
\hat{\beta}_2(Y)=\cos\left(\frac{Y}{2}\right)\Bigl\{\hat{b}_2(Y)\hat{m}_{12}(Y)-\hat{b}_1\hat{m}_{22}(Y)\Bigr\}\\
+\sin\left(\frac{Y}{2}\right)\Bigl\{\hat{b}_2(Y)\hat{m}_{11}(Y)-\hat{b}_1(Y)\hat{m}_{21}(Y)\Bigr\}.
\end{multline}
and the equation for $\alpha$ is
\begin{equation}\hat{\beta}_1(Y)\sin(\alpha(Y))+\hat{\beta}_2(Y)\cos(\alpha(Y))=0.\label{eq:LoadMatEqBasis3}
\end{equation}
which leads to
\begin{equation}
 X(Y)=\frac{Y}{2}+\alpha(Y)=\frac{Y}{2}+\arctan\left(-\frac{\hat{\beta}_2(Y)}{\hat{\beta}_1(Y)}\right).
\label{eq:LoadAlphaSol}
\end{equation}
Finally, the sum of the components of \eqref{eq:LoadMatEqBasis2} yields the equation
\begin{multline}
\mathbb{G}(R)=\mathbb{G}(c_\mathrm{min})\Biggl(1+\ell_1\\
\times\Biggl(1-\frac{\sum\limits_{i=1}^2\Bigl\{\sin(X(Y))\,\hat{m}_{i1}(Y)+\cos(X(Y))\,\hat{m}_{i2}(Y)\Bigr\}}
{\hat{b}_1(Y)+\hat{b}_2(Y)}\Biggr)\Biggr).\label{eq:LoadRadiusSol}
\end{multline}
For a conical or parabolical indenter the scaled radius $R_\mathrm{sca}=R/c_\mathrm{min}$ follows from this result because
\begin{equation}
R_\mathrm{sca}=\begin{dcases}\Bigl\{\mathbb{G}(R)/\mathbb{G}(c_\mathrm{min})\Bigr\}^{1/2}&\enspace\text{(cone),}\\
\Bigl\{\mathbb{G}(R)/\mathbb{G}(c_\mathrm{min})\Bigr\}^{1/3}&\enspace\text{(parabola).}\end{dcases}
\end{equation}
For these indenter types, the equations \eqref{eq:LoadAlphaSol} and \eqref{eq:LoadRadiusSol} constitute a parameterized description of the scaled contact radius during a single period, i.e.,
\begin{equation}
(t,\frac{c(t)}{c_\mathrm{min}})=\begin{dcases}(\tau_1,\frac{c(\tau_1)}{c_\mathrm{min}})=(t_n+\frac{Y}{2}+
\alpha(Y),R_\mathrm{sca}(Y))\enspace\text{(ascending part),}\\
(\tau_2,\frac{c(\tau_2)}{c_\mathrm{min}})=(t_n-\frac{Y}{2}+\alpha(Y),R_\mathrm{sca}(Y))\enspace \text{(descending part).}
\end{dcases}\label{eq:ParameterizedCurve2}
\end{equation}
For other indenter types, \eqref{eq:LoadRadiusSol} is rewritten as
\begin{multline}
\mathbb{G}(R)=\frac{\ell_\mathrm{min}}{4\phi(\infty)}\Biggl(1+\ell_1\\
\times\Biggl(1-
\frac{\sum\limits_{k=1}^2\Bigl\{\sin(X(Y))\,\hat{m}_{k1}(Y)+\cos(X(Y))\,\hat{m}_{k2}(Y)\Bigr\}}
{\hat{b}_1(Y)+\hat{b}_2(Y)}\Biggr)\Biggr),
\end{multline}
and inversion of $\mathbb{G}$ generates $R(Y)$, i.e., the contact radius as function\footnote{Note that $\phi(\infty)$ is not an independent parameter, it is  a function of $\tilde{p}$.} of $Y$, $\tilde{p}$ and the set of load parameters $\boldsymbol{\ell}=\{\ell_\mathrm{min},\ell_1\}$. For interpretation of these results it should be kept in mind that the set $\tilde{p}$ of material parameters and the set $\mathbf{\ell}= \{\ell_\mathrm{min},\ell_1\}$ of load parameters also appear in these descriptions  because $X=X(Y;\tilde{p})$ and $R=R(Y;\tilde{p},\mathbf{\ell})$.

Just like \eqref{eq:AlphaSol} is the result \eqref{eq:LoadAlphaSol} independent of the amplitude parameters $\ell_\mathrm{min}$ and $\ell_1$ of the load; $X=\tau_1-\tmin_n$ depends only on $Y$ and the material parameters $\tilde{p}$. At  $t_n=\tmin_{n}+X(2\pi,\tilde{p})$ the contact radius reaches a maximum and the depth is also maximal here (see \ref{sec:MinOfLoadAndRadius}) whereas the load is maximal at $\tmin_n+\pi$. Therefore, the time difference $\pi-X(2\pi,\tilde{p})$ is again a measure for the phase shift between depth and load and this shift depends only on the material constants $\tilde{p}$, but, as already mentioned several times, from a physics point of view the shift also depends on the period of the actually applied load function.

An important difference between depth and load control is that no loss of contact occurs as long as $\ell(t)$ is positive, which is always the case as $\ell_\mathrm{min}$ is positive; a function comparable to $\mathcal{D}(\tilde{p})$ from Chap.~\ref{sec:DynamicsGeneralVisco} (see \eqref{eq:BbbSolution}) does not exist for load control.

\subsection{Load control: the 'standard linear solid'}\label{Sec:LoadControlledSLE}
For load control the functions $N_{2k+1}$ are needed and for standard linear material behaviour it was shown in \ref{sec:SLEkernels}
that these functions are proportional to $N_1=\dot{\varphi}$ because the rates  $\dot{\phi}$ and $\dot{\varphi}$ entering the transformation formulae are simple exponentials.
Therefore, substitution of \eqref{eq:N2kplus1} in the definitions \eqref{eq:NandSFunctions} of $\mathcal{N}_k$  and $\mathcal{S}_k$ results in
\begin{equation}
\mathcal{N}_k(t)=q\kappa\exp\left(-\kappa(1-q) t+q\kappa(\gamma_{2k+1}-\tau_1)\right) \int\limits_{\tau_{2k+2}}^{\tau_{2k+1}}\exp(\kappa (1-q)s)\,\mathrm{d}s,\label{eq:NjFunctieA}
\end{equation}
and
\begin{multline}
\mathcal{S}_k(t)=q\kappa\exp\left(-\kappa(1-q)t+q\kappa(\gamma_{2k+1}-\tau_1)\right) \\
\times\int\limits_{\tau_{2k+2}}^{\tau_{2k+1}}\exp(\kappa (1-q)s)\cos(s)\,\mathrm{d}s.\label{eq:SjFunctieA}
\end{multline}
\begin{figure}[b]
  \centering
  \includegraphics[scale=1]{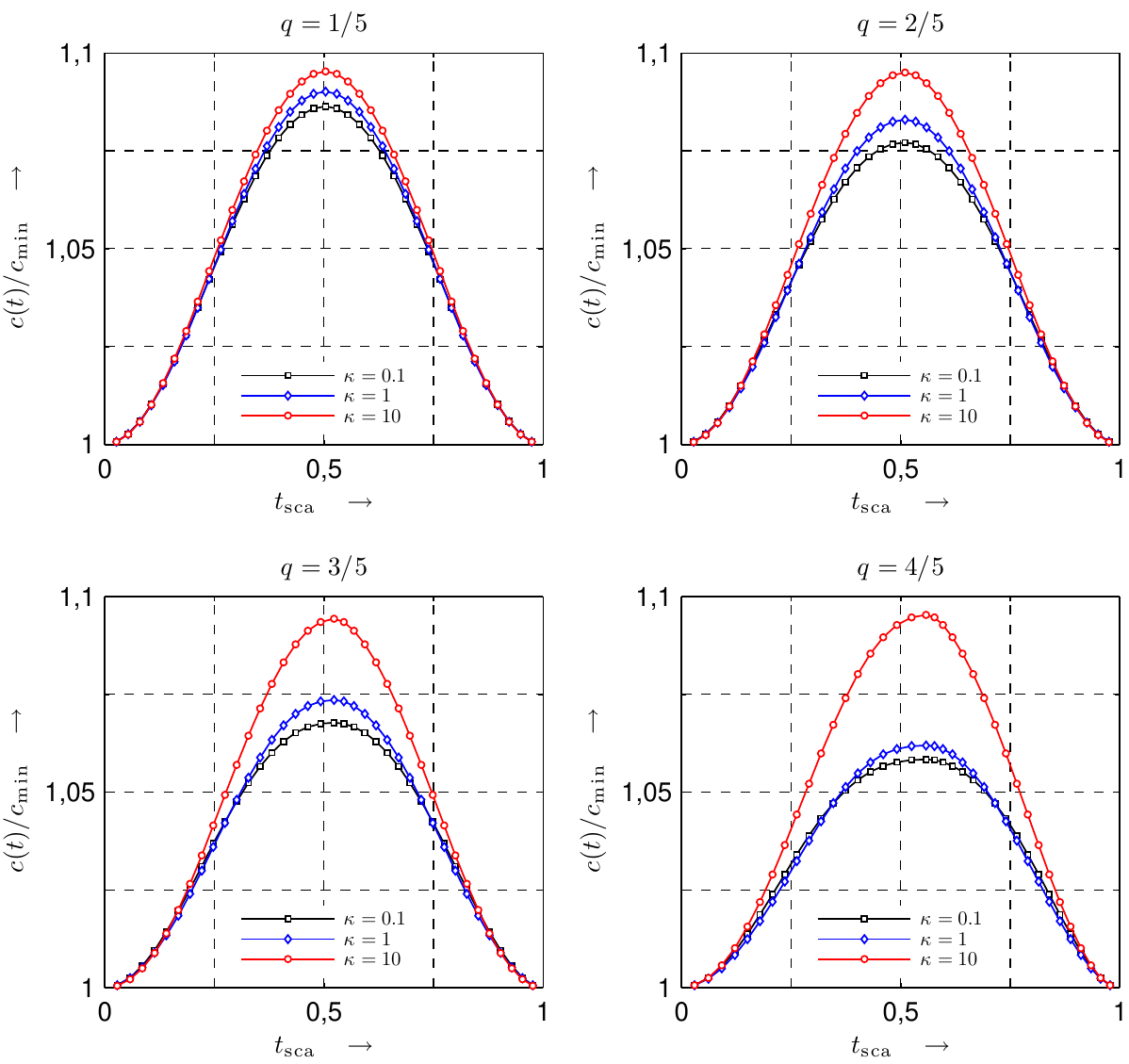}\\
  \caption{Load controlled stationary state, conical indenter. Relative contact radius $c(t)/c_\mathrm{min}$  as function of the scaled time $t_\mathrm{sca}=(t-\tmin_n)/2\pi$ for a complete period $\tmin_n<t<\tmin_{n+1}$. The load parameter is here always $\ell_1=1/10$ and the material parameters are $q=1/5$, $2/5$, $3/5$, $4/5$ and $\kappa=1/10$, $1$, $10$. For an elastic material, the maximal value is always at $t_\mathrm{sca}=1/2$ and equal to$\sqrt{1+2\ell_1}$.  Raising the ordinates to the power 2/3 yields the graphs for a  \emph{parabolic} indenter.}\label{fig:FixedLoadRadiusVersusTime}
\end{figure}
From \eqref{eq:GammaM}, the definition of $\gamma_j$, and the relations between the $\tau$'s for a $2\pi$ periodic contact radius it follows that
\begin{equation}
\gamma_{2k+1}-\tau_1=-k\{2\pi - (\tau_2-\tau_1)\}.
\end{equation}
The integrals \eqref{eq:NjFunctieA} and \eqref{eq:SjFunctieA} were evaluated in Appendix \ref{app:LoadMatrixCoefs} for $t=\tau_1=\tmin_n+X$ and $t=\tau_2=\tau_1-Y$ by using also $\tau_{2j+2}=\tau_2-2j\pi$ and $\tau_{2j+1}=\tau_1-2j\pi$. The resulting expressions  \eqref{eq:Bhat1}, \eqref{eq:Bhat2}, \eqref{eq:hatm11} -- \eqref{eq:hatm22} for the matrix coefficients $\hat{b}_{i}$ and $\hat{m}_{ij}$ appearing in \eqref{eq:LoadMatEqBasis2} are found in this Appendix. The results will be  used to in the next section to investigate the influence of the material parameters $q$ and $\kappa$  on the shape of the contact radius and the time needed to change from a minimum to the next maximum.
\subsubsection{Influence of the parameters, $q$ and $\kappa$ on the contact radius}
Fig. \ref{fig:FixedLoadRadiusVersusTime} shows,  for twelve  different parameter combinations $(q,\kappa)$, graphs of the contact radius of a conical indenter as function of time during an arbitrary period in the stationary state. The relative size $\ell_1$ of the sinusoidal perturbations is always 0.1 in these graphs.

The general features of these graphs are the same as seen earlier in the corresponding plots for depth control. The major difference is, of course, that for a load controlled experiment loss of contact cannot occur as long as the load is positive.

\begin{figure}[h]
  \centering
  \includegraphics[scale=1]{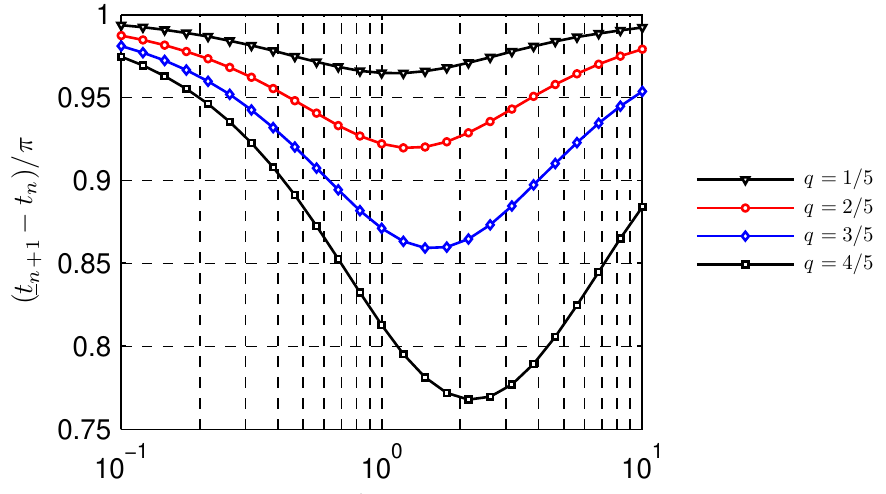}\\
  \caption{Load controlled stationary state, conical indenter. Influence of the material properties of a standard linear solid on the receding time of the contact radius for a sinusoidally perturbed load (case $\ell_1=1/10$). }\label{fig:FixedLoadTminVersusKappa}
\end{figure}

For increasing values of $q$ the maximum of the radius shifts to the right of the period, i.e., in each period the duration of the increasing phase of the contact radius is longer than the duration of the receding phase.  This effect also depends on the values of $\kappa$ as the graphs in Fig.~\ref{fig:FixedLoadTminVersusKappa} show. The largest effect occurs again if $\kappa$ is of order 1, i.e., if the physical relaxation time and the period of the actually applied depth function are of the same order of magnitude. Generally, however, it can be concluded that the contact radius reaches its minimum before the depth does. These graphs are actually the same as those from Fig.~\ref{fig:FixedTminVersusKappa} as a comparison of the numerical data showed.

\section[Determination of the viscoelastic material parameters]{Determination of the viscoelastic material parameters (final step)}\label{sec:FinalStep}
The Secs.  \ref{sec:DynamicsGeneralVisco} to \ref{Sec:LoadControlledSLE} dealt with finding the parameterized representations \eqref{eq:ParameterizedCurve1} and \eqref{eq:ParameterizedCurve2} of the contact radius during a typical period of a sinusoidal controlled stationary state.    The list of known quantities in these representations contains the independent variable $Y$, ranging from 0 to $2\pi$, the parameters of the control variable -- $\mathbf{h}=\{h_\mathrm{max},h_1\}$ for depth and $\boldsymbol{\ell}=\{\ell_\mathrm{min},\ell_1\}$ for load control -- and, albeit in a somewhat hidden form, the period $2\pi/\Omega$ of the actually applied signal as this period serves as a unit of time. The set of material parameters $\tilde{p}$ present in the hypothesized  functional form of the reduced relaxation function $\phi$ is not known and only restricted in quite a general way. To determine the actual value of the material parameters  the  remaining, as yet unused, experimental data -- the function $\ell(t)$ if the depth and $h(t)$ if the load is controlled -- need to be used. The general structure of a (possible) set of equations from which $\tilde{p}$ can be determined numerically is derived next.
\begin{figure}[htb]
  \centering
  \includegraphics[scale=1]{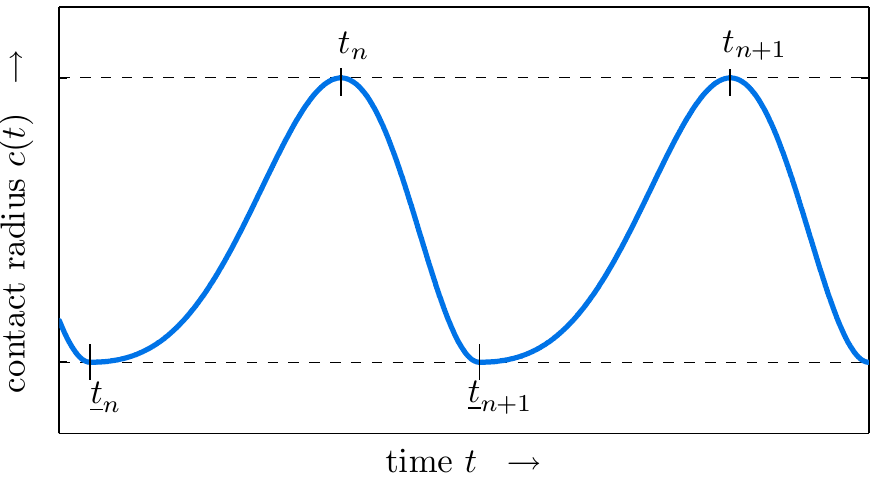}
  \caption{Typical periods of $c(t)$ curve for a step shaped, but sinusoidally perturbed,  depth or load controlled stationary state. At $t=t_n$, both $h$ and $c$ are maximal whereas at $t=\tmin_n$ the load $\ell$ and $c$ are minimal. }\label{fig:GenGplot}
\end{figure}
\subsection{Controlled depth, measured load}
 Since, $h(t)=h_\mathrm{max}\{1+h_1\{\cos t-1\}\}$ and depth, contact radius and load have the same (scaled) period $2\pi$ in the stationary state, the experimental data for the load in the stationary state can be represented by a truncated Fourier series of the form $\bar{a}_0+\sum_{k=1}^{N_\ell} \{\bar{a}_k\cos k t+\bar{b}_k\sin kt\}$. The set of constants in this series, $\bar{\mathbf{a}}=\{\bar{a}_0,\ldots,\bar{a}_{N_\ell},\bar{b}_1,\ldots,\bar{b}_{N_\ell}\}$, is fixed by fitting the series to the measured load data; the approximation
\begin{equation}
\bar{\ell}(t,\bar{\mathbf{a}})=\bar{a}_0+\sum\limits_{k=1}^{N_\ell} \{\bar{a}_k\cos k t+\bar{b}_k\sin kt\}\label{eq:LoadParmDetermination}
\end{equation}
is used for the load $\ell(t)$. The representation \eqref{eq:LoadParmDetermination} implies $\bar{\ell}(t,\bar{\mathbf{a}})=\bar{\ell}(t\pm t_n,\bar{\mathbf{a}})$.
The  parameterized representation \eqref{eq:ParameterizedCurve1} of the contact radius from Chap.~\ref{sec:DynamicsGeneralVisco} originated from 'hill' equations \eqref{eq:GenGRadius7} and  \eqref{eq:GenGRadius6}. Therefore, the corresponding 'hill' equations \eqref{eq:GenGLoad6aa} and \eqref{eq:GenGLoad6b} are used for the load:
\begin{gather}
\ell(\tau_0)-\sum_{k=1}^\infty\mathcal{P}_k(\tau_0,R)=4\mathbb{G}(R)\left(1+\sum_{k=0}^\infty\mathcal{U}_k(\tau_0,R)\right),\enspace
\tau_0=\tau_0(R),\label{eq:GenGLoad6aa1}\\
\ell(\tau_1)-\sum_{k=1}^\infty\mathcal{P}_k(\tau_1,R)=4\mathbb{G}(R)\left(1+\sum_{k=1}^\infty\mathcal{U}_k(\tau_1,R)\right),\enspace
\tau_1=\tau_1(R).\label{eq:GenGLoad6b1}
\end{gather}
This choice is convenient because the terms  on the right-hand sides of \eqref{eq:GenGLoad6aa1} and \eqref{eq:GenGLoad6b1} are already known at this stage. Indeed, comparison of \eqref{eq:DepthMatEqBasis} with \eqref{eq:DepthMatEqBasis1} shows that
$1+\sum_{0}^\infty \mathcal{U}_j(\tau_0)=b_1(Y)$ and $1+\sum_1^\infty \mathcal{U}_j(\tau_1)=b_2(Y)$. Moreover, $\mathbb{G}(R)$ follows from substituting the inverse $R=R(Y)$ of \eqref{eq:RadiusSol} in the characteristic function $\mathbb{G}$.
Obviously, as in Chap.~\ref{sec:DynamicsGeneralVisco},  $Y= \tau_0-\tau_1$,  $\tau_0=t_n+X(Y)$ and $\tau_1=t_n+X(Y)-Y$ with $t_n$ a multiple of $2\pi$. Consequently, the load terms on the left-hand sides of  \eqref{eq:GenGLoad6aa1} and \eqref{eq:GenGLoad6b1} become  $\ell(\tau_0)\approx \bar{\ell}(X(Y)$ and $\ell(\tau_1)\approx\bar{\ell}(X(Y)-Y)$.
The sum terms on the left-hand sides of \eqref{eq:GenGLoad6aa1} and \eqref{eq:GenGLoad6b1}, i.e.,
$\mathcal{P}_k(\tau_i)=\int\ell(s)T_{2k-1}(\tau_i,s)\mathrm{d}s$, with the integration variable ranging from $\tau_{2k}$ tot $\tau_{2k-1}$,
is rewritten using $\bar{\ell}$ for the load and  $s=\tau_{2k}+x=\tau_0-2k\pi+x$ with the new integration variable $x$ now ranging from 0 tot $2\pi-Y$.

As argued in Chap.~\ref{sec:DynamicsGeneralVisco}, the kernels $T_{2k-1}$ depend on the new integration variable $x$ and further on $Y$ and $\tilde{p}$ but not on $X$.
However,  since $\bar{\ell}(s,\bar{\mathbf{a}})=\ell(X(Y,\tilde{p})+x,\bar{\mathbf{a}})$, the functions $\mathcal{P}_k$ and, by extension, the left-hand sides of \eqref{eq:GenGLoad6aa1} and \eqref{eq:GenGLoad6b1} depend on the load constants $\bar{\mathbf{a}}$, the material constants $\tilde{p}$ and the time difference $Y$. With the shorthand notation $\varsigma_{i+1}=\ell(\tau_{i})-\sum_k\mathcal{P}_k(\tau_i)$ for $i=0$ and 1,  \eqref{eq:GenGLoad6aa1} and \eqref{eq:GenGLoad6b1}, become
\begin{equation}
\begin{bmatrix}\varsigma_1(Y,\tilde{p},\mathbf{h},\bar{\mathbf{a}})\\
\varsigma_2(Y,\tilde{p},\mathbf{h},\bar{\mathbf{a}})
\end{bmatrix}=4\mathbb{G}(R(Y,\tilde{p},\mathbf{h})
\begin{bmatrix}b_1(Y,\tilde{p})\\
b_2(Y,\tilde{p})
\end{bmatrix}\enspace\text{for}\enspace\forall Y\in(0,2\pi).\label{eq:LoadNumerical}
\end{equation}
The constants $\mathbf{h}$ and $\bar{\mathbf{a}}$ have a known value and $Y$ ranges from 0 to $2\pi$. The numerical determination of the material constants $\tilde{p}$ is a problem of nonlinear parameter identification. Indeed, given the  depth and and load constants $\mathbf{h}$ and $\bar{\mathbf{a}}$, respectively, the set of constants $\tilde{p}$ is to be estimated such that \eqref{eq:LoadNumerical} is met -- exactly or in some other suitable respect -- for all values of $Y\in(0,2\pi)$.

\subsection{Controlled load, measured depth}
The equations \eqref{eq:LoadNumerical} apply to the case of depth control but similar formulae exist for the case of load control. In the latter case, $\ell(t)=\ell_\mathrm{min}\{1-\ell_1(\cos t-1)\}$ server as control variable, a function with minima at $t=\tmin_n=2 n \pi$.  The matrix equation \eqref{eq:LoadMatEqBasis2} in  Chap.~\ref{sec:DynamicGeneralVisco2} from which first $\tau_1=\tmin_n+X(Y)$, $\tau_2=\tmin_n+X(Y)-Y$ and then the contact radius $R=R(Y)$ is calculated is  based on \eqref{eq:GenGLoad3A1} and \eqref{eq:GenGLoad4aa2}. Both are 'valley' equations and it is convenient to use the 'valley' equations \eqref{eq:GenGRadius4} and \eqref{eq:GenGRadius4a} connecting depth and contact radius for the present purpose, i.e.,
\begin{gather}
h(\tau_1)-\sum\limits_{k=1}^\infty\mathcal{W}_k(\tau_1,R)=
\mathbb{L}'(R)\left(1-\sum\limits_{k=1}^\infty\mathcal{V}_k(\tau_1,R)\right),\enspace \tau_1=\tau_1(R),\label{eq:GenGLoad3A2}\\
h(\tau_2)-\sum\limits_{k=1}^\infty\mathcal{W}_k(\tau_2,R)=
\mathbb{L}'(R)\left(1-\sum\limits_{k=1}^\infty\mathcal{V}_k(\tau_2,R)\right),\enspace \tau_2=\tau_2(R).\label{eq:GenGLoad4aa3}
\end{gather}
The factors between parentheses on the right-hand sides are already known once the matrix equation \eqref{eq:LoadMatEqBasis2} is established. Indeed, from \eqref{eq:ColumnRelations} and the right-hand sides of \eqref{eq:LoadMatEqBasis1} and \eqref{eq:LoadMatEqBasis2}, it follows that
$1-\sum_k\mathcal{V}_k(\tau_i)=\phi(\infty)\hat{b}_i(Y)$ for $i=t$ and 2. The depth data, needed for evaluation of the left-hand sides of \eqref{eq:GenGLoad3A2} and \eqref{eq:GenGLoad4aa3}, are represented by a set of constants, $\bar{\mathbf{d}}=\{\bar{d}_0,\ldots,\bar{d}_{N_h},
\bar{e}_1,\ldots,\bar{e}_{N_h}\}$, generated by fitting the experimental data to a Fourier series:
\begin{equation}
h(t)\approx \bar{h}(t,\bar{\mathbf{d}}) =\bar{d}_0+\sum_{k=1}^{N_h}\{\bar{d}_k\cos kt  +\bar{e}_k\sin kt\}.
\label{eq:DepthParmDetermination}
\end{equation}
This representation guarantees automatically that $h(\tau_1)\approx \bar{h}(X,\bar{\mathbf{d}})$ and $h(\tau_2)\approx \bar{h}(X-Y,\bar{\mathbf{d}})$ because $\bar{h}(t,\bar{\mathbf{d}})=\bar{h}(t-\tmin_n,\bar{\mathbf{d}})$ and $\tmin_n$ is a multiple of $2\pi$.. The sum terms $\mathcal{W}_k(\tau_i)$ on the left-hand sides of \eqref{eq:GenGLoad3A2} and \eqref{eq:GenGLoad4aa3} equal $\int h(s)N_{2k}(\tau_i,s)\mathrm{d}s$ and they are rewritten by using $\bar{h}(s,\bar{\mathbf{d}})$ for the depth and by changing the integration variable from $s$ to $x$ via $s=\tau_{2k+1}+x$. The integration limits of the integrals $\mathcal{W}_k$ are $\tau_{2k+1}=\tau_1-2k\pi$ and $\tau_{2k}=\tau_0-2k\pi$ because the system is in a stationary state. Therefore, the new integration variable $x$ ranges from 0 to $2\pi-Y$ and the transformation from $s$ to $x$ changes $\bar{h}(s,\bar{\mathbf{d}})$ to $\bar{h}(X(Y,\tilde{p})+x,\bar{\mathbf{d}})$. As in Chap.~\ref{sec:DynamicGeneralVisco2}, the kernels $N_{2k}$ depend on the new integration variable $x$ and further on $Y$ and $\tilde{p}$ but not on $X$. The upshot of all these consideration is that eventually an equation comparable to \eqref{eq:LoadNumerical} is found:
\begin{equation}
\begin{bmatrix}\hat{\varsigma}_1(Y,\tilde{p},\boldsymbol{\ell},\bar{\mathbf{d}})\\
\hat{\varsigma}_2(Y,\tilde{p},\boldsymbol{\ell},\bar{\mathbf{d}})
\end{bmatrix}=\phi(\infty)\mathbb{L}'(R(Y,\tilde{p},\boldsymbol{\ell})
\begin{bmatrix}\hat{b}_1(Y,\tilde{p})\\
\hat{b}_2(Y,\tilde{p})
\end{bmatrix}\enspace\text{for}\enspace \forall Y\in(0,2\pi).\label{eq:DepthNumerical}
\end{equation}
The functions $\hat{\varsigma}_i$ ($i=1,2$) on the left-hand side are shorthand for $h(\tau_i)-\sum_k\mathcal{W}_k(\tau_i)$. Note that \eqref{eq:DepthNumerical} is similar to \eqref{eq:LoadNumerical} and the way to compute the material constants $\tilde{p}$ is -- apart from obvious differences -- expected to be also similar.

\chapter{Conclusions}\label{sec:Discussion}
A load depth sensing experiment is characterised by  five functions: the load $p(t)$; the depth $h(t)$; the contact radius $c(t)$;  the material function $\omega(t)$ or its Stieltjes inverse $\omega^\text{i}=\varpi(t)$ and, finally, the indenter shape  function $f(r)$. The shape function $f(r)$ is known in advance and later, that is after finishing an experiment,  the data $p(t)$ and $h(t)$ are  also known functions as they are recorded during the experiment. Therefore, two equations are always necessary to obtain a mathematically solvable system, i.e., two equations that link the data and the indenter shape to the remaining unknowns, which are the contact radius $c(t)$ and the material function $\omega(t)$, where the determination of the latter function is the ultimate aim of the experiment.
\section{The link between experimental data,  contact size,  material properties and indenter shape}
This link can be described by different but equivalent equations (see \eqref{eqPart1:radiusequation} to \eqref{eqPart2:loadequation3e} in Chap. \ref{sec:ContactRadiusEquations}). The two most informative ones are an equation relating depth and contact radius,
\begin{gather}
[L'\orc\omega](r,t)|_{r=c(t)}=[h\orc\omega](t),\label{eq:radiusequationA}\\
\intertext{and an equation connecting load, depth and contact radius,}
p(t) =4\Bigl\{c(t)[h\orc\omega](t)-[L\orc\omega](r,t)|_{r=c(t)}\Bigr\}.\label{eq:loadequation3eA}
\end{gather}
 They apply to elastic and viscoelastic materials alike.

 For elastic materials, $\omega(t)=\omega\young\hea{t}$ and the depth-contact relation \eqref{eq:radiusequationA} implies that $h=\mathbb{L}'(c)$. As the function $\mathbb{L}'$ is characteristic for the indenter (see Appendix \ref{app:CharacteristicFunction}), the depth-contact relation depends \emph{only} on the indenter shape. This means that $c$ can be eliminated from \eqref{eq:loadequation3eA} once and for all in favour of the depth $h$. Doing so results in the classic load-depth formula for elastic materials: $p=4\omega\young\mathbb{F}(h)$. Also the function $\mathbb{F}$ is also characteristic for the indenter (see Table \ref{tab:CharFun}).

When the substrate is viscoelastic the permanent elimination of the contact radius from the load equation is only possible in two special cases: a strictly advancing or a strictly receding contact from the time the loading is switched on to the current time.

\section{Strictly advancing or receding contact}
A contact is strictly advancing if $c(t)$ increases  or remains constant from the time the experiment started and it recedes if the contact, after an initial jump, decreases afterwards. The equations governing these cases differ in two aspects.
The first is about the relation between the the depth $h$ and the contact radius $c$ because (see \eqref{eq:radius1} and \eqref{eq:radius2})
\begin{align}
h=[h\orc\omega]\orc\omega^\text{i}&=\mathbb{L}'(c)\enspace\Rightarrow\enspace c=\mathbb{C}(h)\quad\text{(strictly advancing),}\label{eq:radius11}\\
 [h\orc\omega]\,\omega^{-1}&=\mathbb{L}'(c)\enspace\Rightarrow \enspace c=\mathbb{C}\left(\frac{h\orc\phi}{\phi}\right)\quad\text{(strictly receding).}\label{eq:radius21}
\end{align}
If the contact is advancing, the quantity $h\orc\omega$ must be convoluted with the Stieltjes inverse of $\omega$ to obtain  $\mathbb{L}'(c)$, whereas for a receding contact multiplication by the reciprocal $\omega^{-1}$ suffices.
This difference renders the contact radius for a strictly advancing contact completely independent of the material function $\omega$ whereas a strictly receding contact radius is dependent of the material function -- actually only on the reduced relaxation function $\phi$ because the initial elastic constant $\omega_0$ drops out of the expression.
The second difference is about the relation between the load $p$ and the contact radius $c$ as can be seen from (\ref{eq:advancingcontact2a})
and (\ref{eq:precede3}), respectively, when written as follows:
\begin{align}
p\orc \omega^\text{i}&=4\{ c\mathbb{L}'(c)-\mathbb{L}(c)\}=4\mathbb{G}(c)\quad\text{(strictly advancing),}\label{eq:advancingcontact2a1}\\
p\omega^{-1}&=4 \{c\mathbb{L}'(c)-\mathbb{L}(c)\}=4\mathbb{G}(c)\quad\text{(strictly receding).}\label{eq:precede31}
\end{align} If the contact strictly advances, $p$ must be convoluted with the Stieltjes inverse of $\omega$ to obtain the right hand side $4\mathbb{G}(c)$ but for a receding contact multiplication by de reciprocal $\omega^{-1}$ suffices. Together, \eqref{eq:radius11}, \eqref{eq:radius21}, \eqref{eq:advancingcontact2a1} and \eqref{eq:precede31} show that in both cases the contact radius can be eliminated form the load-depth relation but  the final formulae differ essentially;
\begin{align}
p\orc \omega^\text{i}&=4\mathbb{F}(h)&\text{(strictly advancing),}\label{eq:advancingcontact2a2}\\
p\omega^{-1}&=4 \mathbb{F}\left(\frac{h\orc\phi}{\phi}\right)&\text{(strictly receding).}\label{eq:precede32}
\end{align}

During a classic relaxation experiment where a step sized depth, i.e., $h(t)=h_0\hea{t}$, is prescribed the contact radius is found to be advancing -- actually only non-decreasing. As the load response  $p(t)=4\mathbb{F}(h_0)\omega(t)$, this type of experiment  can be used to determine the stress relaxation function $\omega$.   The contact radius is also increasing during a classic creep experiment, an experiment where step sized load, $p(t)=p_0\hea{t}$, is applied.  The depth response is now according to $4\mathbb{F}(h(t))=p_0\varpi(t)$ and thus this experiment can be used  to measure the creep function $\varpi=\omega^\text{i}$. However, due to practical experimental side effects, like temperature drift and plastic deformation,  which accompany indentation tests, particularly at low loads and small depths,  these test types are not the methods of choice to determine even the initial elastic, let alone the long term viscoelastic material properties.
\section{Distinguishing between initial and long term response}
The basic equations \eqref{eq:radiusequationA} and \eqref{eq:loadequation3eA} do not change if they are divided by $\omega_0$ and the variable $\ell(t)=p(t)/\omega_0$ and the reduced relaxation function $\phi(t)=\omega(t)/\omega_0$ are used. The result is
\begin{gather}
[L'\orc\phi](r,t)|_{r=c(t)}=[h\orc\phi](t),\label{eq:radiusequationB}\\
\ell(t) =4\Bigl\{c(t)[h\orc\phi](t)-[L\orc\phi](r,t)|_{r=c(t)}\Bigr\}.\label{eq:loadequation3eB}
\end{gather}
This suggests to split the determination of $\omega$  in two parts, starting with the finding of the initial elastic constant $\omega_0$ and proceeding with the  determination of the reduced relaxation function $\phi$.
\section{The initial elastic response constant $\omega_0$}
For elastic materials, the normal procedure is to calculate the elastic constant $\omega\young$ from the contact stiffness, i.e., the slope $S=\mathrm{d}p/\mathrm{d}h=4\omega\young c$ of the load versus depth curve, taken at the start of the unload section of the graph. As mentioned in Chap.~\ref{sec:intro} the slope is additionally used to estimate the value of the contact radius from the depth by correcting for plastic deformation.  Historically, researchers faced a problem when transferring this method to the viscoelastic realm because in many cases the famous 'nose' appeared in the curve thus leading to a negative value for the contact stiffness whereas this stiffness must be positive in order to find physically acceptable values for the initial elastic response $\omega_0$. The appearance of a  'nose' signals that there is a time difference between the occurrence of the maxima of load and depth. While for elastic materials this time difference is always zero -- for elastic materials, extrema of load, depth and contact radius are always synchronized -- it is quite normal when indenting viscoelastic materials that this time difference is present and often non-negligible as was shown in Chap.~\ref{sec:Nose}. Although a judicious choice of the testing conditions surely can render the magnitude of this time difference practically zero, there is no need to do so if the experiment is driven in such a way that kinks (rate jumps) in the input variable occur.
\section{The importance of rate jumps for finding $\omega_0$.}
Kinks in the input variable always result in simultaneous kinks in the other variables and essential is that the corresponding rates exhibit jumps here. Rate jumps are always synchronized and it was found in Chap.~\ref{sec:JumpConditions} that at the jump time $T$ the relation
\begin{equation}
\frac{\langle\dot{p}\rangle_T}{\langle\dot{h}\rangle_T}=4\omega_0 c(T)\label{eq:loadanddepthjumpA}
\end{equation}
applies and this relation is similar to the 'elastic' relation between the contact stiffness, contact radius and elastic constant  Moreover, an estimate of the contact radius $c(T)$ can also be made from the depth $h(T)$ by correcting this depth for plastic deformation using the effective indenter theory (see Chap.~\ref{sec:PlasticInfluence} and Appendix \ref{app:Plastic}) and the rate jump ratio $\langle\dot{p}\rangle_T/\langle\dot{h}\rangle_T$. Thus a procedure for estimating $\omega_0$ that parallels the conventional procedure to find $\omega\young$ for elastic materials is established and the results of \citet[][p. 609--610]{DeWith2006} and \citet{NganAndTang2009} are corroborated.
\section{The long term response,: classic approach and an alternative}
To determine viscoelastic properties,  the conventional procedure to asses the frequency dependent storage and loss moduli of the material has several limitations related to the requirement that the contact radius must be increasing at all times. This limits the allowed frequency and/or accuracy (see Chap.~\ref{sec:perturbation}).

A completely different approach to find $\phi$ is through the decomposition method for hereditary integrals  of \cite[][p. 63--68 ]{GoldenGraham1988}\footnote{A summary of this theory using the notation of this report is found in \ref{sec:GenGAppendix}.} This method considers the two functions  $a(t)$ and $b(t)$, which are connected  by Stieltjes convolutions, that is, by $a=b\orc\phi$ and $b=a\orc\varphi$,  but  each only known on sets of disjoint, but abutting, intervals; on any particular interval is either $a$ known and $b$ remains to be determined or it is the other way around.

\begin{figure}[htb]
  \centering
  \includegraphics[scale=1]{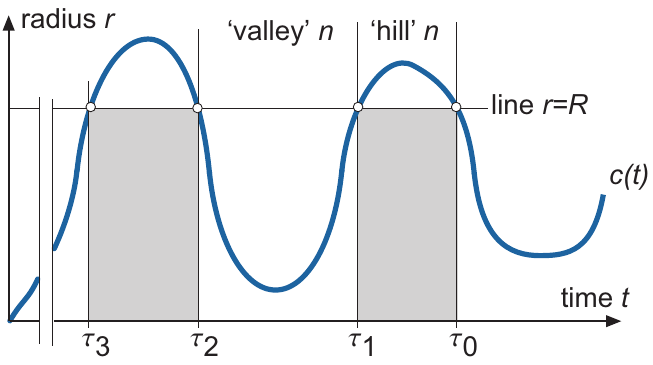}\\
  \caption{Numbering of 'hills' and 'valleys' and the intersection times. For the current time located below 'hill' $n$ or above the previous adjoining 'valley' $n$, the intersection times run from  $\tau_0$  back in time. Note that $c(\tau_j)=R$ for $j=0,1,2\ldots$}\label{fig: VarstFluctatingContactb}
\end{figure}

The application of the decomposition method starts from the observation that the equations \eqref{eq:radiusequationB} and \eqref{eq:loadequation3eB} linking depth, contact radius and load are restrictions of the more general set of equations \eqref{eq:loadequation3} and \eqref{eq:loadequation3d} -- all valid in the whole radius versus time  plane -- to the curve $r=c(t)$ in this plane. The restriction is essentially to the set of surface points currently constituting the edge of the contact region. During dynamic load depth sensing, any particular point  at the substrate surface might move many times in and out the contact region. Considering the history of such a surface  point, it can be said that it moves, metaphorically speaking, along a straight 'road'  -- a line of constant radius $r$, say $R$,  to the indenter tip -- crossing a mountainous region in the $r$-$t$ plane (Fig.~\ref{fig: VarstFluctatingContactb}). The graph of the contact radius can be envisaged as the 'surface profile of this mountainous region' and the point moving along the line $R=$constant is either always above the $c(t)$ curve, i.e., always outside the contact region for large enough values of $R$,  or successively above a 'valley', that is: $R>c(t)$, and under a 'hill', i.e., $R<c(t)$.
At the intersection times, $R$ equals the current contact radius. It was shown in Chap.~\ref{sec:GenGSep1} (see \eqref{eq:GenGRadius} and \eqref{eq:GenGLoad} and lines preceding these equations) that the general equations mentioned above take the form
\begin{gather}
\varrho_c(r,t)=[\mathcal{L}'\orc\phi](r,t)
\genfrac{}{}{0pt}{}{\text{\scriptsize{convolution}}}{\leftrightarrows}
\mathcal{L}'(r,t)=[\varrho_c\orc\varphi](r,t),\label{eq:GenGRadiusA}\\
q_c(r,t)=4[\mathcal{L}\orc\phi](r,t)
\genfrac{}{}{0pt}{}{\text{\scriptsize{convolution}}}{\leftrightarrows}
4\mathcal{L}(r,t)=[q_c\orc\varphi](r,t).\label{eq:GenGLoadA}
\end{gather}
In these equations is $\mathcal{L}(r,t)=L(r,t)-rh(t)$ and $\mathcal{L}'(r,t)=L'(r,t)-h(t)$ and each line consists of two hereditary integrals that are related by a convolution because $\phi$ and $\varphi$ are each other Stieltjes inverse. When the radius in  \eqref{eq:GenGRadiusA} and \eqref{eq:GenGLoadA} is confined  to a constant value $R$, every set is mathematically as  required for the decomposition theorem for hereditary integrals.
Indeed, above a 'valley' $\varrho_c(R,t)=0$ and $q_c(R,t)=\ell(t)$ while $\mathcal{L}'(R,t)$ and $\mathcal{L}(R,t)$ are unknown  whereas under a 'hill' the functions $\mathcal{L}'(R,t)$ and $\mathcal{L}(r,t)$ are equal to $\mathbb{L}'(R)-h(t)$ and $\mathbb{L}(R)-Rh(t)$, respectively, and the unknown functions are now $\varrho_c(R,t)$ and $q_c(R,t)$.
It is outside the scope of the present section to reiterate all relevant formulae; only a general description of the content of Chap.~\ref{sec:GenGSep} is given below.

Decomposition of \eqref{eq:GenGRadiusA}, leads to expressions for $h(\tau_2)$, $h(\tau_1)$ and $h(\tau_0)$ as sums of integral transforms of $h(s)$ and the unit function over all previous 'hill' intervals (see Chap.~\ref{sec:ContactRadius}). The factor $\mathbb{L}'(R)$ also enters these equations because $R=c(\tau_2)=c(\tau_1)=c(\tau_0)$ but the load is not present in the final results.

Similarly, decomposition of \eqref{eq:GenGLoadA} yields formulae for $\ell(\tau_2)$, $\ell(\tau_1)$ and $\ell(\tau_0)$ as sums of integral transforms of $\ell(s)$ over all previous 'valley' intervals and of the unit function over all previous 'hill' intervals (Chap.~\ref{sec:loadformulae} and \ref{sec:SimplifiedLoad}). The value for $R$  is now present through the factor $\mathbb{G}(R)$ but the depth is not.

The material behaviour determines the kernels of these integral transforms,  firstly because the kernel functions $N_j(t,s)$  and $T_j(t,s)$ are recursively defined using $\dot{\varphi}(t-s)$  and $\dot{\phi}(t-s)$ as starting functions and, secondly, because it is used in the recursion itself (see \eqref{eq:NEven} and \eqref{eq:NUneven} for the $N_j$ and  \eqref{eq:TEven} and \eqref{eq:TUneven} for the $T_j$ kernels).  Additionally the kernels also depend on the intersection times $\tau_0, \tau_1,\ldots$. In view of the nature of the recursive formulae plus the fact that the functions $\dot{\varphi}$ and $\dot{\phi}$ depend only through $t-s$ on $t$ and $s$ it is known at a general level -- that is, without more specific information about the behaviour of $\phi$ -- how the kernels depend on the intersection times $\tau_j$, and  the transformation variable $s$ (see Chap. \ref{sec:ArgumentsOfTandN}). Obviously, more detailed information about the kernels becomes  available if the  mathematical form, the reduced relaxation function is assumed to have, is specified in more detail. For example, if a material behaves according to the standard linear element, closed form expressions for the kernels exist and were derived in Appendix \ref{sec:SLEConfiguration}. Often, the hypothesis is that the reduced relaxation function is  a Prony series   (\ref{sec:Prony}). It was shown in \ref{sec:Pronykernels} that all kernels are sums of exponentials with recursively defined coefficients.

The value of the decomposition method of \cite{GoldenGraham1988} is primarily suited for two types of indentation experiments.
The first is the case of a single load and unload, the case $n=1$, and it is shown in Chap.~\ref{sec:SingleLOadUnload} how the decomposition results can be applied to this experiment and the reduced relaxation function be calculated from the experimental data. The results were applied to the case of 'standard linear element' material behaviour for a prescribed depth linearly increasing and later linearly decreasing with speeds $v_1$ and $-v_2$, respectively.
The second is the situation described in Chap.~\ref{sec:hereditary}, where a sinusoidal perturbation is superposed on a step shaped depth or load and the system has reached a stationary state. The actual execution of the latter type of experiment is the same as for classic dynamic load-depth sensing but the processing of the results is different. The basic difference is  that every decomposed equation connects either depth and radius $R$ or load and radius $R$.
Processing of the data now always involves an intermediate step, namely the determination of $R$ as function of the control data, and the results of this step is finally combined with the remaining data to calculate the material constants one is looking for.
\section{The long term response: dynamic load-depth sensing}
\begin{figure}[hbt]
  \centering
  \includegraphics[scale=1]{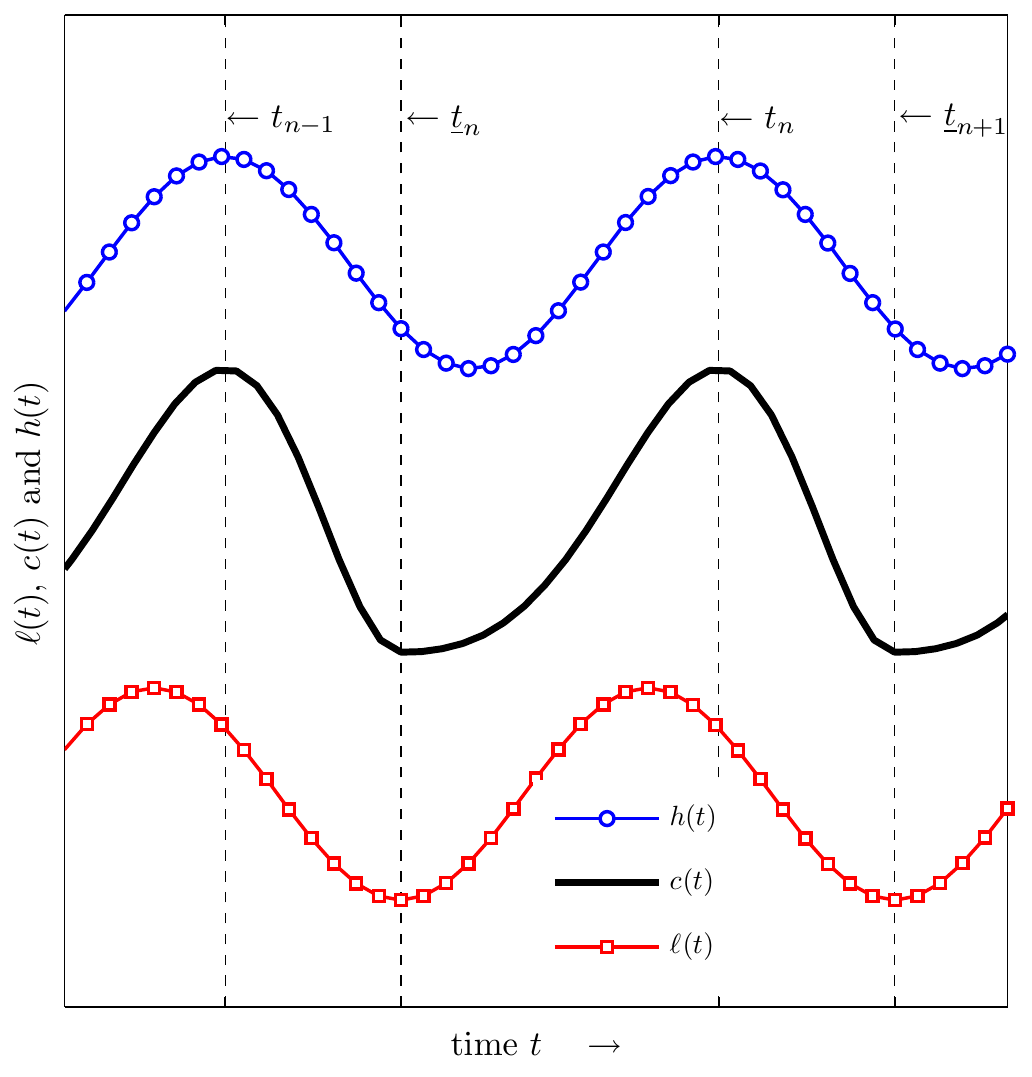}
  \caption{Qualitative picture of the shapes of depth, contact radius and load. At the times $t=\ldots t_{n-1}, t_n,\ldots$ the maxima of $h(t)$ and $c(t)$ coincide. At the times $t=\ldots,\underline{t}_{n},\underline{t}_{n+1},\ldots$ the minima of $c(t)$ and $\ell(t)$ conicide.}\label{fig:AllCurves}
\end{figure}

The application of the decomposition method to dynamic load-depth sensing uses that in the stationary state the maxima of the  contact radius and the depth occur simultaneously, whereas  minima of the contact radius coincide in  time with  the minima of the load This is schematically shown in Fig.~\ref{fig:AllCurves},. Moreover, at the time, $t_n$, where the contact radius is maximal, the relation between $c$ and $h$ is momentarily the same as for an advancing contact: $h(t_n)=\mathbb{L}'(c(t_n))$ or, equivalently, $c(t_n)=\mathbb{C}(h(t_n))$. At the time $\tmin_n$ of a minimum in the contact radius,  matters are more complicated because  the relation between $c$ and $\ell$ also involves the asymptotic value $\phi(\infty)$: $\ell(\tmin_n)/\phi(\infty)=4\mathbb{G}(c(\tmin_n))$.

For a depth controlled stationary state, i.e., a prescribed depth according to\footnote{The period of the actually prescribed sinusoidal perturbation is taken as the unit of time and relaxation times present in the material behaviour are also scaled by this time unit. So, the depth function \eqref{eq:DepthFunctionSec10} has frequency 1.}
\begin{equation}
h(t)=h_\mathrm{max}\{1+h_1(\cos t-1)\},\quad 0\leq h_1 < 1,\label{eq:DepthFunctionSec10}
\end{equation}
it is shown in Chap.~\ref{sec:DynamicsGeneralVisco} that  a matrix equation of the following type
\begin{equation}
h_1\begin{bmatrix}m_{11}(Y;\tilde{p})&m_{12}(Y;\tilde{p})\\
m_{21}(Y;\tilde{p})&m_{22}(Y;\tilde{p})
\end{bmatrix}\begin{bmatrix}\sin(X)\\
\cos(X)
\end{bmatrix}=\mathbb{A}(R,c_\mathrm{max},h_1)
\begin{bmatrix}b_1(Y;\tilde{p})\\
b_2(Y;\tilde{p})
\end{bmatrix}.\label{eq:DepthMatEqBasis4}
\end{equation}
governs the descending part of any period of the contact radius.  The parameter $\tilde{p}$ in \eqref{eq:DepthMatEqBasis4} indicates the set of material constants appearing in the (assumed) mathematical form of $\phi$. The independent variable $Y=\tau_0-\tau_1$ ranges  from 0 to $2\pi$ because the periods of the contact radius is the same as that of the depth. The independent variable $X=\tau_0-t_n$ follows from \eqref{eq:DepthMatEqBasis4}, as shown by \eqref{eq:AlphaSol}. It ranges from 0 to $\tmin_{n+1}-t_n$ when $Y$ ranges from 0 to $2\pi$ and  $X$ depends on $\tilde{p}$ but \emph{not} on $R$ and also not on the depth parameters $h_\mathrm{max}$ and $h_1$. So, $X=X(Y;\tilde{p})$, implying that  the phase shift $\pi-X(2\pi;\tilde{p})$ between depth and load depends only on $\tilde{p}$. After elimination of $X$ in favour of $Y$ with $X(Y,\tilde{p})$ in \eqref{eq:DepthMatEqBasis4}, the solution of the resulting equation yields  $R(Y;\tilde{p},\mathbf{h})$, i.e., a description of the contact radius as function of $Y$, $\tilde{p}$ and the depth parameters $\mathbf{h}=\{h_\mathrm{max},h_1\}$. To determine the magnitudes of the material constants in the set $\tilde{p}$, the remaining equations, those for the load in this case, need to be used. This again leads to a matrix equation of the following type:
\begin{equation}
\begin{bmatrix}\varsigma_1(Y;\tilde{p},\mathbf{h},\bar{\mathbf{a}})\\
\varsigma_2(Y;\tilde{p},\mathbf{h},\bar{\mathbf{a}})
\end{bmatrix}=4\mathbb{G}(R(Y;\tilde{p},\mathbf{h})
\begin{bmatrix}b_1(Y;\tilde{p})\\
b_2(Y;\tilde{p})
\end{bmatrix}\enspace\text{for}\enspace\forall Y\in(0,2\pi).\label{eq:LoadNumerical1}
\end{equation}
The components $\varsigma_1$ and $\varsigma_2$ are shorthand for
$\ell(\tau_{1})-\sum_k\mathcal{P}_k(\tau_1)$ and $\ell(\tau_{2})-\sum_k\mathcal{P}_k(\tau_2)$, respectively.  The load data $\ell(t)$ present in these components are assumed to be approximated by a truncated Fourier series and $\bar{\mathbf{a}}$ indicates the parameter set of  the Fourier coefficients. The numerical computation of the constants in $\tilde{p}$ is a nonlinear parameter identification problem.

If, instead of the depth, the load is controlled according to $\ell(t)=\ell_\mathrm{min}\{1-\ell_1 (cos t-1)\}$, it was shown in Chap.~\ref{sec:DynamicGeneralVisco2} that again a matrix equation of the same type as \eqref{eq:DepthMatEqBasis4} exists. However, now is $Y$ the difference  $\tau_1-\tau_2$ (see Fig.~\ref{fig:GenGplotV1}) and $X=\tau_1-\tmin_n$ . If $Y=0$ also $X=0$, that is $\tau_1=\tau_2=\tmin_n$, and $X=t_n-\tmin_n$, i.e., $\tau_1=t_n$ if $Y=2\pi$. So, this case describes the ascending part of an arbitrary period and the analysis now utilizes -- as shown in Fig.~\ref{fig:AllCurves} -- the coincidence of the minima of load and the contact radius. As the solution for $X$ is again a function of $Y$ and $\tilde{p}$ only; the phase shift  $\pi-X(2\pi,\tilde{p})$ depends only on $\tilde{p}$. Compared to the depth controlled case, the construction of the contact radius differs somewhat because the relation between the minima  of contact  and  load depends on the asymptotic value $\phi(\infty)$. The computation of the material constants in $\tilde{p}$ is again a non-linear parameter identification problem.

For SLS-material behaviour,  $\phi(t)=1+q\{\exp(-\kappa t)-1\}$, and the influence of the material parameters $q$ and $\kappa$ (Chap. \ref{Sec:DepthControlledSLE} and \ref{Sec:LoadControlledSLE}) on the contact radius was investigated for depth and load control, respectively. Generally it was found that in any period of the stationary state the duration of the receding phase of the contact is smaller than that of the advancing phase and this effect occurs for both sinusoidal depth or load control. Moreover, this difference in duration times depends only on the material properties and the frequency of the applied control -- it does not matter whether the depth or the load is controlled. This difference is largest if the period of the control variable is of the same order of magnitude as the time scale present in the reduced relaxation function. If the experiment is depth controlled, this difference in duration times may lead to loss of contact, i.e., $c=0$ without fully retracted indenter ($h>0$), if the amplitude of the perturbation is too large compared to the size of the carrier step.

\appendix
\chapter{Stieltjes convolution}\label{app:convolution}
The most important parts of the extensive paper written by Gurtin and Sternberg \cite{GurtinAndSternberg1962} are summarized here.

A function $f$ is in Heaviside class $H^N$ if it is zero for $t\in (-\infty,0)$ and of class
$C^N$ for $t\in[0,\infty)$. The material functions $\lambda$ and $\mu$ are assumed to be in $H^2$, whereas the depth $h$ and the load $p$ are, at least, in $H^0$.

For $f\in H^0$ and $g\in H^1$, the Stieltjes integral
$
\int_{-\infty}^t f(t-\tau)\,\text{d}g(\tau)
$
is called the {\sl Stieltjes convolution} of $f$ and $g$ and it is written down as
$f\orc g$; in full
\begin{equation}
[f\orc g](t)=\begin{cases} 0 & t<0~,\\
 g(0+)f(t)+\int\limits_{0+}^tf(t-\tau)\dot{g}(\tau)\text{d}\tau & t\geq 0~.
 \end{cases}
\label{eq:a1}
\end{equation}
The integral on the right-hand side  of \eqref{eq:a1} is the conventional convolution usually denoted by the $\star$ symbol only; so $f\orc g =g_0f+f\star\dot{g}$ for $t > 0$.

The {\sl Stieltjes inverse} of $f$ is a function  $k$  such that $ [f\orc k]$ equals the Heaviside step function $\Hea(t)$ and it is denoted by \inv{f}. For $f\in H^2$ a necessary and sufficient condition that \inv{f} exists is that $f_0\neq 0$.

Let $g\in H^0$ and $f$ and $k \in H^1$ then $g \orc f \in H^0$ and $f \orc k \in H^1$ and the most important properties are
\begin{itemize}
\item $g \orc f= f \orc g$ and $g\orc\text{H}=g$,
\item $g \orc(f \orc k)=(g \orc f)\orc k$ and $g \orc(f+k)= g \orc f + g\orc k$,
\item $ g \orc f = 0$ $\forall t$ implies $f = 0$ $\forall t$ or $g = 0$ $\forall t$,
\item $\inv{(f \orc k)}=\inv{f}\orc \inv{k}$ provided \inv{f} and \inv{k} exist.
\end{itemize}
Consider $g \in H^0$ and $f$ at least in $H^2$. Differentiation of $g\orc f$ gives
\begin{gather}
\frac{\mathrm{d}}{\mathrm{d}t}[g\orc f](t)=f_0\dot{g}(t)+\dot{f}_0
g(t)+[g\star\ddot{f}](t)\qquad t> 0,
\label{eq:appdf1}\\
\intertext{or, alternatively}
\frac{\mathrm{d}}{\mathrm{d}t}[g\orc f](t)=f_0\dot{g}(t)+g_0\dot{f}(t)+[\dot{g}\star\dot{f}](t)\qquad t> 0.\label{eq:appdf2}
\end{gather}
The last two terms in \eqref{eq:appdf1} -- also equal to $g\orc\dot{f}$ -- are continuous for $t>0$, so jumps in
$\dot{g}$ lead to jumps in the derivative of $g\orc f$.

Take in \eqref{eq:appdf2} for $f$ a reduced relaxation function  and for $g$ the corresponding creep function, i.e., $g=\inv{f}$ such that $f\orc\inv{f}=\Hea$. Equation \eqref{eq:appdf2} then yields
\begin{equation}
\dot{f}(t)+\dot{g}(t)+\int\limits_0^t\dot{f}(t-s)\dot{g}(s)\mathrm{d}s=0,\enspace\text{if $t>0$, $f_0=g_0=1$ and $g=\inv{f}$}.\label{eq:DotFStarG}
\end{equation}
and from this that $\dot{f}_0=-\dot{g}_0$.
Differentiation of \eqref{eq:DotFStarG} gives for $t>0$
\begin{equation}
\ddot{f}(t)+\ddot{g}(t)+\dot{f}_0\dot{g}t)+\int\limits_0^t\ddot{f}(t-s)\dot{g}(s)\mathrm{d}s=0,
\label{eq:DDotFStarG}
\end{equation}
because  $f_0=g_0=1$ and $g=\inv{f}$. 
\chapter{The indenter characteristic functions}\label{app:CharacteristicFunction}
The {\sl characteristic functions} of the indenter, $\mathbb{L}$, its derivatives $\mathbb{L}'$ and $\mathbb{L}''$ and the auxiliary function $\mathbb{G}$ are defined by
\begin{gather}
\mathbb{L}(r)=\int\limits_0^{r}f'(x)\sqrt{r^2-x^2}\,\mathrm{d}x,\quad
\mathbb{L}'(r)=r\int\limits_0^r\frac{f'(x)\mathrm{d}x}{\sqrt{r^2-x^2}},\label{eq:FirstDerivative}\\
\mathbb{L}''(r)=
\int\limits_0^r\frac{\{f'(x)+xf''(x)\}\mathrm{d}x}{\sqrt{r^2-x^2}},\label{eq:SecondDerivative}\\
\mathbb{G}(r)=r\mathbb{L}'(r)-\mathbb{L}(r)=\int\limits_0^r\frac{x^2f'(x)\,\mathrm{d}x}{\sqrt{r^2-x^2}}.\label{eq:HFunction}
\end{gather}
The inverse $\mathbb{C}$ of $\mathbb{L}'$ exists because \eqref{eq:SecondDerivative} shows that $\mathbb{L}'(r)$ is a monotonic increasing function as $\mathbb{L}''$ is non-negative on account of the properties $f'$ and $f''$ are assumed to have. See table~\ref{tab:CharFun} for these functions calculated for a cone, a parabola and a sphere.
\begin{table}[htb]
\begin{center}
\begin{tabular}{|c|c|c|c|}
\hline
 & Cone & Parabola  & Sphere  \\
   \hline
  $f(r)$ & $k_1r$ & $k_2r^2$ & $k_3\bigl\{1-\sqrt{(1-\dfrac{r^2}{k_3^2})}\bigr\}$ \\
  $\mathbb{L}(c)$ & $k_1 \pi  c^2/4$ & $2k_2c^3/3$ & $\dfrac{ck_3}{2}+\dfrac{(k_3^2-c^2)}{4}\ln\bigl(\dfrac{k_3-c}{k_3+c}\bigr)$ \\
  $\mathbb{L}' (c)$ & $k_1\pi c/2$ & $2k_2c^2$ &$\dfrac{c}{2}\ln\Bigl(\dfrac{k_3+c}{k_3-c}\Bigr) $ \\
  $\mathbb{C}(h)$ & $2h/(k_1\pi)$ & $\sqrt{h/(2k_2)}$&$\sqrt{hk_3}\Bigl(1-\dfrac{h}{6k_3}-\dfrac{h^2}{360k_3^2}+\cdots\Bigr) $ \\
  $\mathbb{F}(h)$& $h^2/(k_1\pi)$&$2h^{3/2}/(3\sqrt{2k_2)})$&
  $ \dfrac{2}{3}\sqrt{h^3k_3}\Bigl(1-\dfrac{h}{10k_3}-\dfrac{h^2}{840k_3^2}+\cdots\Bigr)$\\
   & & & \\
  \hline
\end{tabular}
\caption{Characteristic functions for three indenter types. The parameters $k_1$ and $k_2$ are constants, $k_3$ is the radius of the sphere and $\mathbb{F}(h)=h\mathbb{C}(h)-\mathbb{L}(\mathbb{C}(h))$. Note that $\mathbb{F}'(h)=\mathbb{C}(h)$.}\label{tab:CharFun}
\end{center}
\end{table}
If the indenter profile $f(r)$ is a positively homogeneous function of its argument, i.e., if $f(\alpha r)=\alpha^nf(r)$ for \for all $\alpha >0$, then $\mathbb{L}$ is a power law function, that is, $\mathbb{L}(r)=r^{n+1}\mathbb{L}(1)$, and Euler's equation  $r\mathbb{L}'(r)=(n+1)\mathbb{L}(r)$ applies.

\chapter{Rate jump conditions}\label{app:JumpConditionsAanalysis}
The material functions $\omega$ and $\varpi$ are at least twice differentiable to time for $t>0$. The depth $h$, the load $p$, the radius $c$ and $u'$ are continuous functions of time, but they might exhibit rate jumps at some time, say $T$, i.e., they are piecewise differentiable but the left and right limits of the rates exist at $T$. Since $h\orc\dot{\omega}$ is still continuous for $t>0$, application of \eqref{eq:appdf1} shows that  the rate jump of $h\orc\omega$  at time $T$ is
\begin{equation}
\langle\frac{\mathrm{d}}{\mathrm{d}t}[h\orc\omega]\rangle_T=\omega_0\langle\dot{h}\rangle_T~.
 \label{eq:ratejump1}
\end{equation}
Analogously, as $L(c(t),t)=\mathbb{L}(c(t))$, $L'(c(t),t)=\mathbb{L}'(c(t))$, the continuity of $L\orc\dot{\omega}$ and that of $L'\orc\dot{\omega}$  for $t>0$ shows
\begin{gather}
\langle\frac{\mathrm{d}}{\mathrm{d}t}[L\orc\omega]\rangle_T(r)|_{r=c(T)}=\langle\dot{c}\rangle_T[L'\orc\omega](r,T)|_{r=c(T)},\label{eq:ratejump0}\\
\langle\frac{\mathrm{d}}{\mathrm{d}t}[L'\orc\omega]\rangle_T(r)|_{r=c(T)}=\langle\dot{c}\rangle_T[L''\orc\omega](r,T)|_{r=c(T)}~.
 \label{eq:ratejump2}
\end{gather}
Next, consider the influence of a rate jump on the second of \eqref{eq:loadequation3}, i.e., on \begin{equation}
\vartheta_c(r,t)=\theta(r,t)\hea{c(t)-r}=\frac{2}{\pi}[\{L'-h\}\orc\omega](r,t)~.\label{eq:loadequation3a}
\end{equation}
Take $r < c(T)$ but fixed and note that due to the continuity of $c$ at $T$, there exists a small positive number, say $\delta$ ($0<\delta \ll 1$), such that $r<c(t)$, for $T-\delta T\leq t \leq T+\delta T$. Consequently, $\hea{c(t)-r}=1$ and $\vartheta_c=\theta$ in this time range and thus the rate jump of $\theta(r,t)$, the left-hand side of \eqref{eq:loadequation3a}, becomes
\begin{equation}
\langle\frac{\partial\vartheta_c}{\partial t}\rangle_T(r)=\langle\dot{\theta}\rangle_T(r)\quad \text{for $r < c(T)$}~.
\label{eq:jumplhs}
\end{equation}
For the given value of $r$ and the indicated time range is $L'(r,t)=\mathbb{L}'(r)$ and thus $\dot{L}'(r,t)=0$. The jump in the time derivative of the right-hand side of \eqref{eq:loadequation3a} is therefore
\begin{equation}
\frac{2}{\pi}\langle
\frac{\partial}{\partial t}[\{L'-h\}\orc\omega]\rangle_T(r)=-\frac{2}{\pi}\omega_0\langle\dot{h}\rangle_T
\quad\text{for $ r < c(T)$}~. \label{eq:jumprhs}
\end{equation}
and hence
\begin{equation}
\langle\dot{\theta}\rangle_T(r)=-\frac{2}{\pi}\omega_0\langle\dot{h}\rangle_T~.
\label{eq:thetajump}
\end{equation}
As the right-hand side of this equation is independent of $r$, the rate jumps in $\theta(r,t)$ are also independent of $r$ up to and including $c(T)$.

Differentiation of the identity $\theta(c(t),t)=0$ with respect to time gives
\begin{equation}
\dot{c}(t)\theta'(c(t),t)=-\dot{\theta}(c(t),t)
\label{eq:thetadiff}
\end{equation}
with $\theta'(c(t),t)=\lim_{r\uparrow c(t)}\theta'(r,t)$ because $\theta$ is only defined for $r\leq c(t)$. As $\theta(r,t)$ is a continuous function of time -- the jumps are only in the rates -- it follows that $\theta'(r,t)$ is also continuous in time and this includes the limit $r\uparrow c(t)$.
Then from \eqref{eq:thetadiff} a jump condition at $t=T$ can be constructed, reading
\begin{equation}
\langle\dot{c}\rangle_T \theta'(c(T),T)=
-\langle\dot{\theta}\rangle_T(c(T))~.
\label{eq:jumpa}
\end{equation}
Together \eqref{eq:thetajump} and \eqref{eq:jumpa} combine to
\begin{equation}
\omega_0\frac{\langle\dot{h}\rangle_T}{\langle\dot{c}\rangle_T}=\frac{\pi}{2}\theta'(c(T),T)
\label{eq:jumpb}
\end{equation}

Analogously, from the first equation of \eqref{eq:loadequation3e} with the aid of \eqref{eq:radiusequation}, \eqref{eq:ratejump1} and \eqref{eq:ratejump2} the following jump condition for the loading rate can be derived
\begin{equation}
\frac{\langle\dot{p}\rangle_T}{\langle\dot{h}\rangle_T}=4\omega_0c(T)
\label{eq:jumpc}
\end{equation}

\chapter{Effective indenter theory for viscoelastic materials}\label{app:Plastic}
The same assumptions as for the elastic case are used and the analysis applies to a situation where the contact radius at the jump  time $T$ exceeds all previous values, i.e., in this period the contact radius is assumed to be increasing. The pertinent load equation is now the first of \eqref{eq:advancingcontact2a}, i.e., $[p_\mathrm{e}\orc\varpi](T)=4\{c_\mathrm{e}(T)h_\mathrm{e}(T)-\mathbb{L}_\mathrm{e}(c_\mathrm{e}(T))\}$, and for the same reason $h_\mathrm{e}(T)=\mathbb{L}'_\mathrm{e}(c_\mathrm{e}(T))$. The requirement that the 'effective indenter' shape, $f_\mathrm{e}(r)$ is a power law can be weakened  to the requirement that $f_\mathrm{e}$ is a positively homogeneous function, say of degree $n$, of $r$. This leads to
\[
\mathbb{L}_\mathrm{e}(r)=r^{n+1}\mathbb{L}_\mathrm{e}(1),\enspace r\mathbb{L}_\mathrm{e}'(r)/(n+1)=\mathbb{L}_\mathrm{e}(r)\enspace
\text{and to}\enspace
\mathbb{L}_\mathrm{e}'(yc_\mathrm{e}(T))=y^nh_\mathrm{e}(T).
\]
These relations and $\varpi=\varphi/\omega_0$ show that
\begin{equation}
[p_\mathrm{e}\orc\varphi](T)=4\omega_0 c_\mathrm{e}(T)\frac{nh_\mathrm{e}(T)}{n+1}=
\frac{\langle\dot{p_\mathrm{e}}\rangle_T}{\langle\dot{h}_\mathrm{e}\rangle_T}\frac{nh_\mathrm{e}(T)}{n+1}\label{eq:effectiveVisco1}
\end{equation}
Substitution of the second of \eqref{eq:loadequation3} in \eqref{eq:theta2} with the result taken for  $t=T$ and $r=c(T)$, results in
\begin{equation}
u_\mathrm{e}(c_\mathrm{e}(T),T)=\frac{2}{\pi}\int_0^{c_\mathrm{e}(T)}
\frac{\mathbb{L}'(x)-h_\mathrm{e}(T)}{\sqrt{c_\mathrm{e}^2(T)-x^2}}\mathrm{d}x
=-k_nh_\mathrm{e}(T),\label{eq:effectiveVisco2}
\end{equation}
with
\begin{equation}
\quad\text{with}\quad k_n=\Bigl\{\frac{2}{\pi}\int_0^1
\frac{1-y^n}{\sqrt{1-y^2}}\mathrm{d}y\Bigr\}.
\end{equation}
Combination of \eqref{eq:effectiveVisco1} and \eqref{eq:effectiveVisco2} leads to
\begin{equation}
u_\mathrm{e}(c_\mathrm{e}(T),T)=-\varepsilon_n [p_\mathrm{e}\orc\varphi](T)\frac{\langle\dot{h_\mathrm{e}}\rangle_T}{\langle\dot{p}_\mathrm{e}\rangle_T}\quad\text{with}\quad
\varepsilon_n=k_n\{1+\frac{1}{n}\}.\label{eq:effectiveVisco3}
\end{equation}
The experimental value for the rate jump ratio implies $c_\mathrm{e}(T)=c(T)$ and with the experimental data for the load in \eqref{eq:effectiveVisco3}, this leads to the intermediate result that
\begin{equation}
u_\mathrm{e}(c(T),T)=-\varepsilon_n [p\orc\varphi](T)\frac{\langle\dot{h}\rangle_T}{\langle\dot{p}\rangle_T}.\label{eq:effectiveVisco4}
\end{equation}
Finally, with the assumption $u(c(T),T)=u_\mathrm{e}(c(T),T)$  and because in the actual experiment the contact depth always equals $h(t)+u(c(t),t)$, \eqref{eq:effectiveVisco4} results in
\begin{gather}
h_\mathrm{c}(T)=h(T)-\varepsilon_n p(T)\frac{\langle\dot{h}\rangle_T}{\langle\dot{p}\rangle_T}
\mathcal{R}(T)\label{eq:effectiveVisco5b}\\
\intertext{with the function $\mathcal{R}$ defined by} \mathcal{R}(T)=1+\int_0^T\frac{p(\tau)}{p(T)}\dot{\varphi}(T-\tau)\mathrm{d}\tau.
\label{eq:effectiveVisco5a}
\end{gather}

\chapter{Decomposing hereditary integrals}\label{sec:GenGAppendix}
This appendix is completely based on the theory of Golden and Graham \cite[][p. 63--67]{GoldenGraham1988} for the problem mentioned  in Chap.~\ref{sec:GenGSep1}.

\begin{figure}[htb]
  \centering
  \includegraphics[scale=1]{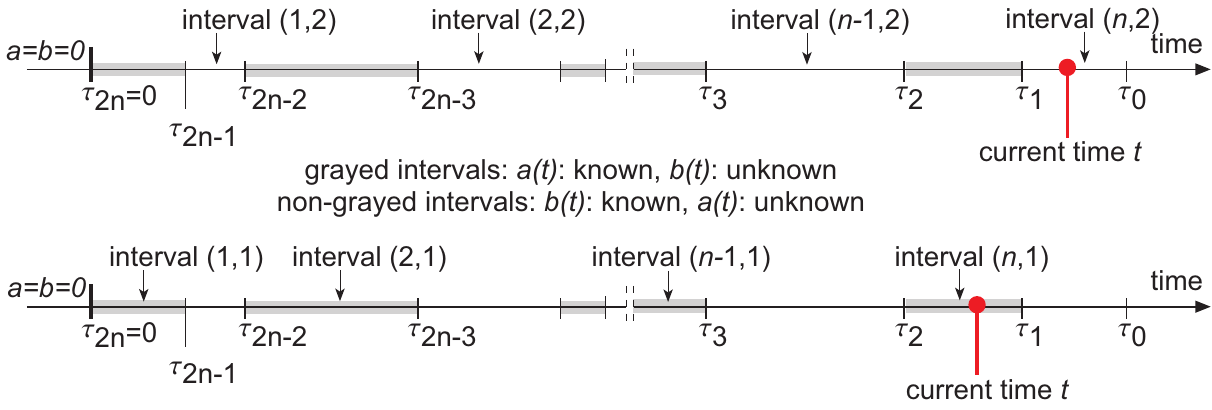}\\
  \caption{Time axes for two different cases of the current time $t$. }\label{fig:VarstGenG1}
\end{figure}

Two continuous and causal functions $a(t)$ and $b(t)$ are related by $a=b\orc\phi$ and $b=a\orc\varphi$ with -- obviously -- $\phi\orc\varphi=\Hea$. Peculiar is that the functions $a$ and $b$ are only known on sets of disjoint, but abutting, intervals; on any particular interval is either $a$ known and $b$ remains to be determined or it is the other way around.
This is shown in Fig.~\ref{fig:VarstGenG1} and it is assumed -- no loss of generality involved here -- that in the first interval $a$ is given and $b$ is to be determined here. The intervals where $a$ is known are the grayed intervals in Fig.~\ref{fig:VarstGenG1} and they are numbered starting at the first from the left and identified by the tuples (1,1), (2,1),(3.1),\dots etcetera. On the other intervals, the non-greyed ones in this picture,  is $b$ given and $a$ to be determined and these intervals are also numbered starting at the first from the left and these intervals are identified by (1,2), (2,2), \dots and so on.
The numbering of the interval boundaries times depends on the interval where the current time is located. Generally, the current time is in any of the union of two adjacent interval types: $t\in \{\mathrm{Interval(n,1)}\cup\mathrm{Interval(n,2)}\}$ and the value of $n$ determines the numbering of the boundary times. As shown in Fig.~\ref{fig:VarstGenG1}, it starts at the upper boundary of $\mathrm{Interval(n,2)}$, i.e., at $\tau_0$ and proceeds subsequently backwards in time as $\tau_1$, $\tau_2$, \dots until the origin  of the time axis, $\tau_{2n}=0$, is reached. So, the numbering depends on $n$ but the actual time values do not because $\tau_j(n)=\tau_{j+2}(n+1)$.

Depending on the interval type the current time is located in, one of the following  two equations is the starting point:
\begin{gather}
a(t)=b(t)+\int_0^tb(\tau)\dot{\phi}(t-\tau)\,\mathrm{d}\tau \quad\text{if current time $t\in\mathrm{Interval}(n,2)$,}\label{eq:GenGa}\\
\intertext{or}
b(t)=a(t)+\int_0^ta(\tau)\dot{\varphi}(t-\tau)\,\mathrm{d}\tau \quad\text{if current time $t\in \mathrm{Interval}(n,1)$.}\label{eq:GenGb}
\end{gather}

 \section{Current time between $\tau_1(n)$ and $\tau_0(n)$}\label{sec:HillPeriod}
Rewrite the starting equation \eqref{eq:GenGa} as
\begin{equation}
a(t)=b(t)+\int\limits_{\tau_1(n)}^t \dot{\phi}(t-s)b(s)\,\mathrm{d}s+\int\limits_0^{\tau_1(n)}\dot{\phi}(t-s)b(s)\,\mathrm{d}s\label{eq:GenGa1}
\end{equation}
and use \eqref{eq:GenGb} to obtain
\begin{multline}
a(t)=b(t)+\int\limits_{\tau_1(n)}^t \dot{\phi}(t-s)b(s)\,\mathrm{d}s\\
+\int\limits_0^{\tau_1(n)}\dot{\phi}(t-s)\Bigl\{a(s)+\int\limits_0^{s}\dot{\varphi}(s-u)a(u)\,\mathrm{d}u\Bigr\}\,\mathrm{d}s.
\end{multline}
Interchange the order of integration of the double integral and define
\begin{equation}
T_0(t,s)=\dot{\phi}(t-s)\label{eqPart1:Transform1}
\end{equation}
and
\begin{equation}
T_1(t,\tau,n)=T_0(t,s)+\int\limits_{s}^{\tau_1(n)}T_0(t,u)\dot{\varphi}(u-s)\,\mathrm{d}u,
\label{eqPart2:Transform1}
\end{equation}
to arrive at
\begin{equation}
a(t)=b(t)+\int\limits_{\tau_1(n)}^tT_0(t,s)b(s)\,\mathrm{d}s+\int\limits_0^{\tau_1(n)}T_1(t,s,n)a(s)\,\mathrm{d}s.\label{eq:ANIs1}
\end{equation}
If $n=1$ the analysis ends here because the $a$ and $b$ functions occurring in the right-hand side of \eqref{eq:ANIs1} are known for all time values occurring here.

Note that  \eqref{eq:ANIs1} is \eqref{eq:GenGa1} but merely written in a different way. So, comparison of the two expressions shows that
\begin{equation}
\int\limits_0^{\tau_1(n)}T_1(t,s,n)a(s)\,\mathrm{d}s=\int\limits_0^{\tau_1(n)}T_0(t,s,n)b(s)\,\mathrm{d}s,\enspace \tau_1(n)<t<\tau_0(n).\label{eq:GenGa2}
\end{equation}

For  $n\geq 2$ the analysis proceeds in the same vein and leads eventually to
\begin{multline}
a(t)=b(t)+\int\limits_{\tau_1(n)}^t T_0(t,s)b(s)\,\mathrm{d}s+
\sum\limits_{k=1}^n\int\limits_{\tau_{2k}(n)}^{\tau_{2k-1}(n)}T_{2k-1}(t,s,n)a(s)\,\mathrm{d}s\\
+\sum\limits_{k=1}^{n-1}\int\limits_{\tau_{2k+1}(n)}^{\tau_{2k}(n)}T_{2k}(t,s,n)b(s)\,\mathrm{d}s\qquad n\geq 2~.\label{eq:ANIsn}
\end{multline}
So, the functions $T_j$ start with $T_0(t,s,n)=\dot{\phi}(t-s)$ and are generated for odd subscripts according to
\begin{multline}
T_{2k+1}(t,s,n)=T_{2k}(t,s,n)\\
+\int\limits_s^{\tau_{2k+1}(n)}T_{2k}(t,u,n)\dot{\varphi}(u-s)\,\mathrm{d}u\qquad k=0,1,2,\ldots\label{eq:TUneven}
\end{multline}
For even subscripts
\begin{multline}
T_{2k}(t,s,n)=
T_{2k-1}(t,s,n)\\
+\int\limits_s^{\tau_{2k}(n)}T_{2k-1}(t,u,n)\dot{\phi}(u-s)\,\mathrm{d}u\qquad k=1,2,\ldots\label{eq:TEven}
\end{multline}
 \section{Current time between $\tau_2(n)$ and $\tau_1(n)$}\label{sec:VallyPeriod}
 Here, the function $a$ is known and $b$ is to determined and the starting point is now \eqref{eq:GenGb} written as
\begin{multline}
 b(t)=a(t)+\int\limits_{\tau_2(n)}^t \dot{\varphi}(t-s)a(s)\,\mathrm{d}s\\
 +\int\limits_0^{\tau_2(n)}\dot{\varphi}(t-s)\Bigl\{b(s)+\int\limits_0^s\dot{\phi}(s-u)b(u)\,\mathrm{d}u\Bigr\}\,\mathrm{d}s.
\end{multline}
If $n=1$ the second integral is not present and, with the definition $N_1(t,s)=\dot{\varphi}(t-\tau)$, one has
\begin{equation}
b(t)=a(t)+\int\limits_0^t N_1(t,s)a(s)\,\mathrm{d}s  \label{eq:BNIs1}
\end{equation}
With basically the same procedure as before one finds for $n\geq 2$:
\begin{multline}
b(t)=a(t)+\int\limits_{\tau_2(n)}^t N_1(t,s)a(s)\,\mathrm{d}s+\sum\limits_{k=1}^{n-1}
\int\limits_{\tau_{2k+1}(n)}^{\tau_{2k}(n)}N_{2k}(t,s,n)b(s)\,\mathrm{d}s\\
+\sum\limits_{k=1}^{n-1}
\int\limits_{\tau_{2k+2}(n)}^{\tau_{2k+1}(n)}N_{2k+1}(t,s,n)a(s)\,\mathrm{d}s \qquad n\geq 2~.\label{eq:BNIsn}
\end{multline}
The functions $N_j$ start with $N_1(t,s,n)=\varphi(t-s)$ and for even subscripts
\begin{multline}
N_{2k}(t,s,n)=
N_{2k-1}(t,s,n)\\
+\int\limits_s^{\tau_{2k}(n)}N_{2k-1}(t,u,n)\dot{\phi}(u-s)\,\mathrm{d}u\qquad k=1,2,\ldots,
\label{eq:NEven}
\end{multline}
and for odd subscripts
\begin{multline}
N_{2k+1}(t,s,n)=
N_{2k}(t,s,n)\\
+\int\limits_s^{\tau_{2k+1}(n)}N_{2k}(t,u,n)\dot{\varphi}(u-s)\,\mathrm{d}u\quad k=1,2,\ldots
\label{eq:NUneven}
\end{multline}
\section{The time derivatives of the functions $T_j$, $N_j$}\label{sec:KernelDerivatives}
Inspection of the definitions of $T_j(t,s,n)$ and $N_j(t,s,n)$ shows that the partial derivatives to the first time argument obey the same transformation rules as the functions itself.
With
\begin{equation}
\dot{T}_j(t,s,n)=\frac{\partial T_j(t,s,n)}{\partial t}\enspace\text{and}\enspace \dot{N}_j(t,s,n)=\frac{\partial N_j(t,s,n)}{\partial t},
\end{equation}
As, obviously, $\dot{T}_{0}(t,s,n)=\ddot{\phi}(t-s)$ it follows that for odd subscripts the derivatives are generated according to
\begin{multline}
\dot{T}_{2k+1}(t,s,n)=\dot{T}_{2k}(t,s,n)\\
+\int\limits_s^{\tau_{2k+1}(n)}\dot{T}_{2k}(t,u,n)\dot{\varphi}(u-s)\,\mathrm{d}u\qquad k=0,1,2,\ldots,\label{eq:TUneven1}
\end{multline}
and for even subscripts according to
\begin{multline}
\dot{T}_{2k}(t,s,n)=\dot{T}_{2k-1}(t,s,n)\\
+\int\limits_s^{\tau_{2k}(n)}\dot{T}_{2k-1}(t,u,n)\dot{\phi}(u-s)\,\mathrm{d}u\qquad k=1,2,\ldots
\label{eq:TEven1}
\end{multline}
In the same way, starting with $\dot{N}_1(t,s,n)=\ddot{\varphi}(t-s)$,
\begin{multline}
\dot{N}_{2k}(t,s,n)=
\dot{N}_{2k-1}(t,s,n)\\+\int\limits_s^{\tau_{2k}(n)}\dot{N}_{2k-1}(t,u,n)\dot{\phi}(u-s)\,\mathrm{d}u\qquad k=1,2,\ldots
\label{eq:NEven1}
\end{multline}
and
\begin{multline}
\dot{N}_{2k+1}(t,s,n)=
\dot{N}_{2k}(t,s,n)\\
+\int\limits_s^{\tau_{2k+1}(n)}\dot{N}_{2k+1}(t,u,n)\dot{\varphi}(u-s)\,\mathrm{d}u\quad k=1,2,\ldots
\label{eq:NUneven1}
\end{multline}

\chapter{The standard linear solid and Prony series}
\section{The standard linear solid (SLS)} \label{sec:SLEConfiguration}
\subsection{The material functions}\label{sec:SLE}
Historically, the term 'standard linear solid' denotes an assembly of two linear but different springs and a damper.
\begin{figure}[htb]
  \centering
  \includegraphics[scale=1]{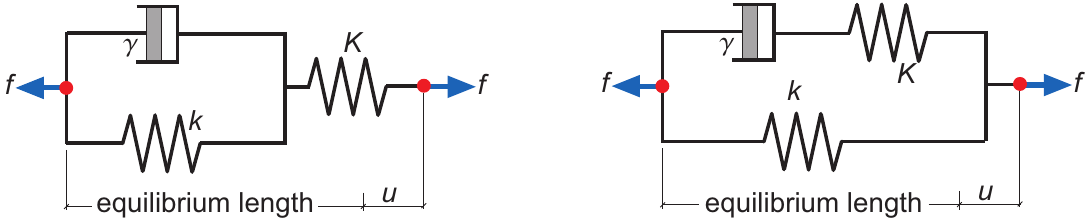}\\
  \caption{Standard linear solid or Poynting-Thomson body \cite[][p. 307]{Kuiken1994}. Two possible configurations, i.e., Kelvin (left) and Maxwell (right) reprsentation.  }\label{fig:SLE}
\end{figure}
The relation between the load $f$ and a step-shaped elongation $u$ is the relaxation function. The creep function is the elongation $u$ if a step in the load is applied. The two configuration shown in Fig. \ref{fig:SLE} are mathematically identical and the reduced relaxation function $\phi(t)$ and the reduced creep function $\varphi(t)$  are
\begin{gather}
\phi(t)=\left\{1+q(\mathrm{e}^{-\kappa t}-1)\right\}\hea{t},\enspace 0<q<1,\enspace \kappa>0,\label{eq:SleReduced}\\
\varphi(t)=\frac{1}{1-q}\left\{1-q\mathrm{e}^{-(1-q)\kappa t})\right\}\hea{t},\enspace 0<q<1,\enspace \kappa>0.\label{eq:SleReduced1}
\end{gather}
How the constants $q$ and $\kappa$ depend on the spring and damper constants is not relevant whatsoever. Note that $\phi(\infty)=1-q$, $\dot{\phi}(\infty)=0$, $\varphi(\infty)=1/(1-q)$ and $\dot{\varphi}(\infty)=0$.
\subsection{The $T_j$ and $N_j$ functions of the SLS}\label{sec:SLEkernels}
Consider the $T_j$ and $N_j$ functions for a standard linear solid\footnote{In a different form, these results, were derived earlier by Golden and Graham \cite{GoldenAndGraham1987}.}.
The sequence of times $\tau_j$ is ordered according to
\begin{equation}
\tau_0(R,n)>\tau_1(R,n)>\tau_2(R,n)>\ldots>\tau_{2n-1}(R,n)>\tau_{2n}(R,n)=0.\label{eq:TauSequence}
\end{equation}
First of all
\begin{equation}
T_0(t,s)=\dot{\phi}(t-s)=-q\kappa\exp\{-\kappa(t-s)\}
\end{equation}
and application of the defining formulae  \eqref{eq:TEven} and \eqref{eq:TUneven} shows that
\begin{equation}
T_1(t,s,R,n)=-q \kappa \exp\{\kappa\tau_1\}\exp\{-\kappa(t-(1-q)s)\},
\end{equation}
and, for $T_2$ and $T_3$, that
\begin{gather}
T_2(t,s,R,n)=T_0(t,s)\exp\Bigl(q\kappa(\tau_1-\tau_2)\Bigr),\\
T_3(t,s,R,n)=T_1(t,s,R,n)\exp\Bigl(-q\kappa(\tau_2-\tau_3)\Bigr).
\end{gather}
The $T_j$ functions with \emph{even} subscript are related by
\begin{equation}
T_{2k+2}(t,s,R,n)=T_{2k}(t,s,R,n)\exp\Bigl(q\kappa(\tau_{2k+1}-\tau_{2k+2})
\end{equation}
with $k=0,1,\ldots,n-2$. For those with  \emph{odd} subscripts the relations
\begin{equation}
T_{2k+1}(t,s,R,n)=T_{2k-1}(t,s,R,n)\exp\Bigl(-q\kappa(\tau_{2k}-\tau_{2k+1})\Bigr)
\end{equation}
apply for $ k=1,2,\ldots$.

For \emph{even} subscripts it is found that eventually
\begin{equation}
T_{2k}(t,s,R,n)=T_0(t,s)\exp\{q\kappa\gamma_{2k}(R,n)\}\enspace k=1, 2, 3,\ldots,n-1\label{eq:T2k}
\end{equation}
with $\gamma_{2k}$ only dependent on  the sequence $\tau_1,\ldots,\tau_{2k}$. Specifically,
\begin{equation}
\gamma_{m}(R,n)=-\sum\limits_{j=1}^{m}(-1)^j\tau_j(R,n)\enspace m=1,2,3,\ldots\label{eq:GammaM}
\end{equation}
Note that $\gamma_{2k}>0$ because
\begin{equation}
-\sum\limits_{j=1}^{2k}(-1)^j\tau_j=\tau_1-\tau_2+\tau_3-\tau_4 \cdots +\tau_{2k-1}-\tau_{2k}>0,
\end{equation}
on account of the properties of the $\tau_j$-sequence \eqref{eq:TauSequence}.
Similarly, for the \emph{odd} subscripts the following results apply
\begin{equation}
T_{2k+1}(t,s,R,n)=T_0(t,s)\exp\{q\kappa(\gamma_{2k+1}(R,n)-s)\}\enspace k=0,1, 2, 3,\ldots\label{eq:T2kplus1}
\end{equation}

The sequence of $N_j$ functions starts with
\begin{equation}
N_1(t,s)=\dot{\varphi}(t-s)=q\kappa\exp\{-(1-q)\kappa(t-s)\}.
\end{equation}
The next function in the $N_j$ sequence is
\begin{equation}
N_2(t,s,R,n)=q\kappa\exp\{-(1-q)\kappa t +\kappa s -q\kappa \tau_2(R,n)\}.
\end{equation}
The $N_j$ functions with \emph{odd} subscripts are related by
\begin{equation}
N_{2k+1}(t,s,R,n)=N_{2k-1}(t,s,R,n)\exp\left(q\kappa(\tau_{2k+1}(R,n)-\tau_{2k}(R,n))\right)
\end{equation}
for the $k$-values \enspace $1,2,3,\ldots$ whereas the $N_j$ functions with \emph{even} subscripts satisfy
\begin{equation}
N_{2k+2}(t,s,R,n)=N_{2k}(t,s,R,n)\exp\left(-q\kappa(\tau_{2k+2}(R,n)-\tau_{2k+1}(R,n))\right),
\end{equation}
and this expression is valid for $ k=1,2,3,\ldots$.
For $k=0,1,2,\ldots$ , the $N_j$ functions with  \emph{odd} subscript are
\begin{equation}
N_{2k+1}(t,s,R,n)=N_1(t,s)\exp\left(q\kappa(\gamma_{2k+1}(R,n)-\tau_1(R,n))\right),\label{eq:N2kplus1}
\end{equation}
and those with \emph{even} subscript
\begin{equation}
N_{2k}(t,s,R,n)=N_1(t,s)\exp\{q\kappa(\gamma_{2k}(R,n)-\tau_1(R,n)+s)\}.\label{eq:N2k}
\end{equation}
\section{Prony series for relaxation and creep}\label{sec:Prony}
\subsection{Basic definition and relation between the parameters}
Often, viscoelastic media are described by discrete spectrum models with the relaxation and creep function represented by \emph{Prony series}. For the reduced relaxation and creep function, this means that they are written as
\begin{gather}
\phi(t)=\frac{\omega(t)}{\omega_0}=1+\sum\limits_{i=1}^M q_i\{\exp(-\kappa_i t)-1\},
\label{eq:Prony10}\\
\varphi(t)=\frac{\varpi(t)}{\varpi_0}=1-\sum\limits_{i=1}^M w_i\{\exp(-\gamma_i t)-1\}.
 \label{eq:Prony11}
\end{gather}
All parameters must be positive and also $0<\phi(\infty)=1-\sum q_i<1$. As $\phi(t)$ and $\varphi(t)$ must each others Stieltjes inverse, the  parameter set $\{w_i,\gamma_i\}$ is a function of the set $\{q_j,\kappa_j\}$ and vice versa. \citet{GoldenGrahamLan1994} derived formulae for the relations between the constants in two Prony series with the property that they are each others Stieltjes inverses. Adapting their results shows that the two parameter sets $\{w_i,\gamma_i\}$ and $\{q_j,\kappa_j\}$ are related by the $2M$ equations
\begin{gather}
\sum_{i=1}^M\frac{\kappa_iq_i}{\kappa_i-\gamma_k}=1,\quad
\gamma_k w_k\sum\limits_{i=1}^M \frac{\kappa_iq_i}{(\kappa_i-\gamma_k)^2}=1,\quad k=1,2,\ldots,M,\label{eq:Prony12a}
\intertext{or, alternatively, by } \sum_{i=1}^M\frac{\gamma_iw_i}{\kappa_k-\gamma_i}=1,\enspace\kappa_k q_k\sum\limits_{i=1}^M\frac{\gamma_i w_i}{(\kappa_k-\gamma_i)^2}=1,\enspace k=1,2,\ldots,M .\label{eq:Prony12b}
\end{gather}
\subsection{The $T_j$ and $N_j$ functions for the Prony series}\label{sec:Pronykernels}
The kernel functions $T_j(t,s,R,n)$ or $N_j(t,s,R,n)$ start at
\begin{gather}
T_0(t,s)=\dot{\phi}(t-s)=-\sum\limits_{i=1}^Ma_i\exp\{-\kappa_i(t-s)\},\enspace \tau_1\leq t\leq \tau_0,\enspace s<t,\label{eq:PronyTStart}\\
N_1(t,s)=\dot{\varphi}(t-s)=\sum\limits_{i=1}^Mb_i\exp\{-\gamma_i(t-s)\},\enspace \tau_2\leq t\leq \tau_1,\enspace s<t.\label{eq:PronyNStart}
\end{gather}
To obtain more concise formulae, the short-hand notations $a_i=q_i\kappa_i$ and $b_i=w_i\gamma_i$ were introduced.

The transformation formula \eqref{eqPart2:Transform1} to generate $T_1$ from $T_0$ and $\dot{\varphi}$ is now
\begin{multline}
T_1(t,s,R,n)=-\sum\limits_{i}a_{i}\exp\{-\kappa_{i}(t-s)\}\\
-\sum\limits_{i}\sum\limits_{j}a_{i}b_{j}
\int\limits_{u=s}^{\tau_1}\exp\{-\kappa_{i}(t-u)-\gamma_{j}(u-s)\}\mathrm{d}u.
\end{multline}
Evaluate  the integral with use of the first relation in \eqref{eq:Prony12b}, to obtain
\begin{equation}
T_1(t,s,R,n)=\sum\limits_j b_j\left(\sum\limits_i \frac{-a_i\exp\{-\kappa_i(t-\tau_1)\}}{(\kappa_i-\gamma_j)}\right)
\exp\{-\gamma_j(\tau_1-s)\}.
\end{equation}
The definition
$Q_j^{(1)}(t-\tau_1)=-b_j\sum_i a_i\exp\{-\kappa_i(t-\tau_1)\}/(\kappa_i-\gamma_j)$ reveals that $T_1$ is again a Prony series in the variable $\tau_1-s$. Generation of the other $T_j$'s  -- with at each step use of the first relation in either \eqref{eq:Prony12a} or \eqref{eq:Prony12b} to simplify the results -- results in
\begin{multline}
T_k(t,s,R,n)=
-\sum_i\biggl(Q_i^{(k)}(t-\tau_1,\tau_1-\tau_2\ldots,\tau_{k-1}-\tau_k)\\
\times\exp\{-\kappa_i(\tau_k-s)\}\biggr),
\quad \text{$k$: even and $k\geq 2$,}
\end{multline}
and
\begin{multline}
T_k(t,s,R,n)=
+\sum_i\biggl(Q_i^{(k)}(t-\tau_1,\tau_1-\tau_2\ldots,\tau_{k-1}-\tau_k)\\
\times\exp\{-\gamma_i(\tau_k-s)\}\biggr)\quad\text{$k$: odd and $k\geq 1$.}
\end{multline}
The functions $Q_j^{(k)}$ are defined recursively starting at
\begin{equation}
Q_j^{(1)}(t-\tau_1)=-b_{j}\left(\sum_i\frac{a_{i}}{(\kappa_i-\gamma_{j})}\exp\{-\kappa_{i}(t-\tau_1)\}\right),
 \end{equation}
and proceeding with
\begin{multline}
Q_j^{(2k)}(t-\tau_1,\tau_1-\tau_2,\ldots,\tau_{2k-1}-\tau_{2k})=\\
a_{j}\left(\sum_i\frac{Q_i^{(2k-1)}}{(\gamma_{i}-\kappa_j)}\exp\{-\gamma_{i}(\tau_{2k-1}-\tau_{2k})\}\right)\enspace k=1,2,3,\ldots,
\end{multline}
for even superscripts and
\begin{multline}
Q_j^{(2k+1)}(t-\tau_1,\tau_1-\tau_2,\ldots,\tau_{2k}-\tau_{2k+1})=\\
-b_{j}\left(\sum_i\frac{Q_i^{(2k)}}{(\kappa_i-\gamma_j)}\exp\{-\kappa_i(\tau_{2k}-\tau_{2k+1})\}\right)
\enspace k=1,2,3,\ldots \end{multline}
for odd ones.

The functions $N_j$ follow similar rules starting from $N_1$ as given in \eqref{eq:PronyNStart}. Specifically,
\begin{multline}
N_k(t,s,R,n)=
              -\sum_i\biggl(W_i^{(k)}(t-\tau_2,\tau_2-\tau_3,\ldots,\tau_{k-1}-\tau_k)\\
              \times \exp\{-\kappa_i(\tau_k-s)\}\biggr),\quad \text{$k$: even and $k\geq 2$,}
\end{multline}
and
\begin{multline}
        N_k(t,s,R,n)=
              +\sum_i\sum_i\biggl(W_i^{(k)}(t-\tau_2,\tau_2-\tau_3,\ldots,\tau_{k-1}-\tau_k)\\
              \times \exp\{-\gamma_i(\tau_k-s)\}\biggr),\quad
               \text{$k$: odd and $k\geq 3$ .}
\end{multline}
The functions $W_i^{(k)}$ are also defined recursively. They start at
\begin{equation}
W_i^{(2)}(t-\tau_2)=a_i\sum_j\frac{b_j}{\gamma_j-\kappa_i}\exp\{-\gamma_j(t-\tau_2)\} \textcolor[rgb]{1.00,0.00,0.00}{,}
\end{equation}
and are generated according to
\begin{multline}
W_i^{(2k+1}(t-\tau_2,\tau_2-\tau_3,\ldots,\tau_{2k}-\tau_{2k+1})=\\
-b_i\left(\sum_j\frac{W_j^{(2k)}}{\kappa_j-\gamma_i}\exp\{-\kappa_j(\tau_{2k}-\tau_{2k+1})\}\right)\enspace k=1,2,\ldots
\end{multline}
and
\begin{multline}
W_i^{(2k)}(t-\tau_2,\tau_2-\tau_3,\ldots,\tau_{2k-1}-\tau_{2k})=\\
a_i\left(\sum_j\frac{W_j^{(2k-1)}}{\gamma_j-\kappa_i}\exp\{-\gamma_j(\tau_{2k-1}-\tau_{2k})\}\right)\enspace k=2,3,\ldots
\end{multline}

\chapter{Crossing local extrema of the contact radius}\label{app:CrossingExtrema}
One of the problems in the Golden and Graham method \cite{GoldenGraham1988} is that the number of 'hills' and 'valleys' including the associated numbering of the intersection times $\tau_j(R,n)$
and the mathematical form of $T_j$ and $N_j$ might depend strongly on the actual value of the considered radius $R$.
\begin{figure}[htb]
  \centering
  \includegraphics[scale=1]{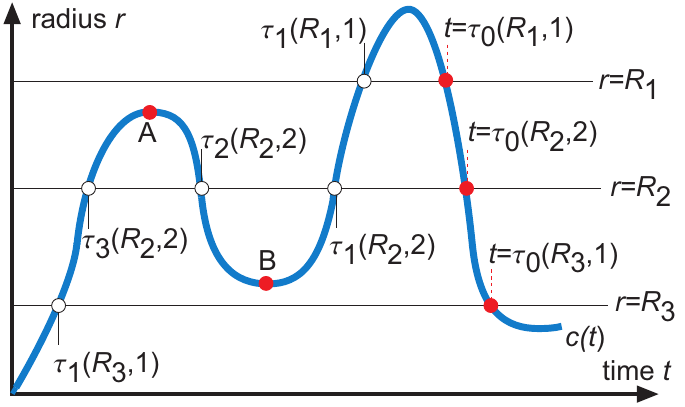}\\
  \caption{Contact radius, lines of constant radius and intersection times.}\label{fig:CrossingExtrema}
\end{figure}
This is graphically demonstrated in Fig.~\ref{fig:CrossingExtrema} where along the line $r=R_1$ only \emph{one} 'valley' and \emph{one} 'hill' is encountered in the time range $0\leq t\leq\tau_0(R_1,1)$ and the situation is similar along the line $r=R_3$ for $0\leq t\leq\tau_0(R_3,1)$ although the 'hill' is in the latter case not a simple one. One the other hand, the line $r=R_2$ -- a line below the local maximum A but above the local minimum B -- cuts through \emph{two} 'valleys'  and \emph{two} 'hills' on the time interval $[0,\tau_0(R_2,2)]$.
Consequently, if the line $r=R=$ moves from zero to a large value 'valleys' and 'hills' might originate, disappear or coalesce  in quite a complicated manner if local extrema are crossed in the process.

The functions $\mathcal{U}_k$, $\mathcal{V}_k$, $\mathcal{H}_k$, $\mathcal{W}_k$, $\mathcal{P}_k$ and $\mathcal{Q}_k$, introduced in Chap.~\ref{sec:GenGSep} are integral operators with the $T_j$ or the $N_j$ functions as kernel and the mathematical form of these functions also depends\footnote{To simplify the notation this dependence is in this section only implied.} on the value of the radius $R$. For even indices these functions are defined below a 'hill' and for odd indices they are defined above a 'valley'. Consequently, if, for example,  a 'valley' disappears and the adjacent 'hills' coalesce one would expect that the associated $T_j$ and $N_j$ functions also disappear or coalesce.

\begin{figure}
  \centering
  \includegraphics[scale=1]{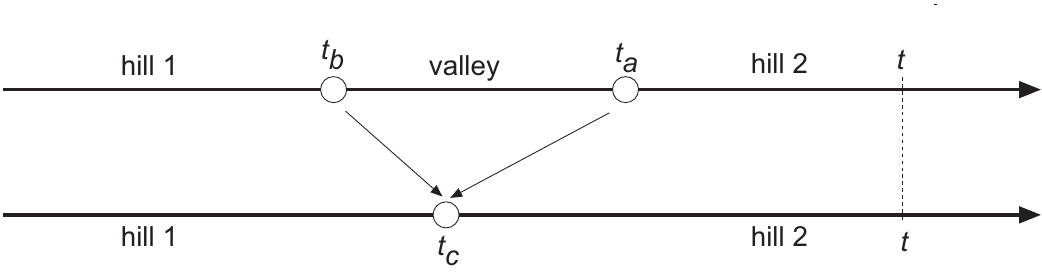}\\
  \caption{Disappearance of the 'valley' and coalescence of the 'hills' due to the limits $t_a\downarrow t_c$ and $t_b\uparrow t_c$}\label{fig:HillCoalescence}
\end{figure}

To see whether this expectation is correct consider the following problem. The starting point is shown in the top of Fig.~\ref{fig:HillCoalescence}. Let $A(t,s)$ be any function with $t\in \mathrm{hill~2}$ and $s<t$. In the valley is $B(t,s)$ defined by
\begin{gather}
B(t,s)=A(t,s)+\int\limits_{u=s}^{t_a}A(t,u)\dot{\varphi}(u-s)\mathrm{d}u.\label{eq:BfromA}
\intertext{In hill 1 is $C(t,s)$ constructed according to}
C(t,s)=B(t,s)+\int\limits_{u=s}^{t_b}B(t,u)\dot{\phi}(u-s)\mathrm{d}u.\label{eq:CfromB}
\end{gather}
The function $C$ can be linked directly to $A$ by substituting \eqref{eq:BfromA} in \eqref{eq:CfromB} and the result is
\begin{multline}
C(t,s)=A(t,s)+\int\limits_{x=s}^{t_a}A(t,x)\dot{\varphi}(x-s)\mathrm{d}s\\
+\int\limits_{x=s}^{t_b}\dot{\phi}(x-s)\Bigl\{A(t,x)+\int\limits_{y=x}^{t_a}A(t,y)\dot{\varphi}(y-x)\mathrm{d}y\Bigr\}\mathrm{d}x.
\end{multline}
Take the limits $t_b\uparrow t_c$ and $t_a\downarrow t_c$, which shows at first
\begin{multline}
\bar{C}(t,s)=A(t,s)+\int\limits_{x=s}^{t_c}A(t,x)\Bigl\{\dot{\varphi}(x-s)+\dot{\phi}(x-s)\Bigr\}\mathrm{d}s\\
+\int\limits_{x=s}^{t_c}\Bigl\{\int\limits_{y=x}^{t_c}A(t,y)\dot{\varphi}(y-x)\dot{\phi}(x-s)\mathrm{d}y\Bigr\}\mathrm{d}x.
\end{multline}
The diacritic '~$\bar{}$~' was added above $C$ to express that the limits were taken. After changing the order of integration in the double integral the final result is
\begin{multline}
\bar{C}(t,s)=A(t,s)+\int\limits_{x=s}^{t_c}A(t,x)\Bigl\{\dot{\varphi}(x-s)+\dot{\phi}(x-s)\\
+\int\limits_{y=0}^{x-s}\dot{\varphi}(x-s-y)\dot{\phi}(y)\mathrm{d}y\Bigr\}\mathrm{d}x.
\end{multline}
As the expression between brackets vanishes because $\phi$ and $\varphi$ are each others Riemann-Stieltjes inverse with initial value $\phi_0=\varphi_0=1$ (see \eqref{eq:DotFStarG}) it follows that
\begin{equation}
\bar{C}(t,s)=A(t,s),
\end{equation}
that is,  $\bar{C}$ is the continuation of $A$ to hill 1 after it coalesced with hill 2. The situation where two 'valleys' coalese is basically the same, except that the functions $\phi$ and $\varphi$ must be interchanged in the three equations \eqref{eq:BfromA} to \eqref{eq:CfromB} and that in Fig.~\ref{fig:HillCoalescence} a 'hill' becomes a  'valley' and vice versa.

The functions $A$ and $C$ can represent $T_j$ or $N_j$ with even index whereas $B$ can represent these function for an odd index. Specifically, if $A$ represents $T_{2k}$ for $k=0,1,2\ldots$, then $B$ and $C$ represent $T_{2k+1}$ and $T_{2k+2}$, respectively. For the  $N_j$ kernels the situation is somewhat different as $N_0$ does not exist; the $N_j$ sequence starts with $N_1$. These kernels must be treated differently because the functions $A$, $B$ and $C$ can only represent  $N_{2k}$, $N_{2k+1}$ and $N_{2k+2}$ kernels, respectively, starting at the value $k=1$

For the $N_1$ and $N_2$ kernels the current time $t$ is not in 'hill 2' as shown in Fig.~\ref{fig:HillCoalescence}, but in the preceding 'valley' $t_b\leq t\leq t_a$ and without loss of generality the choice $t=t_c$ is made. The $N_j$ sequence  starts with $N_1=\dot{\varphi}(t_c-s)$. Now
\begin{equation}
\begin{split}
N_2(t_c,s)&=\dot{\varphi}(t_c-s)+\int\limits_{u=s}^{t_b}\dot{\varphi}(t_c-u)\dot{\phi}(u-s)\mathrm{d}u\\
&=\dot{\varphi}(t_c-s)+\int\limits_{0}^{t_b-s}\dot{\varphi}(t_c-s-u)\dot{\phi}(u)\mathrm{d}u,
\end{split}
\end{equation}
and the limit $t_a\downarrow t_c$ plus $t_b\uparrow t_c$ shows -- again because $\phi$ and $\varphi$ are each others Riemann-Stieltjes inverse with initial value $\phi_0=\varphi_0=1$ (see \eqref{eq:DotFStarG}) --  that
\begin{equation}
\begin{split}
\bar{N}_2(t_c,s)&=\dot{\varphi}(t_c-s)+\int\limits_{0}^{t_c-s}\dot{\varphi}(t_c-s-u)\dot{\phi}(u)\mathrm{d}s\\
&=-\dot{\phi}(t_c-s)=-T_0(t_c,s).
\end{split}
\end{equation}
Apart from its sign is $\bar{N}_2$ equal to $T_0$ and this is as it should be, because the integral operators in which the $N_j$ kernels are present always occur in the formulae for the 'valley' intervals with a minus sign, whereas in the corresponding formulae for the 'hill' intervals the operators with $T_j$ kernels have a plus sign, cf.  \eqref{eq:GenGRadius4}, \eqref{eq:GenGRadius4a}, \eqref{eq:GenGLoad3}, \eqref{eq:GenGLoad4} with \eqref{eq:GenGRadius6}, \eqref{eq:GenGRadius7}, \eqref{eq:GenGLoad6}, \eqref{eq:GenGLoad6a}.
Special attention must also be paid to the behaviour of $\partial N_2(t,s)/\partial t$ when the next adjacent 'valley' disappears.
In \ref{sec:KernelDerivatives}  it was shown that the derivatives of the kernels transform in the same way as the kernels itself and the starting kernel for the derivatives
$\partial N_j(t,s)/\partial t$ is $\ddot{\varphi}(t-s)$, i.e., the derivative of $N_1$. Then, with $t=t_c$ in the 'valley', the derivative of $N_2$ is
\begin{equation}
\frac{\partial N_2(t_c,s)}{\partial t}=\ddot{\varphi}(t_c-s)+\int\limits_{u=s}^{t_a}\ddot{\varphi}(t_c-u)\dot{\phi}(u-s)\mathrm{d}u.
\end{equation}
The limits $t_b\uparrow t_c$ and $t_a\downarrow t_c$ indicate that
\begin{equation}\label{eq:DotN2}
\begin{split}
\lim_{t_b\uparrow t_c,\,\,t_a\downarrow t_c}\frac{\partial N_2(t_c,s)}{\partial t}&=\ddot{\varphi}(t_c-s)+\int\limits_{u}^{t_c-s}\ddot{\varphi}(t_c-s-u)\dot{\phi}(u)\mathrm{d}u\\
&=-\ddot{\phi}(t_c-s)+\dot{\phi}_0\dot{\phi}(t_c-s).
\end{split}
\end{equation}
The validity of the last equality sign is guarantied by \eqref{eq:DDotFStarG} as $\phi$ and $\varphi$ are each others Riemann-Stieltjes inverse with initial value $\phi_0=\varphi_0=1$, and $\dot{\phi}_0=-\dot{\varphi}_0$.

\chapter{Extremal contact in the stationary phase}\label{sec:Extrema}
\section{Maxima of contact radius and depth coincide }\label{sec:MaxOfDepthAndRadius}
\begin{figure}[htb]
  \centering
  \includegraphics[scale=1]{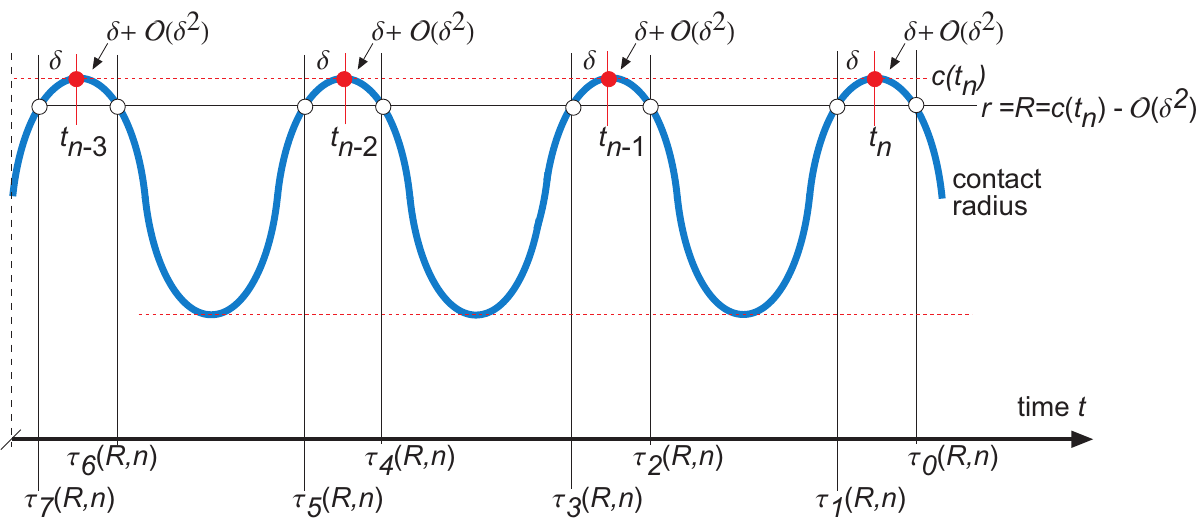}\\
  \caption{Periodic contact radius during the stationary phase. Line $r=R$ near maxima.}\label{fig:PeriodicMaximum}
\end{figure}

Fig.~\ref{fig:PeriodicMaximum} depicts a typical part of a smooth and periodic $c(t)$-curve during the stationary phase and the period in this phase phase is $\Lambda$.
Consider a value for $R$ just below the maximum of $c(t)$ at $t_n$. Because of the periodicity,  $c(t)$ has the same maximum value at $t_{n-j}=t_n-j\Lambda$ for $j=0,1,\ldots,$. In the limit $R\uparrow c(t_n)$ the equations
\eqref{eq:GenGRadius7A} and \eqref{eq:GenGRadius6A} are equal.. Moreover, the summations cancel because, the functions $\mathcal{U}_j$ and $\mathcal{H}_j$ are integrals between $\tau_{2j+1}(R,n)$ and $\tau_{2j}(R,n)$ and the lower and upper integration boundaries become equal in the limit $R\uparrow c(t_n)$ thus rendering all integrals zero and, therefore,
\begin{equation}
h(t_n)=\mathbb{L}'(c(t_n)\enspace \Rightarrow c(t_n)=\mathbb{C}(h(t_n)).\label{eq:Htn}
\end{equation}

Next consider the time derivative that can be constructed from the two equations \eqref{eq:GenGRadius7A} and \eqref{eq:GenGRadius6A}. For fixed $R$, but $\tau_0$ and $\tau_1$ near $t_n$,  and $j=1,2,\cdots$ the following expansions apply:
\begin{gather}
h(\tau_0)-h(\tau_1)=\dot{h}(t_n)(\tau_0-\tau_1)+\mathcal{O}((\tau_0-\tau_1)^2),\label{eq:Expansionh}\\
\mathcal{U}_j(\tau_0,R,n)-\mathcal{U}_j(\tau_1,R,n)=(\tau_0-\tau_1)\dot{\mathcal{U}}_j(t_n,R,n)+\mathcal{O}((\tau_0-\tau_1)^2),\label{eq:ExpansionU}\\
\mathcal{H}_j(\tau_0,R,n)-\mathcal{H}_j(\tau_1,R,n)=(\tau_0-\tau_1)\dot{\mathcal{H}}_j(t_n,R,n)+\mathcal{O}((\tau_0-\tau_1)^2),\label{eq:ExpansionH}
\end{gather}
with
\begin{gather}
\dot{\mathcal{U}}_j(t,R,n)=\int\limits_{\tau_{2j+1}}^{\tau_{2j}}\frac{\partial T_{2j}(t,s,R,n)}{\partial t}\mathrm{d}s, \\ \dot{\mathcal{H}}_j(t,R,n)=\int\limits_{\tau_{2j+1}}^{\tau_{2j}}\frac{\partial T_{2j}(t,s,R,n)}{\partial t}h(s)\mathrm{d}s.
\end{gather}
The terms $\mathcal{U}_0(\tau_0,R,n)$ and $\mathcal{H}_0(\tau_0,R,n)$ in \eqref{eq:GenGRadius7A} are rewritten using the mean value theorem for integrals, i.e., there exist $\xi_1\in(\tau_1,\tau_0)$ and $\xi_2\in(\tau_1,\tau_0)$ such that
\begin{gather}
\mathcal{U}_0(\tau_0,R,n)=(\tau_0-\tau_1)T_0(\tau_0,\xi_1,R,n)=(\tau_0-\tau_1)\dot{\phi}(\tau_0-\xi_1)\label{eq:ExpansionU0}\\
\mathcal{H}_0(\tau_0,R,n)=(\tau_0-\tau_1)T_0(\tau_0,\xi_2,R,n)h(\xi_2)=(\tau_0-\tau_1)\dot{\phi}(\tau_0-\xi_1)h(\xi_2)\label{eq:ExpansionH0}
\end{gather}
The difference of \eqref{eq:GenGRadius7A} and \eqref{eq:GenGRadius6A}, with \eqref{eq:Expansionh} to \eqref{eq:ExpansionH0}  used in it and divided by $\tau_0-\tau_1$ yields
\begin{multline}
\mathbb{L}'(R)\left(\dot{\phi}(\tau_0-\xi_1)+\sum\limits_{j=1}^\infty\dot{\mathcal{U}}_j(t_n,R,n)\right)=\dot{h}(t_n)\\
+\dot{\phi}(\tau_0-\xi_1)h(\xi_2)+\sum\limits_{j=1}^\infty\dot{\mathcal{H}}_j(t_n,R,n)+\mathcal{O}((\tau_0-\tau_1)).\label{eq:Difference}
\end{multline}
As $R=c(t_n)-\mathcal{O}(\delta^2)$ and $\mathcal{O}((\tau_0-\tau_1))=\mathcal{O}(\delta)$,  the limit $R\uparrow c(t_n)$, i.e., $\delta\downarrow 0$, of this equation results in
\begin{equation}
\{\mathbb{L}'(c(t_n))-h(t_n)\}\dot{\phi}_0=\dot{h}(t_n)\enspace\Rightarrow\enspace \dot{h}(t_n)=0\label{eq:Vtn}
\end{equation}
because  $\tau_0\rightarrow t_n$, $\tau_1\rightarrow t_n$, $\xi_1\rightarrow t_n$, $\xi_2\rightarrow t_n$, $\mathbb{L}'(c(t_n))\rightarrow h(t_n)$ and the summations in \eqref{eq:Difference} cancel since the integration boundaries of these terms become equal. So, the final conclusion is that during the stationary phase the maxima of depth and contact radius occur at the same time and are related by $h(t_n)=\mathbb{L}'(c(t_n))$, i.e., $c(t_n)=\mathbb{C}(h(t_n))$.
\section{Minima of contact radius and load coincide}\label{sec:MinOfLoadAndRadius}
\begin{figure}[htb]
  \centering
  \includegraphics[scale=1]{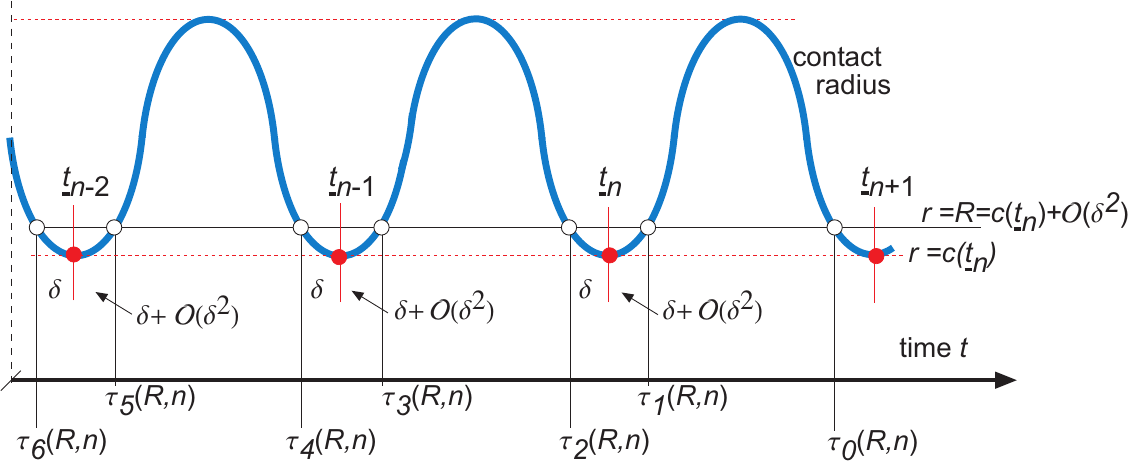}\\
  \caption{Periodic contact radius during the stationary phase. Line $r=R$ near minima.}\label{fig:Periodic minimum}
\end{figure}

The simplified equations \eqref{eq:GenGLoad3A} and \eqref{eq:GenGLoad4aa} at the end,
$\tau_1(R,n)$, and start, $\tau_2(R,n)$, of 'valley' $n$ are
\begin{gather}
\ell(\tau_1)+\sum\limits_{k=0}^\infty\mathcal{Q}_k(\tau_1,R,n)=4\mathbb{G}(R)\left(1-\sum\limits_{k=1}^\infty\mathcal{V}_k(\tau_1,R,n)\right)
\label{eq:GenGLoad3aa}\\
\intertext{and}
\ell(\tau_2)+\sum\limits_{k=1}^\infty\mathcal{Q}_k(\tau_2,R,n)=4\mathbb{G}(R)\left(1-\sum\limits_{k=1}^\infty\mathcal{V}_k(\tau_2,R,n)\right),
\label{eq:GenGLoad4a}
\end{gather}
respectively.
Consider \eqref{eq:GenGLoad3aa} and \eqref{eq:GenGLoad4a} along a line $r=R$ just above  the minima of $c(t)$ (Fig.~\ref{fig:Periodic minimum}). In the limit of $R$ to the minimum values of $c(t)$ at $\tmin_n$, $\tmin_{n-1}, \ldots$ the intersection times $\tau_1$ and $\tau_2$ both approach $\tmin_n$ and the equations become identical. The sums on the left-hand sides of \eqref{eq:GenGLoad3aa} and \eqref{eq:GenGLoad4a}  cancel if  $R\downarrow c(\tmin_n)$,  because the upper and lower integration limits of all $\mathcal{Q}_k$ integrals become equal.
The previous 'valleys' $n-1, n-2,\ldots $ disappear and the 'hills' $n-1,n-2,\ldots$ coalesce when the limit is taken. Heuristically one would expect that this means that all 'hills' now must be counted as a single large 'hill' and that the kernels of the integral operators $\mathcal{V}_j$  become equal to $N_2$. In Appendix \ref{app:CrossingExtrema} it is shown that this expectation is correct and also shown there is that in this limit $N_2\rightarrow -T0=-\dot{\phi}$.  Since $\tau_1(R,n)\rightarrow \tmin_n$ and as in the stationary phase all minima of the contact radius are equal, it is found that
\begin{gather}
\lim_{R\downarrow c(\tmin_n)}\mathcal{V}_k(\tau_1,R,n)=
-\int\limits_{\tmin_{n-k}}^{\tmin_{n+1-k}}\dot{\phi}(\tmin_{n}-s)\mathrm{d}s=\phi((k-1)\Lambda)-\phi(k\Lambda)\\
\intertext{resulting in}
 \lim_{R\downarrow c(\tmin_n)}\left(1-\sum\limits_{k=1}^\infty\mathcal{V}_k(\tau_1,R,n)\right)=
 \phi(\infty).\label{eq:DepthAtMin0}
\end{gather}
The previous results lead to the conclusion that in  the stationary phase the minima of the contact radius are related to the current load values by
\begin{equation}\label{eq:LoadAtMin}
 \ell(\tmin_n)=4\mathbb{G}(c(\tmin_n))\phi(\infty).
\end{equation}
For the rate of change of the load at $\tmin_n$ the difference of \eqref{eq:GenGLoad3aa} and \eqref{eq:GenGLoad4a} is taken and the rest of the procedure is similar to that followed in \ref{sec:MaxOfDepthAndRadius}. Expansion of this difference in powers of $\tau_1-\tau_2$ at constant $R$ leads for $k=1,2,\ldots$  to
\begin{gather}
  \ell(\tau_1)-\ell(\tau_2)=\dot{\ell}(\tmin_n)(\tau_1-\tau_2)+\mathcal{O}((\tau_1-\tau_2)^2),\label{eq:ExpansionEll}\\
\mathcal{Q}_k(\tau_1,R,n)-\mathcal{Q}_k(\tau_2,R,n)=(\tau_1-\tau_2)\dot{\mathcal{Q}}_k(\tmin_n,R,n)+\mathcal{O}((\tau_1-\tau_2)^2),\label{eq:ExpansionQk}\\
\mathcal{V}_k(\tau_1,R,n)-\mathcal{V}_k(\tau_2,R,n)=(\tau_1-\tau_2)\dot{\mathcal{V}}_k(\tmin_n,R,n)+\mathcal{O}((\tau_1-\tau_2)^2).\label{eq:ExpansionVk}
\end{gather}
For $k=0$ the mean value theorem for integrals  ensures that a scalar $\zeta\in(\tau_2,\tau_1)$ exists such that
\begin{equation}
\mathcal{Q}_0(\tau_1,R,n)=(\tau_1-\tau_2)\ell(\zeta)N_1(\tau_1,\zeta,R,n)=(\tau_1-\tau_2)\ell(\zeta)\dot{\varphi}(\tau_1-\zeta).\label{eq:ExpansionQ0}
\end{equation}
The derivatives of the integrals $\mathcal{V}_k$ and $\mathcal{Q}_k$ are given by
\begin{gather}
\dot{\mathcal{V}}_k(t,R,n)=\int\limits_{\tau_{2k+1}}^{\tau_{2k}}\frac{\partial N_{2k}(t,s,R,n)}{\partial t}\mathrm{d}s,\\
\dot{\mathcal{Q}}_k(t,R,n)=\int\limits_{\tau_{2k+2}}^{\tau_{2k+1}}\frac{\partial N_{2k+1}(t,s,R,n)}{\partial t}\ell(s)\mathrm{d}s.
\end{gather}
The expanded difference of \eqref{eq:GenGLoad3aa} and \eqref{eq:GenGLoad4a} divided  by $\tau_1-\tau_2$ and in the limit $R\downarrow c(\tmin_n)$ shows, firstly, that for $k\geq 1$ all terms with $\dot{\mathcal{Q}}_k$ cancel because the integration limits become equal and, secondly, that the contribution of $\mathcal{Q}_0$ reduces to  $\ell(\tmin_n)\dot{\varphi}_0$ (see \eqref{eq:ExpansionQ0}).  Thirdly, the integrands $\partial{N}_{2k}/\partial t$ of $\dot{\mathcal{V}}_k$ become all equal to $\partial N_2/\partial t$ evaluated at $\tmin_n$ and the integration limits become $\tmin_{n-k}$ and $\tmin_{n+1-k}$.
According to \eqref{eq:DotN2}, $\partial N_2(\tmin_n,s,c(\tmin_n),n)/\partial t$ equals $-\ddot{\phi}(\tmin_n-s)+\dot{\phi}_0\dot{\phi}(\tmin_n-s)$. Consequently
\begin{multline}
\lim_{R\downarrow c(\tmin_n)}\left(-\dot{\mathcal{V}}_k(t,R,n)\right)=
\int\limits_{\tmin_{n-k}}^{\tmin_{n+1-k}}\{\ddot{\phi}(\tmin_n-s)-\dot{\phi}_0\dot{\phi}(\tmin_n-s)\}\mathrm{d}s\\
=-\dot{\phi}((k-1)\Lambda)+\dot{\phi}(k\Lambda)+\dot{\phi}(0)\{\phi((k-1)\Lambda)-\phi(k\Lambda)\}.\label{eq:dotVk}
\end{multline}
The sum over $k$ of \eqref{eq:dotVk} with $\phi_0=1$ and $\dot{\phi}(\infty)=0$ reveals that
\begin{equation}
\sum\limits_{k=1}^\infty\left(\lim_{R\downarrow c(\tmin_n)}-\dot{\mathcal{V}}_k(t,R,n)\right)=-\phi_0\phi(\infty).
\end{equation}
All relevant results together lead to
\begin{equation}
\dot{\ell}(\tmin_n)+\ell(\tmin_n)\dot{\varphi}_0=-4\mathbb{G}(c(\tmin_n))\phi_0\phi(\infty).
\end{equation}
This equation combined with \eqref{eq:LoadAtMin} and $\dot{\varphi}_0=-\dot{\phi}_0$ proves that in the stationary phase
\begin{equation}\label{eq:LoadRateAtMin}
  \dot{\ell}(\tmin_n)=0,
\end{equation}
i.e., load and contact radius reach their minimum simultaneously.

\chapter{Matrix coefficients for the examples in chapter 7}
\section{Example 1 : SLS material \&  depth control}\label{app:MatrixCoefs}
The set of inhomogeneous equations \eqref{eq:DepthMatEqBasis} read
\begin{gather}
h_1\left(\cos(\tau_0)+\sum\limits_{j=0}^\infty\mathcal{F}_j(\tau_0)\right)=\mathbb{A}(R,c_\mathrm{max},h_1)\left(1+\sum\limits_{j=0}^\infty\mathcal{U}_j(\tau_0)\right),
        \label{eq:GenGRadius7A2}\\
        h_1\left(\cos(\tau_1)+\sum\limits_{j=1}^\infty\mathcal{F}_j(\tau_1)\right)=\mathbb{A}(R,c_\mathrm{max},h_1)\left(1+\sum\limits_{j=1}^\infty\mathcal{U}_j(\tau_1)\right).
       \label{eq:GenGRadius6A2}
\end{gather}
The functions $\mathcal{U}_j$ and $\mathcal{F}_j$ (see \eqref{eq:UjFunctieA} and \eqref{eq:FjFunctieA}), evaluated at $t=\tau_0=t_n+X$, where $t_n$ is a multiple of $2\pi$ and $t=\tau_1=\tau_0-Y$ with the use of
\begin{gather*}
\gamma_{2j}=-j(\tau_0-\tau_1)+2j\pi =-j Y+2j\pi, \\
\tau_{2j}=\tau_0-2j\pi=t_n+X(Y)-2j\pi,\\
\tau_{2j+1}=\tau_1-2j\pi=t_n+X(Y)-Y-2j\pi,\\
t_n=\text{multiple of $2\pi$,}
\end{gather*}
for $j=0,1,\ldots,$ shows that for standard linear solid material behaviour
\begin{gather}
\mathcal{U}_j(\tau_0)=-q \exp\{-\kappa Y\}\mathcal{A}(Y) Z^j(Y),\label{eq:UjFunctieC2}\\
\mathcal{U}_j(\tau_1)=-q\mathcal{A}(Y) Z^j(Y),\label{eq:UjFunctieC2A}\\
\mathcal{F}_j(\tau_0)=-q \exp\{-\kappa Y\}\mathcal{B}(X,Y)Z^j(Y),\label{eq:FjFunctieC1A}\\
\mathcal{F}_j(\tau_1)=-q \mathcal{B}(X,Y)Z^j(Y).\label{eq:FjFunctieC1}
\end{gather}
The functions $\mathcal{A}$ and  $Z$ only depend on $Y$, according to
\begin{gather}
\mathcal{A}(Y)=\int\limits_0^Y\exp(\kappa s)\,\mathrm{d}s=\exp(\kappa Y)-1,\label{eq:ZandAFunction1}\\
Z(Y)=\exp\bigl\{-\kappa\left(qY+2\pi(1-q)\right)\bigr\},\label{eq:ZandAFunction}
\end{gather}
whereas $\mathcal{B}$ additionally depends on $X$ through the trigonometric functions $\sin X$ and $\cos X$ as
\begin{equation}
\mathcal{B}(X,Y)=\kappa \int\limits_0^Y\exp(\kappa s)\cos(X-Y+s)\,\mathrm{d}s
\label{eq:BFunction1}
\end{equation}
which leads to
\begin{multline}
\mathcal{B}(X,Y)=\frac{\kappa}{1+\kappa^2}\Bigl\{\exp(\kappa Y)\{\kappa\cos X +\sin Y\}
-\cos Y\sin X\qquad\\
+\sin Y\cos X-\kappa\{\cos Y\cos X+\sin Y\sin X\}\Bigr\}.\label{eq:BFunction}
\end{multline}

Inspection of \eqref{eq:GenGRadius7A2}, \eqref{eq:GenGRadius6A2} and \eqref{eq:UjFunctieC2} to  \eqref{eq:FjFunctieC1}  shows that all the summations are geometric sequences with the exponential $Z$ as common ratio. The argument of this exponential ranges between $-2\kappa\pi$ and $-2\kappa\pi(1-q)$ and is always negative; the common ratio is smaller than one and
\begin{equation}
\sum_{j=0}^\infty Z^j(Y)=\frac{1}{1-Z(Y)},\qquad \sum_{j=1}^\infty Z^j(Y)=\frac{Z(Y)}{1-Z(Y)}.
\end{equation}
From all these results the next four  relations follow
\begin{gather}
1+\sum\limits_0^\infty\mathcal{U}_j(\tau_0)=1-q \exp(-\kappa Y)\mathcal{A}( Y) \frac{1}{1-Z( Y)},\label{eq:UjFunctieC3}\\
1+\sum\limits_1^\infty\mathcal{U}_j(\tau_1)=1-q\mathcal{A}( Y) \frac{Z( Y)}{1-Z( Y)},\label{eq:UjFunctieC4}\\
\cos\tau_0+\sum_0^\infty\mathcal{F}_j(\tau_0)=\cos X-q \exp(-\kappa Y)\mathcal{B}(X, Y)\frac{1}{1-Z( Y)},\label{eq:FjFunctieC2}
\end{gather}
\begin{multline}
\cos\tau_1+\sum_1^\infty\mathcal{F}_j(\tau_1)=\cos X\cos Y \\
+ \sin X \sin Y-q \mathcal{B}(X, Y)\frac{Z( Y)}{1-Z( Y)},\label{eq:FjFunctieC3}
\end{multline}
The right-hand sides of \eqref{eq:UjFunctieC3} and \eqref{eq:UjFunctieC4} represent the coefficients $b_1( Y)$ and $b_2( Y)$ appearing in the matrix equation \eqref{eq:DepthMatEqBasis1} from Chap.~\ref{sec:DynamicsGeneralVisco}. Upon substitution of the expressions for $\mathcal{A}$ and $Z$ from \eqref{eq:ZandAFunction} it is found that
\begin{gather}
b_1( Y)=1-q\frac{1-\exp(-\kappa Y)}{1-\exp\{-\kappa(q Y+2\pi(1-q))\}}\label{eq:B1Coef}\\
b_2( Y)=1-q\frac{\{\exp(+\kappa Y)\}\exp\{\kappa(q Y+2\pi(1-q))\}}{1-\exp\{\kappa(q Y+2\pi(1-q))\}}\label{eq:B2Coef}
\end{gather}
The other coefficients follow from the observations that the left-hand sides of the basic equations \eqref{eq:GenGRadius7A2} and \eqref{eq:GenGRadius6A2} are linear forms in the variables $\sin\tau_0=\sin X$ and $\cos\tau_0=\cos X$. The coefficients $m_{ij}$ in the matrix equation  \eqref{eq:DepthMatEqBasis1} can then be found by differentiating these forms -- actually the right-hand sides of \eqref{eq:FjFunctieC2} and \eqref{eq:FjFunctieC3} --  to $\sin X$ and $\cos X$, respectively. All this results in
 \begin{gather}
 m_{11}( Y)=\left(\frac{q\kappa}{1+\kappa^2}\right)\left(\frac{\{\cos Y+\kappa\sin Y\}\exp(-\kappa Y)-1}{1-\exp\{-\kappa(q Y+2\pi(1-q))\}}\right),
\label{eq:A11Coef}\\
 m_{12}( Y)=1+\left(\frac{q\kappa}{1+\kappa^2}\right)\left(\frac{\{\kappa\cos Y-\sin Y\}\exp(-\kappa Y)-\kappa}{1-\exp\{-\kappa(q Y+2\pi(1-q))\}}\right),
 \label{eq:A12Coef}\\
 m_{21}( Y)=\sin Y-\left(\frac{q\kappa}{1+\kappa^2}\right)\left(\frac{\cos Y+\kappa\sin Y-\exp(\kappa Y)}{1-\exp\{\kappa(q Y+2\pi(1-q))\}}\right),
 \label{eq:A21Coef}\\
 m_{22}( Y)=\cos( Y)-\left(\frac{q\kappa}{1+\kappa^2}\right)\left(\frac{\kappa\cos Y-\sin Y-\kappa\exp(\kappa Y)}{1-\exp\{\kappa(q Y+2\pi(1-q))\}}\right).
 \label{eq:A22Coef}
 \end{gather}
\section{Example 2 : SLS material \& load control}\label{app:LoadMatrixCoefs}
The starting equation is here \eqref{eq:LoadMatEqBasis1}, i.e.,
\begin{equation}
-\ell_1\begin{bmatrix}\cos\tau_1 +\sum\limits_{k=0}^\infty \mathcal{S}_k(\tau_1)\\
\cos\tau_2+\sum\limits_{k=1}^\infty \mathcal{S}_k(\tau_2)
\end{bmatrix}=\mathbb{B}(R,c_\mathrm{min},\ell_1)\begin{bmatrix}1+\sum\limits_{k=0}^\infty \mathcal{N}_k(\tau_1)\\
1+\sum\limits_{k=1}^\infty \mathcal{N}_k(\tau_2)\end{bmatrix}\label{eq:LoadMatEqBasis1A}
\end{equation}
The integrals \eqref{eq:NjFunctieA} and  \eqref{eq:SjFunctieA} for $\mathcal{N}_k$ and $\mathcal{S}_k$, respectively, taken at  $\tau_1$ and $\tau_2$, and with use of
$\tau_1=\tmin_n+X$, $\tau_2=\tmin_n+X-Y$, $\tau_{2k+2}=\tau_2-2k\pi$ and  $\tmin_n$ a multiple of $2\pi$, become
\begin{gather}
\mathcal{N}_k(\tau_1)=q \kappa Z_\ell(Y)^k\int\limits_0^Y\exp\{\kappa(1-q)(x-Y)\}\mathrm{d}x,\label{eq:NTau1}\\
\mathcal{N}_k(\tau_2)=q \kappa Z_\ell(Y)^k\int\limits_0^Y\exp\{\kappa(1-q)x\}\mathrm{d}x,\label{eq:NTau2}\\
\mathcal{S}_k(\tau_1)=q \kappa Z_\ell(Y)^k\int\limits_0^Y\exp\{\kappa(1-q)(x-Y)\}\cos(X-Y+x)\mathrm{d}x,\label{eq:STau1}\\
\mathcal{S}_k(\tau_2)=q \kappa Z_\ell(Y)^k\int\limits_0^Y\exp\{\kappa(1-q)x\}\cos(X-Y+x)\mathrm{d}x.\label{eq:STau2}
\end{gather}
As the function $Z_\ell(Y)$ which is defined by
\begin{equation}
Z_\ell(Y)=\exp\{-\kappa(2\pi-qY)\}
\end{equation}
is always positive and less than 1, the sums over $k$ are again geometric series with $Z_\ell$ as common factor and
\begin{equation}
\sum\limits_{k=0}^\infty Z_\ell(Y)^k=\frac{1}{1-Z_\ell(Y)}\quad \sum\limits_{k=1}^\infty Z_\ell(Y)^k=\frac{Z_\ell(Y)}{1-Z_\ell(Y)} \label{eq:Zell}
\end{equation}
Combination of \eqref{eq:LoadMatEqBasis1A},\eqref{eq:NTau1}, \eqref{eq:NTau2} and \eqref{eq:Zell} shows that the components $\hat{b}_i$ in \eqref{eq:LoadMatEqBasis2} are equal to
\begin{gather}
\hat{b}_1(Y)=1+\sum\limits_{k=0}^\infty \mathcal{N}_k(\tau_1)=1+\frac{q\Bigl(1-\exp\{-\kappa(1-q)Y\}\Bigr)}{(1-q)(1-Z_\ell(Y))},\label{eq:Bhat1}\\
\hat{b}_2(Y)=1+\sum\limits_{k=1}^\infty \mathcal{N}_k(\tau_2)=1-\frac{qZ_\ell(Y)\Bigl(1-\exp\{\kappa(1-q)Y\}\Bigr)}{(1-q)(1-Z_\ell(Y))}.\label{eq:Bhat2}
\end{gather}
The components $\hat{m}_{ij}$ in \eqref{eq:LoadMatEqBasis2} are determined by the integrals \eqref{eq:STau1} and \eqref{eq:STau2}. Upon evaluating these integrals and by defining
\begin{gather}
\mathcal{B}_s(Y)=\frac{\exp\{\kappa(1-q)Y\}-\cos Y-\kappa(1-q)\sin Y}{1+\kappa^2(1-q)^2}\\
\intertext{and}
\mathcal{B}_c(Y)=\frac{\kappa(1-q)\left(\exp\{\kappa(1-q)Y\}-\cos Y\right)+\sin Y}{1+\kappa^2(1-q)^2},
\end{gather}
it is eventually found that
\begin{gather}
\hat{m}_{11}(Y)=\frac{q \kappa\mathcal{B}_s(Y)\exp\{-\kappa(1-q)Y\}}{1-Z_\ell(Y)},\label{eq:hatm11}\\
\hat{m}_{12}(Y)=1+\frac{q \kappa\mathcal{B}_c(Y)\exp\{-\kappa(1-q)Y\}}{1-Z_\ell(Y)},\label{eq:hatm12}\\
\hat{m}_{21}(Y)=\sin Y+\frac{q \kappa\mathcal{B}_s(Y)Z_\ell(Y)}{1-Z_\ell(Y)},\label{eq:hatm21}\\
\hat{m}_{22}(Y)=\cos Y+\frac{q \kappa\mathcal{B}_c(Y)Z_\ell(Y)}{1-Z_\ell(Y)}. \label{eq:hatm22}
\end{gather}

\backmatter
\bibliographystyle{plainnat}
\bibliography{VarstViscoIndentationMain}

\end{document}